\documentclass[fleqn,usenatbib]{mnras}

\usepackage{newtxtext,newtxmath}

\usepackage[T1]{fontenc}

\DeclareRobustCommand{\VAN}[3]{#2}
\let\VANthebibliography\thebibliography
\def\thebibliography{\DeclareRobustCommand{\VAN}[3]{##3}\VANthebibliography}

\def\mpcoh{\,h^{-1}{\rm Mpc}} 
\def\gpcoh{\,h^{-1}{\rm Gpc}} 

\def\msunoh{\,h^{-1}{\rm M}_\odot}
\newcommand{\lyaf}{Ly-$\alpha$ forest\xspace}
\newcommand{\Lya}{Ly-$\alpha$\xspace}
\newcommand{\lya}{Ly-$\alpha$}

\newcommand{\vega}{\textsc{Vega}\xspace}
\newcommand{\abacus}{\textsc{AbacusSummit}\xspace}

\newcommand{\bk}{\mathbf{k}}
\usepackage{booktabs}

\usepackage{xspace}
\newcommand{\be}{\begin{equation}}
\newcommand{\ee}{\end{equation}}
\newcommand{\bea}{\begin{align}}
\newcommand{\eea}{\end{align}}
\newcommand{\beqa}{\begin{eqnarray}}
\newcommand{\eeqa}{\end{eqnarray}}

\newcommand\GG{\Gamma_3}

\newcommand{\bseq}{\begin{subequations}}
\newcommand{\eseq}{\end{subequations}}

\renewcommand{\ln}{\mathop{\rm ln}\nolimits}

\def\ltsima{$\; \buildrel < \over \sim \;$\xspace}
\def\gtsima{$\; \buildrel > \over \sim \;$\xspace}
\def\simlt{\lower.5ex\hbox{\ltsima}}
\def\simgt{\lower.5ex\hbox{\gtsima}}

\newcommand{\hinvMpc}{\,h^{-1}\, {\rm Mpc}\,}

\newcommand{\hMpcinv}{\,h\, {\rm Mpc}^{-1}\,}
\newcommand{\hinvGpc}{\,h^{-1}\, {\rm Gpc}\,}

\newcommand{\hMpc}{\, h\mathrm{Mpc}^{-1}\, }

\newcommand{\kpch}{\, h^{-1}\mathrm{kpc}\, }
\newcommand{\kmax}{\, k_{\rm max}\, }
\newcommand{\knl}{\, k_{\rm NL}\, }

\newcommand{\dd}{\partial}

\newcommand{\qvec}{\mathbf{q}}

\newcommand{\kvec}{\mathbf{k}}

\newcommand{\kpar}{k_{\parallel}}

\newcommand{\kvperp}{\mathbf{k}_{\perp}}
\newcommand{\apar}{\alpha_{\parallel}}
\newcommand{\aperp}{\alpha_{\perp}}

\usepackage{graphicx}	
\usepackage{amsmath}	

\title[\lyaf BAO shift in \textsc{AbacusSummit}]{Measuring and unbiasing the BAO shift in the Lyman-Alpha forest with \textsc{AbacusSummit}}

\author[B.~Hadzhiyska \& R.~de Belsunce et al.]{Boryana Hadzhiyska,$^{1,2}$\thanks{Both authors contributed equally to this work.}\thanks{\url{boryanah@berkeley.edu}}
Roger de~Belsunce,$^{1,3}$\thanks{\url{rbelsunce@berkeley.edu}}
A.~Cuceu,$^{1,4}$
J.~Guy,$^{1}$
M.~M.~Ivanov,$^{5,6}$
H.~Coquinot$^{1}$
\newauthor
and A.~Font-Ribera$^{7}$
\\
$^{1}$ Lawrence Berkeley National Laboratory, 1 Cyclotron Road, Berkeley, CA 94720, USA\\
$^{2}$ University of California, Berkeley, 110 Sproul Hall \#5800 Berkeley, CA 94720, USA\\
$^{3}$ Berkeley Center for Cosmological Physics, Department of Physics, UC Berkeley, CA 94720, USA\\
$^{4}$ Center for Cosmology and AstroParticle Physics, The Ohio State University, 191 West Woodruff Avenue, Columbus, OH 43210, USA\\
$^{5}$Center for Theoretical Physics, Massachusetts Institute of Technology, 
Cambridge, MA 02139, USA\\
$^{6}$The NSF AI Institute for Artificial Intelligence and Fundamental Interactions, Cambridge, MA 02139, USA\\
$^{7}$ Institut de F\'{i}sica d’Altes Energies (IFAE), The Barcelona Institute of Science and Technology, Campus UAB, 08193 Bellaterra Barcelona, Spain\\
}

\date{Accepted XXX. Received YYY; in original form ZZZ}

\pubyear{2025}

\begin{document}
\label{firstpage}
\pagerange{\pageref{firstpage}--\pageref{lastpage}}
\maketitle

\begin{abstract}
The currently observing Dark Energy Spectroscopic Instrument (DESI) places sub-percent constraints on measurements of the Baryon Acoustic Oscillations (BAO) scaling parameters from the Lyman-$\alpha$ (\Lya) forest. However, no systematic error budget stemming from non-linearities in the three-dimensional clustering of the \Lya forest is included in the DESI-\Lya analysis. In this work, we measure the size of the shift of the BAO peak using large \lyaf\ mocks produced on the $N$-body simulation suite \textsc{AbacusSummit}. Specifically, we measure the \lya\ auto-correlation and the \lya-quasar cross-correlation functions. We use the DESI \Lya forest fitting pipeline, \textsc{Vega}, together with the publicly available covariance matrix from the extended Baryon Oscillation Spectroscopic Survey data release 16 (eBOSS DR16) analysis. To mitigate the noise, we adopt a linear control variates (LCV) technique, reducing the error bars by a factor of up to $\sim \sqrt{50}$ on large scales. From the auto-correlation, we detect a small positive shift in radial direction of $\Delta\apar = 0.35\%$ at the 3$\sigma$ level and virtually no shift in the transverse direction, $\alpha_\perp$. 
From the cross-correlation, we see a similar shift to $\Delta\alpha_\parallel$, albeit with larger error bars, and a small negative shift, $\Delta\aperp=\sim$0.25\%, at the 2$\sigma$ level. 
We also make a connection with the \lyaf\ effective field theory (EFT) framework and find that the one-loop EFT power spectrum yields unbiased measurements of the BAO shift parameters in radial and transverse direction for \Lya auto and the \Lya-quasar cross-correlation measurements. When using the one-loop EFT framework, we find that we can recover the BAO parameters without a shift, which has important implications for future \lyaf\ analyses based on EFT. This work paves the way for novel full-shape analyses of the currently observing DESI and future surveys such as the PFS, WEAVE-QSO and 4MOST.
\end{abstract}

\begin{keywords}
cosmology: theory -- quasars: absorption lines -- methods: numerical
\end{keywords}



\section{Introduction}
\label{sec:intro}
By serving as a `standard ruler', the Baryon Acoustic Oscillations (BAOs) provide a powerful tool to measure cosmic distances and thus the expansion history of the Universe. These cosmological features, seen as an excess clustering at a comoving scale of approximately $100 \, h^{-1} \, {\rm Mpc}$, appear in both the two-point correlation function, as a single peak, and the power spectrum, as a series of peaks and troughs, for all tracers of the matter field. BAOs originate from early Universe acoustic density waves and photon-plasma interactions, and by observing them at various redshifts, we can measure cosmological parameters more precisely, especially when combined with additional probes such as Cosmic Microwave Background (CMB) data. Thus, the BAO scale contributes to refining the standard $\Lambda$CDM model and aids in constraining potential extensions of this framework \citep{Weinberg2013}.

The first independent detections of the BAO peak by the \textit{Sloan Digital Sky Survey} (SDSS) and the \textit{2dF Galaxy Redshift Survey} (2dFGRS) collaborations using galaxy clustering \citep{Eisenstein2005,Cole2005} validated predictions from standard cosmological theory. Since then, multiple detections of the acoustic peak in large-scale structure using various tracers have reinforced this framework \citep{Beutler2011,Anderson2012,Busca2013,Slosar2013,FontRibera2014,Delubac2015,Kitaura2016}. Recently, the Dark Energy Spectroscopic Instrument \citep[DESI,][]{2016arXiv161100036D}, a five-year survey aiming to measure BAO over a wide range of redshifts by obtaining accurate redshifts for over 40 million galaxies and quasars, has published seven BAO measurements at different redshifts as part of the Data Release 1 (DR1) analysis, including one from the \lyaf\ \citep{DESI:2024Lya}.

While the BAO peak position can be accurately predicted in linear theory by means of a Boltzmann solver \citep[e.g.][]{Lewis2000,Blas2011}, the gravitational growth of structure and redshift space distortions (RSD) induce non-linear effects on the BAO peak position, which are more challenging to compute from first principle. In particular, the non-linear evolution of the density field broadens the BAO, thus making the measurement less precise, and introduces a systematic sub-percent shift in the acoustic peak position, the magnitude and sign of which depend on the cosmological tracer under study. To tackle these problems, a plethora of reconstruction techniques, based on shifting cosmic structures to their non-evolved positions, have been developed and shown to successfully reduce the uncertainty on the BAO peak position without biasing it \citep[e.g.][]{Eisenstein2007,Schmittfull2015}. As for the shift to the BAO peak due to non-linear growth, various works have studied it both from a theoretical 
\citep{Crocce:2007dt,Eisenstein:2006nj,Padmanabhan:2009,Sherwin2012,McQuinn:2015tva,Blas:2016sfa,2022JCAP...02..008C,Chen:2024tfp, deBelsunce:2024rvv} and from an empirical \citep{Achitouv2015,Kitaura2016,Neyrinck2018,HernandezAguayo2020} perspective. 

The \lya\ forest, resulting from absorption features in the spectra of quasars due to the presence of neutral hydrogen along the line-of-sight, offers access to the high-redshift Universe: on scales of tens of kiloparsecs, gas pressure dominates the distribution of neutral hydrogen \citep{McQuinn2016}, while on larger scales the \lya\ forest traces the density fluctuations in the matter distribution \citep{Slosar2011}. Thus, through measurements of the BAO scale, it provides an invaluable lever arm of dark energy at cosmic noon. The first BAO measurement from the auto-correlation of the \lyaf\ was presented in \cite{Slosar2013,Busca2013}, whereas the cross-correlation of quasars and the \lyaf\ was presented in \cite{FontRibera2014}. Subsequent Sloan Digital Sky Survey (SDSS) data releases updated these measurements \citep{Delubac2015,Bautista2017,dMdB2017,Blomqvist2019,deSainteAgathe2019,dMdB2020}. The DESI \lyaf\ BAO measurement, which was presented in \citet{DESI:2024Lya}, doubled the number of lines of sight used in the last SDSS measurement and further reduced the uncertainties, bringing to the forefront the question of whether a systematic shift is present in the \lya\ BAO peak. Recently, \citet{2024ApJ...971L..22S} presented simulation-based measurements of the BAO feature using a large suite of 1,000 \Lya forest simulations with $1 \hinvGpc$ in volume centered at redshift $z=2.0$ and found a large negative shift in the BAO peak at the $\sim 1\%$ ($\sim -0.3\%$)-level in redshift (real) space, respectively. In addition, \citet{deBelsunce:2024rvv} analytically derive using perturbation theory how non-linearities in the three-dimensional clustering of the \Lya forest introduce biases in measurements of the BAO scaling parameters. Based on measurements of the bias parameters they \textit{predict} a theoretical error budget of -0.2\% (-0.3\%) for the radial (transverse) parameters at the same redshift. This work aims to bridge the gap between these two measurements. 

In this work, we study the shift in the BAO peak inferred from the clustering of the \lyaf, adopting the high-resolution \lyaf\ mocks built on the full $N$-body simulation suite \textsc{AbacusSummit}. Namely, we perform fits to measurements of the two-point auto- and cross-correlation function of the \lyaf\ and quasars employing the \textsc{Vega} model, used in the analysis of the DESI \lyaf\ data \citep{DESI:2024Lya}. To decrease the noise in our measurements of the correlation function, we also apply a linear control variates technique similar to the Zel'dovich control variates presented in \citet{2022JCAP...09..059K,2023JCAP...02..008D,2023OJAp....6E..38H}. Finally, we make a connection with the effective field theory (EFT) of large-scale structure \citep{Ivanov2024}. We fit the \Lya forest auto and \Lya--quasar cross-spectrum using the  one-loop EFT prescription and demonstrate that the EFT can jointly fit for the bias and BAO scaling parameters in a self-consistent manner.  

This paper is organized as follows. In Section~\ref{sec:meth}, we briefly describe the \textsc{AbacusSummit} \lyaf\ and quasar mocks, the \textsc{Vega} model and fitting procedure, and the linear control variates technique. In Section~\ref{sec:res}, we summarize our main results from performing fits on the auto- and cross-correlation function for a number of scale cut choices. In Section~\ref{sec:eft}, we draw a connection with EFT and test the ability of the model to obtain unbiased constraints on the BAO shift parameters. Finally, in Section~\ref{sec:disc}, we offer a discussion of our results in a broader context and conclude with an outlook to the upcoming fits to \lyaf\ data from DESI.

\section{Methods}
\label{sec:meth}

\subsection{\textsc{AbacusSummit}} \label{sec:abacus}

\textsc{AbacusSummit} is a suite of high-performance cosmological $N$-body simulations, which was designed to meet and exceed the Cosmological Simulation Requirements of the DESI survey \citep{2021MNRAS.508.4017M}. The simulations were run with \textsc{Abacus} \citep{2019MNRAS.485.3370G,2021MNRAS.508..575G}, a high-accuracy cosmological $N$-body simulation code, optimized for GPU architectures and  for large-volume simulations, on the Summit supercomputer at the Oak Ridge Leadership Computing Facility. 

The majority of the \textsc{AbacusSummit} simulations are made up of the \texttt{base} resolution boxes, which house 6912$^3$ particles in a $2\gpcoh$ box, each with a mass of $M_{\rm part} = 2.1 \times 10^9\msunoh$. While the \textsc{AbacusSummit} suite spans a wide range of cosmologies, here we focus on the fiducial outputs with \textit{Planck} 2018 cosmology: $\Omega_b h^2 = 0.02237$, $\Omega_c h^2 = 0.12$, $h = 0.6736$, $A_s = 2.0830 \times 10^{-9}$, $n_s = 0.9649$, $w_0 = -1$, $w_a = 0$. For full details on the simulation products, see \citet{2021MNRAS.508.4017M}. 

In this work, we utilize the \lyaf\ mocks generated on the \textsc{AbacusSummit} suite of simulations \citep{2023MNRAS.524.1008H}.
These mocks are generated on the 6 \texttt{base} boxes \texttt{AbacusSummit\_base\_c000\_ph\{000-005\}}, for which there is full particle outputs at $z = 2.5$, which falls within the redshift of interest of the DESI \lyaf\ analysis ($z \sim 2$ to 3). In Section~\ref{sec:lya_mocks}, we offer a brief summary of the methods for generating these mocks.

\subsection{Quasar catalogue}
\label{sec:quasar}

One of the major observables for \lyaf\ analyses is the cross-correlation function of the \lyaf\ with quasars. Therefore, studying the shift to the BAO peak for this observable is of great importance. In this work, we utilize the quasar catalogs generated as part of the \lyaf\ mock release on \textsc{AbacusSummit} \citep{2023MNRAS.524.1008H}. These are obtained via \textsc{AbacusHOD}, an ornamented halo occupation distribution (HOD) model, which incorporates various extensions affecting both the one- and two-halo terms. The full model is described in detail in \citet{2022MNRAS.510.3301Y}. 

The quasi-stellar object (QSO) mocks used here adopt a simple HOD model without any decorations:
\begin{align}
    \bar{n}_{\mathrm{cent}}^{\mathrm{QSO}}(M) & = \frac{\mathrm{ic}}{2}\mathrm{erfc} \left[\frac{\log_{10}(M_{\mathrm{cut}}/M)}{\sqrt{2}\sigma}\right], \label{equ:zheng_hod_cent}\\
    \bar{n}_{\mathrm{sat}}^{\mathrm{QSO}}(M) & = \left[\frac{M-\kappa M_{\mathrm{cut}}}{M_1}\right]^{\alpha}\bar{n}_{\mathrm{cent}}^{\mathrm{QSO}}(M),
    \label{equ:zheng_hod_sat}
\end{align}
where $M_{\mathrm{cut}}$ characterizes the minimum halo mass to host a central galaxy, $M_1$ the typical halo mass that hosts one satellite galaxy, $\sigma$ the steepness of the transition from 0 to 1 in the number of central galaxies, $\alpha$ the power law index on the number of satellite galaxies, {${\rm ic}$ the incompleteness parameter}, and $\kappa M_\mathrm{cut}$ gives the minimum halo mass to host a satellite galaxy. The parameters we choose for our QSO catalogs are in units of $\msunoh$
\begin{eqnarray}
    \log_{10}{(M_{\rm cut})} = 13.2, \ \
        \kappa = 1.11, \ \
        \sigma = 0.65, \\
        \log_{10}{(M_1)} = 13.8, \ \ 
        \alpha = 0.8, \ \ {\rm ic} = 1.0, \nonumber
\end{eqnarray}
which have been selected so as to yield a linear bias of about $b_{\rm QSO} \approx 3.3$, roughly matching the quasar bias in \citet{dMdB2020}. The number density of the catalog is about $1.75\times 10^{-4} \ [\mpcoh]^{-3}$. 

\subsection{\lyaf\ mocks}
\label{sec:lya_mocks}

Here, we summarize the methods used to generate the \textsc{AbacusSummit} \lyaf\ mocks, the goal of which is to support the full-shape analysis of the \lyaf\ power spectrum, planned to be conducted as part of the DESI Y3 \Lya\  science program. 

These mocks take the approach of painting gas on the fully evolved \textsc{AbacusSummit} $N$-body simulations via the approximate Fluctuating Gunn-Peterson Approximation (FGPA)  \citep{Croft:1998ApJ...495...44C}, calibrated to \lyaf\ skewers extracted from the IllustrisTNG hydro simulation \citep{2022ApJ...930..109Q}. The resolution of the mocks is 6912$^3$ cells per box, corresponding to an average of one particle per cell and a mean interparticle distance of 0.29 $\mpcoh$. The resolution is chosen to be comparable to  the Jeans length at that redshift ($\sim$100 kpc/$h$). Due to the simulation resolution, the power on scales below $\sim$0.3 $\mpcoh$ is suppressed. For this reason, we boost the power spectrum by adding small-scale fluctuations to the density field.

The optical depth is obtained by converting the dark matter density field into a neutral gas density field. Specifically, we adopt two different, but closely related approaches. The first one of them follows the standard FGPA prescription, while the second introduces a small modification to it: in particular, it attaches weights to the particles to construct the optical depth field directly in redshift space. This is in contrast with the first approach, for which we first construct the density field, then from it the optical depth field, and finally, we add the redshift space distortions (RSDs, see \citet{2023MNRAS.524.1008H} for details). In both cases, we add small-scale noise to compensate for the lack of small-scale power. We finally convert them to transmission flux spectra. 
The parameter values of the four models can be found in Table~\ref{tab:models}.

While the methods of \citet{2023MNRAS.524.1008H} are simplistic, they offer a fast and transparent way of connecting the matter density to that of the neutral hydrogen. Examples of more complex techniques include the \lya\ Mass Association Scheme (LyMAS; \citet{Peirani:2014,Peirani:2022}), the Iteratively Matched Statistics (IMS; \citet{Sorini:2016} method, Hydro-BAM \citep{2022ApJ...927..230S}, and cosmic-web-dependent FGPA \citep{2024A&A...682A..21S}. These use a variety of approaches tuned using smaller hydro simulations that range from matching the \lyaf\ probability distribution function and/or power spectrum to using a supervised machine learning method. 
We leave the application of more complex recipes for future work \citep[see e.g.,][]{Irsic:2018JCAP...04..026I}. 


\begin{table*}
\begin{center}
\begin{tabular}{c c c c c c c c c c c c c c}
 \hline\hline
Model \# & Method & Fit & $\langle F \rangle$ & $\sqrt{{{\rm Var}}[F]}$ & $\tau_0$ & $\sigma_\epsilon$ & $\gamma$ & $n$ & $k_1$ & $\chi^2_{{\rm 1D}}$ & $\chi^2_{{\rm 3D}}$ & $b_{{\rm Ly \alpha}}$ & $\beta_{{\rm Ly \alpha}}$ \\ [0.5ex]
 \hline
1 & Method I & P3D & 0.801 & 0.168 & 0.387 & 0.000 & 1.650 & --- & --- & 242.276 & 61.138 & $-$0.146 & 0.920 \\ [1ex]
2 & Method I & P1D+P3D & 0.811 & 0.187 & 0.391 & 0.772 & 1.450 & 1.000 & 0.063 & 27.536 & 104.887 & $-$0.129 & 0.949 \\ [1ex]
3 & Method II & P3D & 0.825 & 0.212 & 0.385 & 1.696 & 1.500 & 1.500 & 1.000 & 770.975 & 25.186 & $-$0.130 & 2.022 \\ [1ex]
4 & Method II & P1D+P3D & 0.810 & 0.187 & 0.654 & 2.116 & 1.550 & 0.000 & --- & 131.403 & 100.800 & $-$0.126 & 2.330 \\ [1ex]
 \hline
 \hline
\end{tabular}
\end{center}
\caption{Specifications of the four models used in the creation of the \lyaf\ synthetic catalogs \citep{2023MNRAS.524.1008H}. In particular, we indicate the model parameters, $\gamma$, $n$, $k_1$, $\tau_0$ and $\sigma_\epsilon$, as well as the clustering bias parameters, $b_{\rm Ly \alpha}$ and $\beta_{\rm Ly \alpha}$.}
\label{tab:models}
\end{table*}

\subsection{Correlation function measurements}
To compute the correlation function efficiently and accurately, we make use of the saved complex fields for each of the four models and 6 simulation boxes. In particular, those are low-pass filtered Fourier transforms of the full-resolution \lyaf\ and QSO fields up to $k_\parallel = 4 \ h/{\rm Mpc}$ and $k_\perp = 2 \ h/{\rm Mpc}$. As such, they contain information down to the Mpc scale. We then compute the 3D auto- and cross-power spectrum of the \lyaf\ with the quasars, zeroing out the $\mathbf{k} = 0$ mode\footnote{We found this step to be very important, as the $\delta_F = F/\bar F -1$ step performed when obtaining the complex $\delta_F$ field has some small vestigial noise due to the piecewise manner in which we obtain $\bar F$, such that $\langle \delta_F \rangle$ is small but non-zero.}. Looping over each mode in the line-of-sight direction, we pad the 3D power spectrum with zeros in the remaining two directions and inverse Fourier transform the field, saving the thus-obtained half-transformed 3D correlation function (also known as the cross-spectrum) only up to $r_\perp = 200 \ {\rm Mpc}/h$. We repeat the same operation for the line-of-sight direction and finally, bin the 3D correlation function into $r_\perp, \ r_\parallel$ bins from 0 to $4 \ {\rm Mpc}/h$ with bin size of $\Delta r = 4 \ {\rm Mpc}/h$. The end product is, thus, a (50, 50) matrix for each of the 4 models, 6 simulations and 2 lines-of-sight directions, with the measured $\xi(r_\perp, \ r_\parallel)$ correlation functions, which matches the format of the \lya\ measurements of both eBOSS and DESI.

Apart from the complex fields, full 3D maps (6912$^3$) of the redshift-space flux are also available. However, we opt not to use them, as that would make our measurements either less efficient or less accurate. In order to brute-force compute the correlation function using the configuration space fields, we would need to downsample the field (due to the large number of cells). However, we found that this induces sharp features into the measured correlation function. Since the fields are on a regular grid, other clever techniques could be employed such as identifying all possible pairs through geometric arguments. However, if the focus is on scales beyond $1 \ {\rm Mpc}/h$, such approaches would still be more computationally expensive than the Fourier transform approach. For measurements that require one to probe the sub-Mpc regime, those might be more suitable, as these small-scale Fourier modes are not saved for the \textsc{AbacusSummit} mocks.

\subsection{\textsc{Vega} model}
\label{sec:vega}

We use the \texttt{Vega}\footnote{Publicly available at \url{https://github.com/andreicuceu/vega}.} package to model the \lyaf\ correlation functions and measure the BAO position. This is the same package used by DESI, and our model is a simplified version of the one used in \cite{DESI:2024Lya}, that only includes \lyaf\ and quasar clustering, and does not include the effect of contaminants. The model is based on a template approach that starts with a linear isotropic matter power-spectrum, $P_{\rm fid}(k)$ computed using a fiducial cosmology based on the Planck CMB results \citep{Planck:2018}. This power-spectrum is decomposed into a peak (or wiggles) component, $P^p_{\rm fid}(k)$, and a smooth (or no-wiggles) component, $P^s_{\rm fid}(k)$, following \cite{2013JCAP...03..024K}. These two components undergo the rest of the modeling process independently and are combined at the end.

To model the effect of broadening of the BAO peak due to non-linear evolution, we follow \cite{2013JCAP...03..024K} and multiply the peak component, $P^p_{\rm fid}(k)$, by a Gaussian smoothing term $\exp[-k_{||}^2\Sigma_{||}^2/2 - k_\bot^2\Sigma_\bot^2/2]$. We treat $\Sigma_{||}$ and $\Sigma_\bot$ as nuisance parameters and allow them to vary in the fits.


The anisotropic model power spectrum is then given by:
\begin{equation}
    P_{A\times B}(k,\mu_k,z)=b_A b_B (1+\beta_A\mu_k^2) (1+\beta_B\mu_k^2) P_{QL}(k,z) F_{NL}(k, \mu_k, z),
\end{equation}
where (A, B) are either (\lya, \lya) for the auto-correlation, or (\lya, QSO) for the cross-correlation. The linear bias and redshift space distortion parameters, $b$ and $\beta$, are treated as nuisance parameters and marginalized over. For the peak component, $P_{QL}$ is given by the product of $P^p_{\rm fid}(k)$ with the Gaussian smoothing term that models the BAO broadening, while for the smooth component $P_{QL}=P^s_{\rm fid}$. The $F_{NL}$ term only applies to the \lya\ auto-correlation functions, and represents the model for small-scale non-linearities. We follow \citep{DESI:2024Lya} and use the empirical model developed by \cite{Arinyo-i-Prats:2015}.

The resulting anisotropic power spectrum model is decomposed into even multipoles up to $\ell=6$, which are then Fourier transformed into the equivalent correlation function multipoles. Finally, these multipoles are combined to compute the correlation function on the 2D grid in $r_{||}$ and $r_\bot$, along with the appropriate window function that reflects the rectangular bins of the 2D grid.

The BAO position is fitted when combining the peak and smooth components into the full correlation function model:
\begin{equation}
    \xi(r_{||},r_\bot) = \xi_s(r_{||},r_\bot) + \xi_p(\alpha_{||}r_{||},\alpha_\bot r_\bot),
\end{equation}
where $\xi_s$ and $\xi_p$ are the model correlation functions for the smooth and peak components, respectively. The BAO parameters, $(\alpha_{||},\alpha_\bot)$, rescale the coordinates of the peak component and are the target parameters we wish to measure.

Besides the physical model described above, we also add broadband polynomials to ensure the robustness of our BAO measurements. We follow \cite{DESI:2024Lya}, and use Legendre polynomials $L_j(\mu)$ of order $j=0,2,4$ and $6$, divided by powers of $r^i$ with $i=0,1,2$ (corresponding to a parabola in $r^2\xi(r)$). These polynomials are added to the final correlation function model, and have an extra 12 free parameters for each correlation function. 

\subsection{Fitting procedure}

Our fitting procedure follows that of \citet{DESI:2024Lya}, as implemented in the \texttt{Vega} package. We use a Gaussian likelihood and the \texttt{iminuit} minimizer to obtain the best fit parameter values. Our parameter vector includes the BAO parameters, $(\alpha_{||},\alpha_\bot)$, and 16 nuisance parameters: 12 associated with the broadband polynomials, and 4 associated with the \lya\ model, $(b_{\mathrm{Ly}\alpha}, \beta_{\mathrm{Ly}\alpha}, \Sigma_{||}, \Sigma_\bot)$. We use wide uninformative priors for all parameters. 

As we do not have estimates of the covariance matrices of our correlation function measurements, we use substitute covariances from real data for the fitting process, and then compute uncertainties using the population of BAO fits we obtain by fitting multiple simulations. The covariance matrices we use were measured by \citet{dMdB2020} from the SDSS DR16 data set. We note that for convenience we normalize the covariance matrix such that $\chi^2/{\rm d.o.f.} \sim 1$, but note that this normalization has no bearings on the constraints we get for the BAO shift parameters, as those are obtained by performing jackknifes of the measured $\alpha$ parameters.

\subsection{Linear control variates}
\label{sec:lcv}

To mitigate the noise of the measured correlation function, we employ an analytical implementation of the linear control variate technique (LCV) following \citet{2023OJAp....6E..38H}, which is a simplified version of the Zel'dovich CV \citep{2022JCAP...09..059K,2023JCAP...02..008D,2023OJAp....6E..38H}. The LCV technique allows us to subtract the noise from an expensive observable of interest as long as we can find an alternative (surrogate) quantity that is highly correlated with the quantity of interest. In our case, the surrogate is the linear density field at the initial conditions of the simulation, whereas the quantity of interest is the auto- and cross-correlation of the \lyaf\ with the quasars. In this work, we pioneer an application of analytical CV to the cross-correlation between \lyaf\ and quasars.

In the case of the auto-correlation of the \lyaf, we start with the LCV equation in Fourier space (as that is easier to manipulate):
\begin{equation}
    \hat{P}^{\ast, tt}_{\ell}(k) = \hat{P}^{tt}_{\ell}(k) - \beta_{\ell}(k) \left(\hat{P}^{ll}_{\ell}(k) - P^{ll}_{\ell}(k)\right) ,
\label{eq:lcv}
\end{equation}
where $\hat{P}^{tt}_{\ell}(k)$ is the measured \lyaf\ power spectrum, whereas $P^{ll}_{\ell}(k)$ and $\hat{P}^{ll}_{\ell}(k)$ are the analytical and measured power spectra predicted by linear theory. 

As done in \citet{2022JCAP...09..059K,2023JCAP...02..008D,2023OJAp....6E..38H}, we adopt the disconnected approximation for $\beta_\ell(k)$:
\begin{equation}
    \beta_{\ell}(k) = \left[ \frac{\hat{P}^{tl}_{\ell}(k)}{\hat{P}^{ll}_{\ell}(k)} \right]^2,
\label{eq:beta}
\end{equation}
where $\hat{P}^{tl}$ is the measured cross-power spectrum between the true and the modeled reconstructed fields. We apply damping on small scales and smoothing with a Savitsky-Golay filter. For estimates of the bias parameters, we use the \textsc{Vega} fits to each raw measurement; $\beta_\ell(k)$ is assumed to be uncorrelated with the measured \lyaf\ and linear theory power spectra.  

To make the linear theory prediction of the \lyaf\ power spectrum, we use the CLASS-generated \citep{2011arXiv1104.2932L} linear power spectrum for the \textsc{AbacusSummit} initial conditions, $P_{\rm lin}(k)$, and compute the theory-predicted multipoles, truncating at the hexadecapole, $\ell = 4$,
\begin{equation}
    P^{ll}_{\ell}(k) = b_{\rm Ly\alpha}^2 C_{\ell}(\beta_{\rm Ly \alpha}) P_{\rm lin}(k)
    \label{eq:p_ell}
\end{equation}
where
\begin{eqnarray}
C_{\ell}(\beta_{\rm Ly \alpha}) &\equiv& \frac{2\ell+1}{2}\int_{-1}^{1} d\mu\, \left(1 + \beta_{\rm Ly \alpha} \mu^2\right)^2 \mathcal{L}_{\ell}(\mu) \nonumber \\
&=&
\begin{cases}
    1 + \frac{2}{3}\beta_{\rm Ly \alpha} + \frac{1}{5}\beta_{\rm Ly \alpha}^2 & \ell = 0 \\
    \frac{4}{3}\beta_{\rm Ly \alpha} + \frac{4}{7}\beta_{\rm Ly \alpha}^2 & \ell = 2 \\
    \frac{8}{35}\beta_{\rm Ly \alpha}^2 & \ell = 4
\end{cases} \; .
\label{eq:c_ell}
\end{eqnarray}
Finally, we need to convert to configuration space in order to obtain the correlation function, $\xi(r_\parallel, r_\perp)$. As in the case of \citet{2023OJAp....6E..38H}, we work with three-dimensional quantities, and express the LCV-reduced power spectrum measurement as follows:
\begin{equation}
    \hat{P}^{\ast, tt}(\bk) = \hat{P}^{tt}(\bk) - \beta(\bk) \left(\hat{P}^{ll}(\bk) - P^{ll}(\bk)\right) ,
\end{equation}
where $\beta(\bk)$ and $P^{ll}(\bk)$ are expanded into the three-dimensional $\bk$-grid from their multipole counterparts. We then apply an inverse Fourier transform to the LCV-reduced three-dimensional field
\begin{equation}
    \hat{\xi}^{\ast, tt}(\mathbf{r}) = {\rm IFT}[\hat{P}^{\ast, tt}(\bk)]
\end{equation}
and bin into $\xi^{tt} =(r_\parallel, r_\perp)$ with $\Delta r = 4 \ {\rm Mpc}/h$ from 0 to $200 \ {\rm Mpc}/h$.

The procedure for obtaining the LCV-mitigated measurements of the cross-correlation function is analogous, with two main differences. The first one is that the linear theory prediction of the cross-power spectrum with QSOs is given by the following equation:
\begin{equation}
    {P}_{\times,\ell}(k) = b_{\rm Ly\alpha} \ b_{\rm QSO} C_{\times,\ell}(\beta_{\rm Ly\alpha}, \beta_{\rm QSO}) P_{\rm lin}(k)
\end{equation}
with
\begin{eqnarray}
C_{\times,\ell}(\beta, \beta_q) \equiv 
\begin{cases}
1 + \frac{1}{3}\beta + \frac{1}{3}\beta_{q} + \frac{1}{5}\beta \ \beta_{q} & \ell = 0 \\
\frac{2}{3}\beta + \frac{2}{3}\beta_{q} + \frac{4}{7}\beta \ \beta_{q} & \ell = 2 \\
\frac{8}{35}\beta \ \beta_{q} & \ell = 4
\end{cases} \;
\label{eq:c_ellq}
\end{eqnarray}
where $\beta \equiv \beta_{\rm Ly\alpha}$ and $\beta_q \equiv \beta_{\rm QSO}$. 

The second difference is that $\beta_{\times, \ell}(k)$, which encodes the amount of correlation between the observable of interest, $P^{ab}(k)$ (with `$a$' being \lya\ and `$b$' QSO), and its surrogate, $P^{cd}(k)$ (with `$c$' being the linear theory prediction for \lya\ and `$d$' the same thing for QSO), takes the following form in the disconnected approximation using the Knox formula \citep{1995PhRvD..52.4307K}:
\begin{equation}
    \beta_{\times, \ell}(k) =  \frac{\hat{P}^{ac}_{\ell}(k) \hat{P}^{bd}_{\ell}(k) + \hat{P}^{ad}_{\ell}(k) \hat{P}^{bc}_{\ell}(k)}{\hat{P}^{cc}_{\ell}(k) \hat{P}^{dd}_{\ell}(k) + \hat{P}^{cd}_{\ell}(k)^2}.
\end{equation}

\section{Results}
\label{sec:res}

In this section, we describe our main findings from performing fits to the \lyaf\ auto- and cross-correlation function with quasars.

\subsection{\lyaf\ auto-correlation scatter} \label{sec:fits_auto}

First, we study the fits to the raw measurements of the \lyaf\ auto-correlation function on the \textsc{AbacusSummit} mock catalogs.

\begin{figure*}
    \centering
        \includegraphics[width=0.45\linewidth]{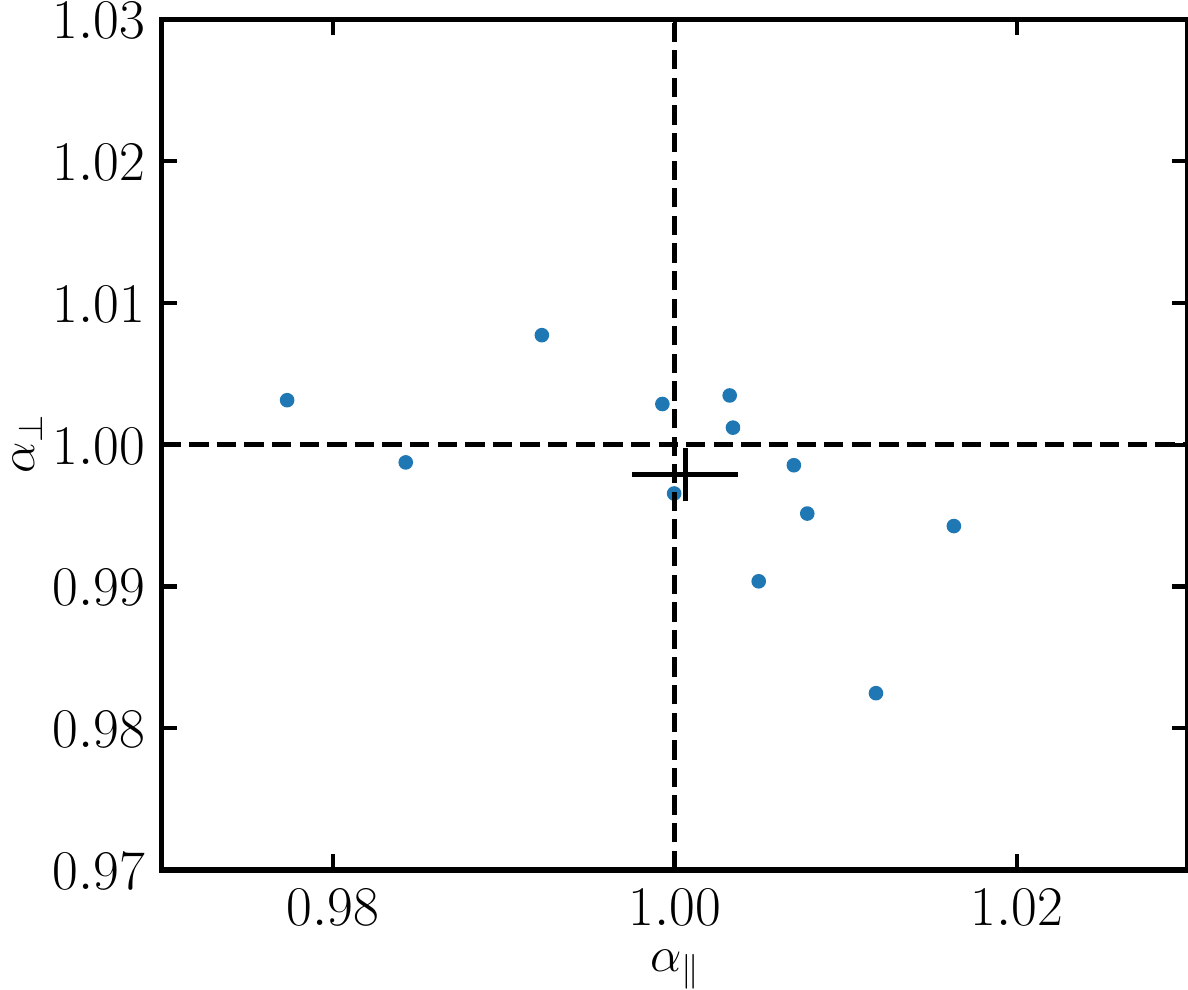}\hfill
        \includegraphics[width=0.45\linewidth]{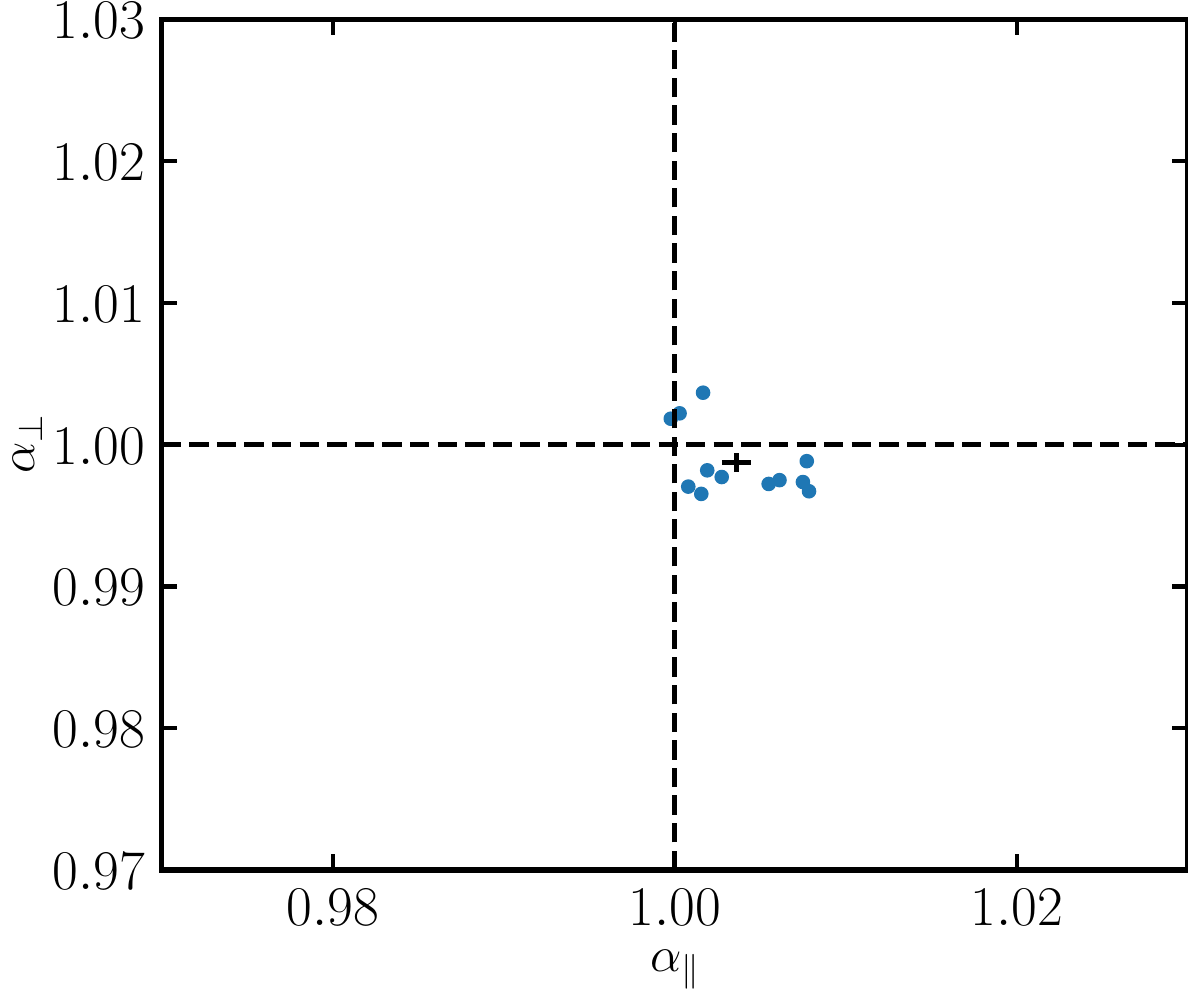}
        \vspace{-0.05in}
    \caption{Scatter plot of the BAO parameters, $\alpha_\parallel$ and $\alpha_\perp$, for Model 1, measured from the six \textsc{AbacusSummit} simulations with lines-of-sight along the $z$ and the $y$ axes. Fits are performed down to minimum isotropic separation of $r_{\rm min} = 30 \ {\rm Mpc}/h$. \textit{Left panel:} We see that the scatter between them is large, but the average value of the two parameters is consistent with no shift, with slight preference for $\alpha_\parallel > 1$ and $\alpha_\perp < 1$. \textit{Right panel:} Same set of fits but using the LCV technique to mitigate the noise on the measurements. The scatter in both parameters is significantly reduced, \textit{i.e.} the error bars on the measured $\alpha$'s are reduced by a factor of $\sim$3.}
    \label{fig:scatter_auto_combined}
\end{figure*}

In the left panel of Fig.~\ref{fig:scatter_auto_combined}, we show a scatter plot of the BAO parameters, $\alpha_\parallel$ and $\alpha_\perp$, for Model 1, measured from the 6 \textsc{AbacusSummit} simulations with lines-of-sight along the $z$ and the $y$ axes, for a total of 12 measurements. Thus, the rough volume of these mocks is $\sim$$100 \ ({\rm Gpc}/h)^3$. We see that the scatter between them is large, but the average value of the two parameters are consistent with 1, i.e. no shift. There seems to be a slight preference for $\alpha_\parallel > 1$ and $\alpha_\perp < 1$, which we examine closer in Section~\ref{sec:lcv_res} by reducing the errors through the use of the linear control variates (LCV) technique. The mean value across all simulations with its uncertainty is shown in black. We note that the size of the error bars, around 0.25\%, which is computed via the jackknife method, is larger than this shift when no noise reduction is applied. 
In this figure, we perform fits of the auto-correlation function of the \lyaf\ using the \textsc{Vega} model down to minimum isotropic separation of $r_{\rm min} = 30 \ {\rm Mpc}/h$. For details on the \textsc{Vega} model, see Section~\ref{sec:vega}.

\begin{table*}
\begin{center}
\begin{tabular}{c c c c c}
 \hline\hline
$\Delta \alpha_\parallel, \ \Delta \alpha_\perp$ & Model 1 & Model 2 & Model 3 & Model 4 \\ [0.5ex]
 \hline
$r_{\rm min} = 10 \ {\rm Mpc}/h$ & 0.43 $\pm$ 0.10\%, 0.20 $\pm$ 0.08\% & 0.42 $\pm$ 0.10\%, 0.18 $\pm$ 0.08\% & 0.38 $\pm$ 0.12\%, 0.38 $\pm$ 0.11\% & 0.36 $\pm$ 0.12\%, 0.41 $\pm$ 0.12\% \\ [1ex]
$r_{\rm min} = 20 \ {\rm Mpc}/h$ & 0.37 $\pm$ 0.10\%, -0.16 $\pm$ 0.13\% & 0.37 $\pm$ 0.10\%, -0.17 $\pm$ 0.08\% & 0.34 $\pm$ 0.12\%, 0.05 $\pm$ 0.11\% & 0.33 $\pm$ 0.12\%, 0.09 $\pm$ 0.12\% \\ [1ex]
$r_{\rm min} = 30 \ {\rm Mpc}/h$ & 0.37 $\pm$ 0.10\%, -0.13 $\pm$ 0.07\% & 0.36 $\pm$ 0.10\%, -0.14 $\pm$ 0.08\% & 0.34 $\pm$ 0.12\%, 0.06 $\pm$ 0.11\% & 0.33 $\pm$ 0.12\%, 0.08 $\pm$ 0.12\% \\ [1ex]
$r_{\rm min} = 40 \ {\rm Mpc}/h$ & 0.37 $\pm$ 0.12\%, -0.13 $\pm$ 0.13\% & 0.37 $\pm$ 0.10\%, -0.15 $\pm$ 0.08\% & 0.37 $\pm$ 0.12\%, 0.02 $\pm$ 0.11\% & 0.36 $\pm$ 0.12\%, 0.02 $\pm$ 0.12\% \\ [1ex]
 \hline
 \hline
\end{tabular}
\end{center}
\caption{Shifts from the linear theory values of the BAO parameters, $\alpha_\perp$ and $\alpha_\parallel$, defined as $\Delta \alpha \equiv \alpha - 1$. A positive value of $\Delta \alpha$, therefore, corresponds to a preference for $\alpha > 1$ and vice versa. Values of the best-fit $\alpha$'s are obtained by using the \textsc{Vega} model down to a minimum isotropic separation of $r_{\rm min}$, with noise mitigation obtained via the LCV technique. We adopt four different models, each with distinct values of the $b_{\rm Ly\alpha}$ and $\beta_{\rm Ly\alpha}$ parameters (see Table~\ref{tab:models}). We see 3$\sigma$ significant shifts to the value of $\alpha_\parallel$ for all values of $r_{\rm min}$ and all models of about 0.35$\pm$0.11\%. 
The shift of $\alpha_\perp$, on the other hand, is consistent with zero. We note that we disregard the $r_{\rm min} = 10 \ {\rm Mpc}/h$ case, as those values are affected by shortcomings of the model on small scales.}
\label{tab:alpha_auto_lcv}
\end{table*}

\begin{table*}
\begin{center}
\begin{tabular}{c c c c c}
 \hline\hline
$\Delta \alpha_\parallel, \ \Delta \alpha_\perp$ & Model 1 & Model 2 & Model 3 & Model 4 \\ [0.5ex]
 \hline
$r_{\rm min} = 10 \ {\rm Mpc}/h$ & 0.42 $\pm$ 0.17\%, -0.09 $\pm$ 0.08\% & 0.41 $\pm$ 0.18\%, -0.09 $\pm$ 0.09\% & 0.39 $\pm$ 0.20\%, -0.12 $\pm$ 0.10\% & 0.38 $\pm$ 0.20\%, -0.12 $\pm$ 0.11\% \\ [1ex]
$r_{\rm min} = 20 \ {\rm Mpc}/h$ & 0.48 $\pm$ 0.17\%, -0.27 $\pm$ 0.08\% & 0.48 $\pm$ 0.18\%, -0.28 $\pm$ 0.09\% & 0.44 $\pm$ 0.20\%, -0.24 $\pm$ 0.10\% & 0.43 $\pm$ 0.23\%, -0.24 $\pm$ 0.12\% \\ [1ex]
$r_{\rm min} = 30 \ {\rm Mpc}/h$ & 0.55 $\pm$ 0.18\%, -0.34 $\pm$ 0.08\% & 0.55 $\pm$ 0.18\%, -0.35 $\pm$ 0.09\% & 0.50 $\pm$ 0.20\%, -0.29 $\pm$ 0.10\% & 0.48 $\pm$ 0.20\%, -0.29 $\pm$ 0.11\% \\ [1ex]
$r_{\rm min} = 40 \ {\rm Mpc}/h$ & 0.56 $\pm$ 0.18\%, -0.32 $\pm$ 0.17\% & 0.63 $\pm$ 0.26\%, -0.36 $\pm$ 0.10\% & 0.54 $\pm$ 0.24\%, -0.29 $\pm$ 0.12\% & 0.51 $\pm$ 0.21\%, -0.31 $\pm$ 0.13\% \\ [1ex]
 \hline
 \hline
\end{tabular}
\end{center}
\caption{Same as Table~\ref{tab:alpha_auto_lcv}, except for the reported values are for the case of fits to the cross-correlation function with noise mitigation obtained via the LCV technique. We see that the $\alpha_\parallel$ shifts are a bit larger compared with the auto-correlation case, but the error barse are also about twice as big.  
The shift of $\alpha_\perp$ is also a bit more pronounced, about -0.3\% at around 3$\sigma$. We note that we disregard the $r_{\rm min} = 10 \ {\rm Mpc}/h$ case, as those values are affected by shortcomings of the model on small scales.}
\label{tab:alpha_cross_lcv}
\end{table*}

\begin{figure*}
    \centering
    \includegraphics[width=0.45\linewidth]{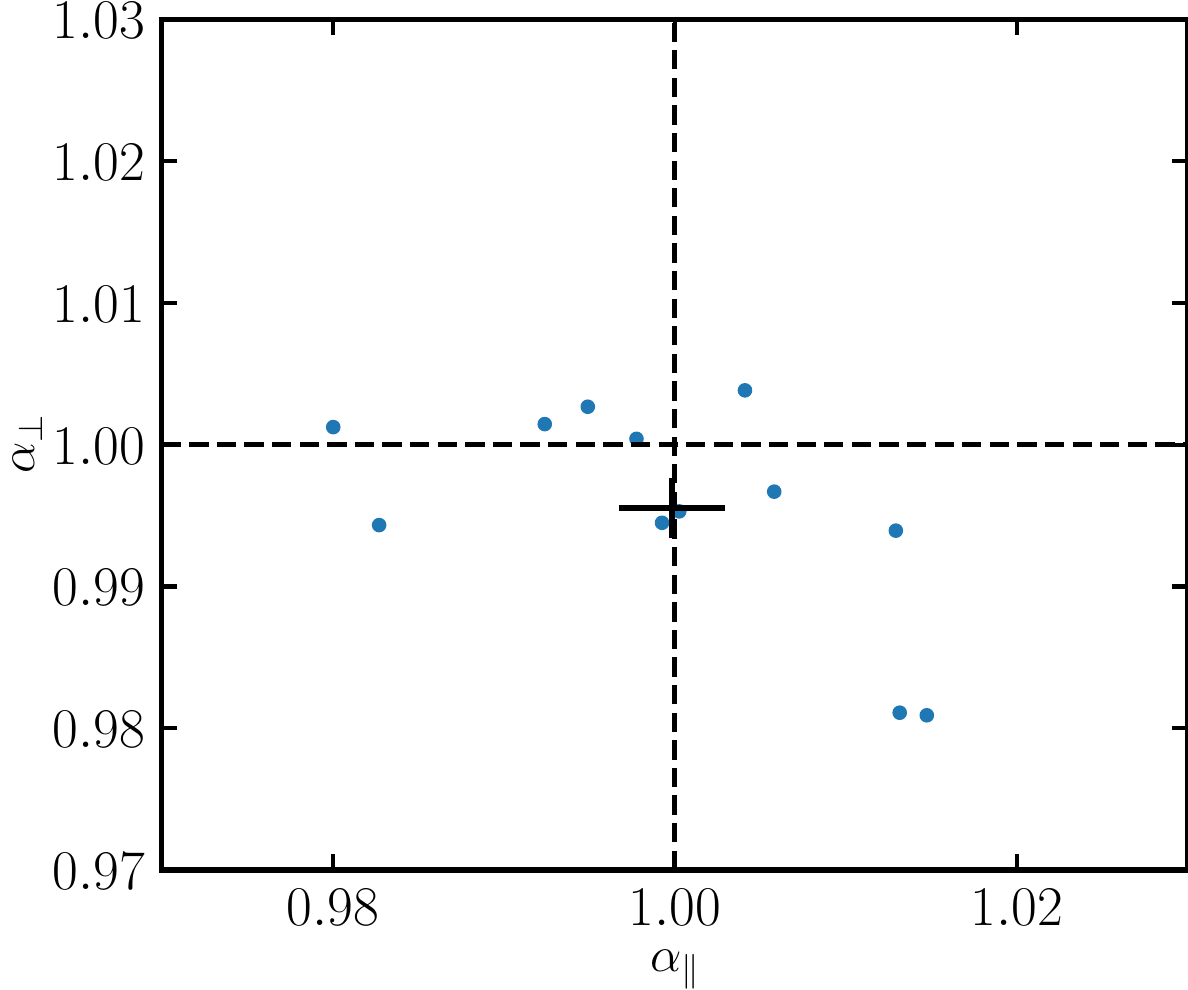}\hfill
    \includegraphics[width=0.45\linewidth]{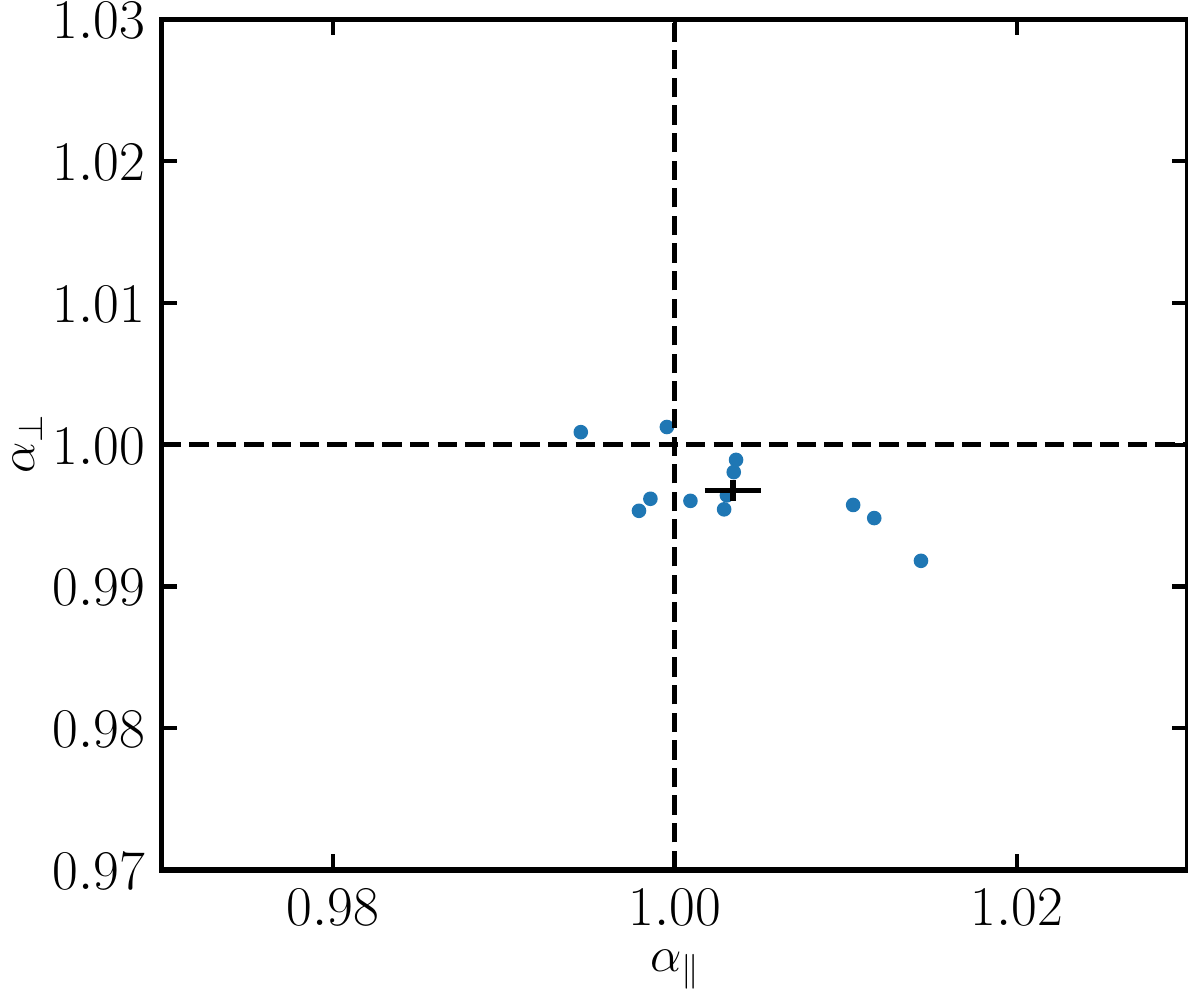}
    \vspace{-0.05in}
    \caption{Similar to Fig.~\ref{fig:scatter_auto_combined} except here, we show results from the cross-correlation function of the \lyaf\ with quasars. Fit to six simulations of model 1 with lines-of-sight $y$ and $z$.  \textit{Left panel:} The BAO parameter values are consistent with there being no shift, with $\alpha_\perp$ preferring slightly lower values, but still within the error bars, and $\alpha_\parallel$ preferring no shift. \textit{Right panel:} Fits of the cross-correlation function of the \lyaf\ with the quasars are done using the LCV technique to mitigate the noise on the measurements. The scatter is significantly reduced, following baseline expectation.}
    \label{fig:scatter_cross_combined}
\end{figure*}

\begin{figure*}
    \centering
    \includegraphics[width=\textwidth]{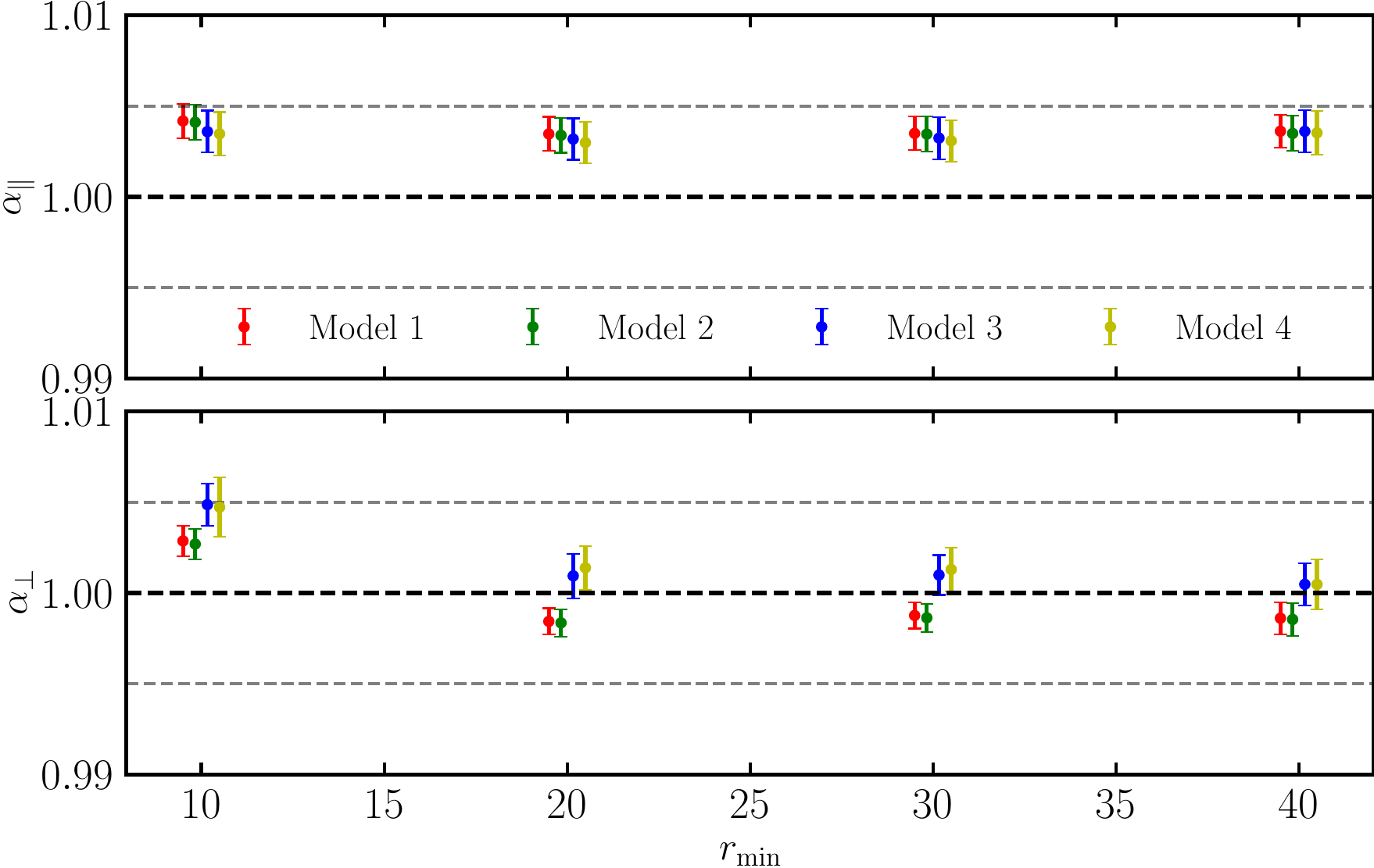}
    \vspace{-0.2in}
    \caption{Constraints on the BAO parameters $\alpha_\parallel$ and $\alpha_\perp$, for all four models, as a function of the minimum separation, $r_{\rm min}$, used in the \textsc{Vega} fits of the auto-correlation function of the \lyaf,  with noise mitigated by the linear control variates technique. 12 measurements are made for each of the 6 \textsc{AbacusSummit} simulations with lines-of-sight along the $z$ and the $y$ axes. There is a $\sim$3$\sigma$ shift of about 0.35\% to the $\alpha_\parallel$, while $\alpha_\perp$ is consistent with 1 within $\sim$1$\sigma$. The mild dependence that we see in $\alpha_\perp$ between $r_{\rm min} = 10 \ {\rm Mpc}/h$ and the larger $r_{\rm min}$ values can be attributed to issues with the small-scale modeling, which is expected to fail at $r \lesssim 10 \ {\rm Mpc}/h$. In App.~\ref{app:lcv}, we test in more detail the LCV outputs and demonstrate that the error bars shrink by almost a factor of 10 on large scales.}
    \label{fig:rmin_auto_lcv}
\end{figure*}


\begin{figure*}
    \centering
    \includegraphics[width=\textwidth]{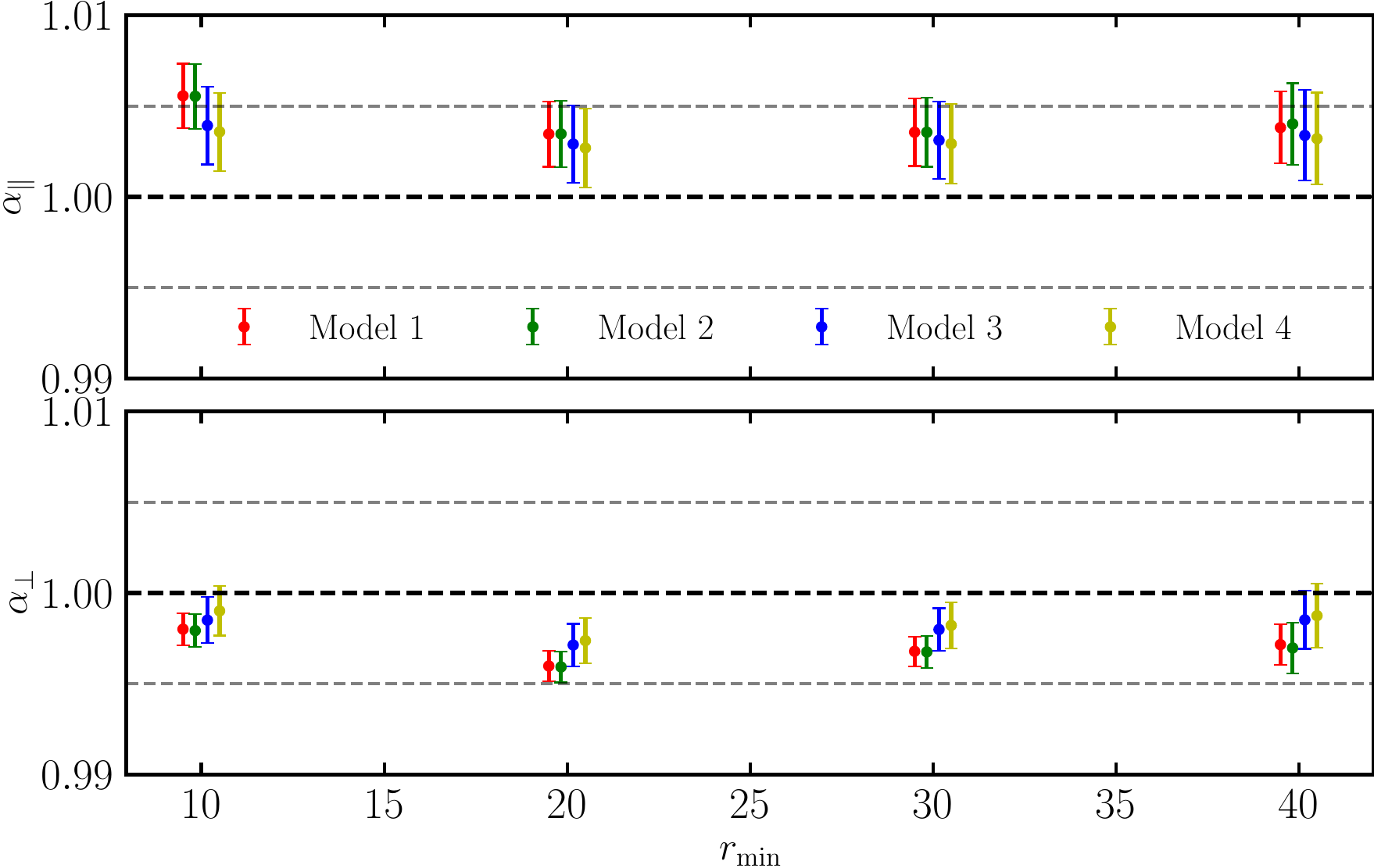}
    \vspace{-0.2in}
    \caption{Similar to Fig.~\ref{fig:rmin_auto_lcv}. The difference is that here, we show the BAO parameters as a function of minimum separation, $r_{\rm min}$, measured from fits of the cross-correlation function of the \lyaf\ with noise mitigated by the linear control variates technique. The size of the error bars has shrunk significantly. There is a small shift towards values larger than one in $\alpha_\parallel$ of about 0.3-0.4\% (though with slightly larger error bars), whereas $\alpha_\perp$ is slightly shifted (at the 0.2-0.3\% level) towards values smaller than one.}
    \label{fig:rmin_cross_lcv}
\end{figure*}

\subsection{\lyaf\ $\times$ QSO cross-correlation scatter}

Next, we move to the cross-correlation between \lya\ and quasars. 
We first study a scatter plot of the BAO parameters for all 12 measurements in the left panel of Fig.~\ref{fig:scatter_cross_combined}. We see that the values of both parameters are consistent with there being no shift, with $\alpha_\perp$ preferring slightly lower values, and $\alpha_\parallel$ preferring no shift. 
%
We note that the shifts of the two BAO parameters in the `cross' case do not need to match those in the `auto' case, as in the case of `cross' they depend on the behavior of both tracers: the \lyaf\ and the quasars, whereas in the case of `auto' they depend solely on the \lyaf\ (through predominantly the so-called $b_2$ bias parameter\footnote{The non-linear shift of the BAO parameters in the \Lya--quasar cross-correlation is driven by the ratio of the linear to the non-linear bias parameter
\begin{equation}
    \Delta \alpha \sim \left( \frac{b_2^q}{b_1^q} + \frac{b_2}{b_1} \right) \sigma^2_d\,.
    \label{eqn:cross_shift_schem}
 \end{equation}
The mode-coupling shift term is given in parentheses and divided by the size of the \textit{linear} BAO signal with the non-linear bias parameters $b_2$ (auto) and $b^q_2$ (cross).}) \citep{Ivanov2024, Chen:2024tfp, deBelsunce:2024rvv}. Nonetheless, the fact that the values of the $\alpha$'s are broadly similar reveals that neither of the two tracers is significantly biased at this level of precision. 

\subsection{Linear control variates fits}
\label{sec:lcv_res}

In this section, we apply the LCV method to the correlation function measurements, as described in Section~\ref{sec:lcv}, to obtain tighter constraints on $\alpha_\parallel$ and $\alpha_\perp$. Comparing the left to the right panel of Fig.~\ref{fig:scatter_auto_combined}, we see that the scatter is much smaller, demonstrating that the LCV technique has successfully reduced the noise in the inferred $\alpha$ parameters.

We next study, in Fig.~\ref{fig:rmin_auto_lcv}, how the constraints on the BAO parameters $\alpha_\parallel$ and $\alpha_\perp$, for all four models, change as a function of minimum separation, $r_{\rm min}$, used in the \textsc{Vega} fits of the auto-correlation function of the \lyaf. Interestingly, the behavior of all four models is largely consistent with each other despite the different values of $b_{\rm Ly\alpha}$ and $\beta_{\rm Ly\alpha}$ that each of them exhibits (see Table~\ref{tab:models}). We also note that Models 3 and 4 are constructed using a modified version of FGPA, but nonetheless, they yield very similar values to those extracted for Models 1 and 2. 

Looking at the $\alpha$ values, we see a $\sim$3$\sigma$ shift to the $\alpha_\parallel$, while $\alpha_\perp$ is consistent with one within $\sim$1$\sigma$. The shift to $\alpha_\parallel$ is in a consistent direction (towards $\alpha_\parallel > 1$) with what we expect from the left panel of Fig.~\ref{fig:scatter_auto_combined}. On average, we see that it is consistent for all four models and it is about 0.35\% with an error of around 0.11\% (see Table~\ref{tab:alpha_auto_lcv}). 

Additionally, we note that the values of the BAO parameters show very weak dependence on $r_{\rm min}$. This is expected, as the BAO peak occurs on much larger scales, $r \sim 100 \ {\rm Mpc}/h$ and therefore should not be sensitive to the choice of $r_{\rm min}$. In App.~\ref{app:bb}, we test that if we switch off the broadband parameters, the dependence on $r_{\rm min}$ is a bit more pronounced. This shows that the broadband model manages to absorb some of the issues on small scales: one of the most prominent one being that we keep the `Arinyo model' parameters fixed. In the future, we plan to study this in more detail through comparisons with more physically motivated frameworks such as EFT \citep{Ivanov2024}.


We see that  $\alpha_\perp$ appears to be mildly more sensitive to the model choice (in particular, between Models 1 and 2 and Models 3 and 4) and this sensitivity is underscored when adopting the LCV method, but still within 1$\sigma$ of each other. The mild dependence that we see in $\alpha_\perp$ between $r_{\rm min} = 10 \ {\rm Mpc}/h$ and the larger $r_{\rm min}$ values can be attributed to issues with the small-scale modeling, which is expected to fail at $r \lesssim 10 \ {\rm Mpc}/h$. For this reason, when quoting the final errors on the shifts, we quote them at $r = 30 \ {\rm Mpc}/h$, which does not appear to be affected by this. In App.~\ref{app:lcv}, we test in more detail the LCV outputs. We note that the effective volume increase is nearly a factor of 50 on large scales (note that the error bars go as $\sqrt V_{\rm eff}$), with this number going down on small scales. 

Next, we move on to applications of the LCV approach to the cross-correlation function with the quasars.  Reassuringly, comparing the left to the right panel in Fig.~\ref{fig:scatter_cross_combined}, we see that the scatter is much smaller, demonstrating that the LCV technique has successfully reduced the noise in the inferred $\alpha$ parameters. Looking at the plot of the BAO parameters as a function of minimum separation, $r_{\rm min}$, the size of the error bars has shrunk significantly in Fig.~\ref{fig:rmin_cross_lcv}. Similarly to the auto-correlation case, we see that there is a small shift towards values larger than one in $\alpha_\parallel$ of about 0.3-0.4\% (though with slightly larger error bars), whereas $\alpha_\perp$ is slightly shifted (at the 0.2-0.3\% level) towards values smaller than one. Neither of these is at a high significance likely due to the larger noise in the cross-correlation (LCV works less well and the quasars yield a noisier correlation function measurement due to their sparsity and larger non-linearities).

\section{Connection to Effective Field Theory}
\label{sec:eft}
The accessible amount of cosmological information from the \Lya forest can be increased by (i) cross-correlating the \Lya forest with quasars \citep{Font-Ribera:2013fha, dMdB2020} and (ii) including information beyond the BAO feature either through compressed statistics such as redshift-space distortions and the Alcock-Pacy\'nski effect (AP) \citep{Cuceu:2021hlk} or by using the broadband shape (\emph{i.e.} the full-shape) of the correlations \citep{Gerardi:2022ncj}. However, only recent advances in theoretical modeling using the effective field theory of large-scale structure (EFT; \citealt{Baumann:2010tm, Carrasco:2012cv}) extended to the \Lya forest \citep{Ivanov2024, Chudaykin:2025gsh} allow for a consistent model spanning large to intermediate scales to jointly extract the BAO information as well as the shape of the power spectrum (or correlation function). 

In the first part of the present work, we quantify the systematic error budget pertaining to the BAO peak shift stemming from using linear-theory templates to fit the BAO peak on \abacus simulations. In the second part (the present section), we introduce the effective field theory (EFT) model for the one-loop \Lya forest auto-correlation and for the cross-correlation with quasars (or halos) in Secs.~\ref{sec:EFT_theory}-\ref{sec:EFT_IR}. For the first time, we use EFT to fit the large and small scales of the \Lya forest directly whilst including information from the BAO feature.\footnote{Previous EFT analyses used hydrodynamical simulations with access to fewer quasi-linear modes than \abacus \citep{Givans:2022qgb,  Ivanov2024,deBelsunce:2024rvv,Chudaykin:2025gsh}.} Therefore, we jointly fit the BAO scaling and EFT parameters for the auto and subsequently the auto- and cross-correlation in Sec.~\ref{sec:EFT_likelihood}-\ref{sec:EFT_alpha}. We present two tests to connect both approaches in this work in Sec.~\ref{sec:EFT_consistency}: First, we use the quasi-linear theory fitting procedure introduced in Sec.~\ref{sec:meth} to fit the best-fit EFT spectra obtained from fits to the \abacus simulations. 
Second, we use EFT to 
\emph{theoretically predict} the physical shift of the BAO feature in the auto-correlation based on measurements of the bias parameters from \textsc{AbacusSummit} simulations \citep{Chen:2024tfp, deBelsunce:2024rvv}. 

\subsection{One-loop \Lya forest EFT model} \label{sec:EFT_theory}
An alternative approach to using phenomenological fitting functions from hydrodynamical simulations to model the \Lya forest, as described in Sec.~\ref{sec:vega}, is to model the \Lya forest using the effective field theory of large-scale structure (EFT)~\citep{Baumann:2010tm,Carrasco:2012cv,Ivanov:2022mrd}. The high redshift ($2 \leq z \leq 4$) of the \Lya forest yields access to more (quasi)-linear modes than, e.g., galaxy surveys, rendering it particularly fruitful for perturbative frameworks such as EFT~\citep[see, e.g.,~][]{McDonald:2009dh,Baumann:2010tm,Carrasco:2012cv}. 
EFT for the \Lya forest fluctuations describes large-scale dynamics using only the equivalence principle and rotations around the line of sight $\hat{z}$, denoted by the $SO(2)$ group~(\citet{2020PhRvD.102b3515G,Chen:2021rnb,Ivanov2024,Belsunce_skewspectrum}). In the following, we briefly summarize the one-loop \Lya EFT model. For a fuller presentation, we refer the reader to \cite{Desjacques:2018pfv,Ivanov2024,2024arXiv240513208I, deBelsunce:2024rvv}. 

Schematically, the one-loop EFT model of the \Lya forest for the auto-correlation consists of four components
\begin{equation} \label{eq:Pmodel}
    P^{\rm th.}(k,\mu) = P^{\rm tree}(k,\mu) + P^{\rm 1-loop}(k,\mu) + P^{\rm ct}(k,\mu) + P^{\rm st.}(k,\mu) \,,
\end{equation}
where $k$ is the Fourier wavenumber, $\mu$ the angle of $k=\{\kpar,\kvperp\}$ to the line-of-sight, $\mu \equiv \kpar/k$, $P^{\rm tree}(k)$ is the infrared resummed linear theory power spectrum\footnote{The linear power spectrum as well as the decomposition into an oscillatory and a smooth part, which are required for IR resummation, are obtained using the Boltzmann solver \texttt{CLASS-PT}~\citep{Diego_Blas_2011,Chudaykin:2020aoj}. We perform an anisotropic one-loop IR resummation of the power spectrum.}\footnote{We remind the reader that our convention of $b_{\eta}$ is related to the literature by a negative sign.} (details of IR resummation will be presented in Sec.~\ref{sec:EFT_IR}), which connects the redshift-space flux power to the linear matter power spectrum through~\citep{Kaiser:1987qv, McDonald:1999dt, McDonald:2001fe} 
\beqa \label{eq:IR-Kaiser}P^{\rm tree}(k,\mu) = K^2_1(\kvec) P_{\rm lin}(k)\,,\,\,\, K_1(\kvec)\equiv (b_1-b_{\eta}f\mu^2)\,,
\eeqa
where $f$ is the (linear) growth rate,
and $P^{\rm 1-loop}(k)$ is the one-loop \Lya power spectrum
\begin{align}
P^{\rm 1-loop}& (k,\mu) 
=2\int_{\qvec} K_2^2(\qvec,\kvec-\qvec)
P_{\text{lin}}(|\kvec-\qvec|)P_{\text{lin}}(q)  \nonumber\\
&\quad + 6 K_1(\kvec)P_{\rm lin}(k)\int_{\qvec} K_3(\kvec,-\qvec,\qvec)P_{\text{lin}}(q)\,.
\end{align} with higher order redshift-space kernels, $K_{2,3}$ \citep[see Eq.~(3.19) in][]{Ivanov2024} and we use the notation $\int_{\qvec}\equiv\int \frac{d^3q}{(2\pi)^3}$ to denote the three-dimensional integral over $\qvec$. The counter terms 
\begin{align}
\label{eq:shoch}
P^{\rm ct}(k,\mu) =
&-2(c_0+c_2\mu^2+c_4\mu^4)K_1(\kvec)k^2 P_{\rm lin}(k)\,,
\end{align} scale as $k^2P_{\rm lin}(k)$ and the stochastic contributions 
\begin{align}
P^{\rm st.}(k,\mu) =
&P_{\text{shot}}+a_0\frac{k^2}{\knl^2}+a_2\frac{k^2\mu^2}{\knl^2}\,,
\end{align}
scale as a constant shot noise piece and a scale-and angle-dependent term  capturing small-scale clustering. The non-linear scale $\knl$  is defined as the point where the dimensionless power spectrum reaches unity, i.e. $\Delta^2 = k^3P_{\rm lin}(k,z)/(2\pi^2)\sim 1$ which corresponds to  $k_{\rm NL} = 4.1 \hMpcinv$ at the redshift of the \abacus simulation. Beyond these scales the one-loop EFT model should be treated as phenomenological. The parameters $c_{0,2,4}$ control the angular dependence of the counter terms and $a_{0,2}$ are the so-called Wilson coefficients canceling the UV sensitivity \citep{Ivanov2024}. The final EFT model is evaluated at the redshift of the simulation, here $z=2.5$. For conciseness we suppress the explicit time dependence. 

The one-loop \Lya power spectrum is described by the following set of nuisance parameters \be \label{eq:nuissance_param} \{b_1, b_\eta, b_2, b_{\mathcal{G}_2}, b_{(KK)_\parallel}, b_{\Pi^{[2]}_\parallel}, b_{\delta \eta}, b_{\eta^2}\}\,,\ee in addition to the cubic EFT terms \be \label{eq:cubic_param} \{b_{\Pi^{[3]}_\parallel},b_{(K\Pi^{[2]})_\parallel},b_{\delta\Pi^{[2]}_\parallel},b_{\eta\Pi^{[2]}_\parallel},b_{\Gamma_3}\}\,,\ee and the counter and stochastic terms \be \label{eq:ct_param}  \{c_{0,2,4},a_{0,2},P_{\rm shot}\}\,,\ee yielding a total of 19 parameters. We adopt the prior choices of table 1 in \citet{deBelsunce:2024rvv} and analytically marginalize over the parameters in Eqs.~\eqref{eq:cubic_param}-\eqref{eq:ct_param}. 

\subsection{EFT model for \Lya -- quasar cross-correlation}\label{sec:EFT_theory_cross}
We extend the one-loop EFT model to include cross-correlations of the \Lya forest with quasars (or halos) \citep{Chudaykin:2025gsh}.\footnote{For \Lya -- halo cross-correlation analyses on hydrodynamic simulations using heuristic fitting functions, see,~\citet{Givans:2022qgb}.} In linear theory, this model has been computed as the geometric mean of the tree-level for both tracers, \textit{i.e.}
\be
P^{\rm tree}_{\times}(k, \mu) = (b_1 - b_{\eta}f \mu^2) (b_1^q + f \mu^2) P_{\rm lin}(k) \, ,
\ee
where the subscript $\times$ denotes the cross-correlation, and superscript $q$ represents the quasar (or halo) tracer, $b_1^q$ is the linear bias  parameter for quasars. To include higher-order contributions, we follow \citet{Chudaykin:2025gsh} and use the same structure as the auto-correlation in Eq.~\eqref{eq:Pmodel} for the one-loop \Lya -- quasar cross-correlation
\begin{align}\label{P1loopX}
P_{\times}^{\rm th.}(& k,\mu) = K_1(\textbf{k})K_1^{\rm q}(\textbf{k})P_{\rm lin}(k) \\
& +2\int_\qvec K_2(\qvec,\textbf{k}-\qvec)K_2^{\rm q}(\qvec,\textbf{k}-\qvec)
P_{\text{lin}}(|\textbf{k}-\qvec|)P_{\text{lin}}(q)  \nonumber\\
&+ 3 P_{\rm lin}(k)\int_\qvec [K_1(\textbf{k}) K_3^{\rm q}(\textbf{k},-\qvec,\qvec) \nonumber\\
&+ K_1^{\rm q}(\textbf{k}) K_3(\textbf{k},-\qvec,\qvec)]P_{\text{lin}}(q)\nonumber\\
&-(c_0+c_2\mu^2+c_4\mu^4)K_1^{\rm q}(\textbf{k})k^2P_{\rm lin}(k)\nonumber\\
&-(c^{\rm q}_0+c^{\rm q}_2\mu^2+c_4^q\mu^4)K_1(\textbf{k})k^2P_{\rm lin}(k)\nonumber\\
& -c_x^q(f\mu k)^4(K^q_1(\kvec))^2 P_{\rm lin}(k)\,,\nonumber
\end{align}
where $K_1^q(\kvec)\equiv (b_1^q+f\mu^2)$ and $K_{2,3}^q$ are the standard redshift space kernels for galaxies (see, e.g., \citet{Bernardeau:2002PhR...367....1B}).\footnote{To recover the expression for galaxies, one can remove the so-called line-of-sight operators from the \Lya one-loop model \citep{Desjacques:2018pfv,Desjacques:2016bnm}, yielding the description for quasars (or galaxies) by setting $b_{\eta}=-1, b_{\eta^{2}}=1,\,b_{\delta\eta}=-b_{1}$, $b_{(KK)_\parallel} = b_{\Pi^{[2]}_\parallel} = 0$, in addition to removing the cubic EFT terms in Eq.~\eqref{eq:cubic_param}.} The last term $c_x^q$ marginalizes over the estimated uncertainty due to the fingers-of-God modeling \citep{2020JCAP...05..042I,2021PhRvD.103d3525C}. The cross-correlation analysis is described by including three additionally sampled quasar bias parameters \be \label{eq:nuissance_param_qso} \{b_1^q,\, b_2^q,\, b_{\mathcal{G}_2}^q\}\,.\ee in addition to EFT terms that are analytically marginalized over, \textit{i.e.} \be \label{eq:cubic_param_qso} \{b_{\GG}^q,\, c_{0,2,4}^q,\,P_{\rm shot}^q, c_x^q\}\,,\ee yielding a total of nine parameters. For the cross-bias terms, we follow the prior choices of \cite{Chudaykin:2022nru}, \textit{i.e.} we use $b_1^q \in \mathcal{U}(0,4)$ and $\{b_{\mathcal{G}_2}^q,\, b_{\GG}^q\} \in \mathcal{N}(0, 2^2)$. We additionally, impose a simulation-based prior for $b_2^q \in \mathcal{U}(1.5, 1^2)$ \citep{Ivanov:2024_astrid}. For the counter terms we use $c_{0,2,4}^q/[h^{-1}\text{Mpc}]^2 \in \mathcal{N}(0, 10^2)$, and for the stochastic terms $P_{\rm shot}^q ,a^q_{0,2}\in \mathcal{N}(0, 1^2)\times \bar n^{-1}$ (note that we subtract the Poisson piece from the power spectrum, i.e. $P_{\rm shot}$
captures corrections to the Poisson sampling). 

\subsection{BAO scaling parameters}\label{sec:EFT_AP}
The key quantity of interest in the present work is to measure the position of the BAO peak \citep[see, e.g.,~][]{DESI:2024uvr, DESI:2024Lya}. By fitting (linear) BAO templates to two-point clustering measurements and re-scaling these templates in radial and transverse directions \citep{Padmanabhan:2008ag}, we extract the BAO scaling parameters, parameterized by $\boldsymbol{\alpha}\equiv \{\apar
, \aperp\}$,\footnote{These can also be recast into an isotropic and anisotropic component.}
\begin{align}
\alpha_{\parallel} \equiv \frac{H^{{\rm fid}}(z) r_s^{{\rm fid}}(z_d)}{H(z) r_s(z_d)}\,,\quad \alpha_{\perp} \equiv \frac{D_A(z) r_s^{{\rm fid}}(z_d)}{D_A^{{\rm fid}}(z) r_s(z_d)}\,,
\end{align}
where the superscript ``fid'' denotes the fiducial value at a reference cosmology. 
The BAO scaling parameters 
originate from a mismatch between the fiducial and true cosmologies giving rise to additional anisotropies when converting observed to projected coordinates, i.e.~the so-called Alcock-Paczy\'nski effect~\citep{Alcock:1979mp}. Both parameters measure distortions in radial and transverse directions encoding the Hubble parameter $H(z)$, the sound horizon at the baryon-drag epoch $r_s(z_d)$ and the angular diameter distance $D_A(z)$ at the effective redshift $z$ of the sample. This amounts to performing a re-mapping for the observed wavenumber
\beqa \label{eq: coord-rescaling-k}
	k \to k' \equiv k\left[\left(\frac{H_{\rm true}}{H_{\rm fid}}\right)^2\mu^2+\left(\frac{D_{A,\rm fid}}{D_{A,\rm true}}\right)^2(1-\mu^2)\right]^{1/2}\,,
\eeqa
and for the angle to the line-of-sight
\beqa \label{eq: coord-rescaling-mu}
	\mu \to\mu' \equiv \mu\left(\frac{H_{\rm true}}{H_{\rm fid}}\right)\left[\left(\frac{H_{\rm true}}{H_{\rm fid}}\right)^2\mu^2+\left(\frac{D_{A,\rm fid}}{D_{A,\rm true}}\right)^2(1-\mu^2)\right]^{-1/2}\,,
\eeqa
where unprimed quantities are those measured observationally. In the present analysis, this corresponds to fitting the power spectrum predictions as a function of $k'$ and $\mu'$, \textit{i.e.}, 
\begin{align}\label{eq:Pmodtogrid}
    P(k,\mu)&= \frac{1}{\aperp^2\apar} P(k',\mu')
\end{align} 
which is evaluated at the fiducial redshift of the simulations ($z=2.5$). 
The power spectrum decomposition into a smooth (nw) and oscillatory (w) component is done using a peak-average split in time-sliced perturbation theory which we introduce in the following~\citep{Blas:2016sfa,Chudaykin:2020aoj}.

\subsection{Non-linear evolution
of the BAO and IR-resummation} \label{sec:EFT_IR}

Eulerian perturbation
theory fails to capture the non-linear suppression of the 
BAO wiggles in the statistics 
of clustering observables. 
This happens because 
the Eulerian perturbative expansion contains 
the linear displacement,
which is enhanced 
by the soft (infrared, IR)
modes, making the  
convergence of the perturbative
series slow \citep{Crocce:2007dt}. 
Individual diagrams 
in Eulerian standard perturbation
theory contain IR 
enhanced terms, which cancel 
when added together \citep{Blas:2013bpa}. The cancellation
happens due to the equivalence 
principle, which 
dictates that physical 
observables are free from
IR divergences \citep{Blas:2015qsi}. The cancellation is, however, not complete
because of the BAO feature~\citep{Baldauf:2015xfa,Blas:2016sfa}. 
The procedure 
to perform resummation of the large 
IR terms in the BAO is called
``IR resummation''~\citep{Senatore:2014via}.
IR resummation depends on
the line-of-sight because 
the mapping from the rest frame
of the tracer to the 
observer's frame (commonly known as the redshift space mapping)
depends 
on the displacement 
along the line-of-sight,
which is IR enhanced. 
In the context of the \lyaf, the line-of-sight
operators are present
in the bias expansion to begin
with, and therefore 
it may not be immediately obvious 
that IR resummation will only
depend on operators generated
by the redshift space mapping. 
The bias expansion for the 
\lyaf, however,
depends only on the operators allowed by the equivalence 
principle, and these operators 
are IR safe. The
line-of-sight displacements relevant for IR resummation 
are IR unsafe, and therefore they
can only stem from the redshift 
space mapping of the rest-frame 
Lyman-$\alpha$ forest field. 
Thus, IR resummation for the Lyman-$\alpha$
forest depends on the same 
operators as a usual bias tracer
without selection effects. 
This means that the non-linear damping of the BAO in the 
Lyman-$\alpha$ forest will have 
the same structure as 
that of galaxies. 
Specifically, using \textit{time-sliced perturbation theory} at one-loop order we obtain~\citep{Ivanov:2018gjr} 
\be 
\label{eq:irres_rsd}
\begin{split}
P_{FF}=
& P^{FF}_{\rm tree}[e^{-\mathcal{S}(k,\mu)}(1+\mathcal{S}(k,\mu))P_{ w} + P_{nw}] \\
& +P^{FF}_{\rm 1-loop}[e^{-\mathcal{S}(k,\mu)}P_{ w} + P_{nw}]\,,
\end{split}
\ee 
where the damping functions are given by: 
\be 
\label{eq:damp_rsd}
\begin{split}
& \mathcal{S}(k,\mu) = k^2 \left(\Sigma^2(1+f\mu^2(2+f)) + \delta\Sigma^2f^2\mu^2(\mu^2-1) \right)\,,\\
& \Sigma^2 \equiv \frac{1}{6\pi^2}\int_0^{\Lambda_{\rm IR}}~dp P_{\rm lin}(p,z)[1-j_0(r_s p)+2j_2(r_s p)]\,,\\
&\delta \Sigma^2  = \int_0^{\Lambda_{\rm IR}} \frac{dp}{2\pi^2} P_{\rm lin}(p,z) j_2(p r_s)\,.
\end{split}
\ee 
where $r_s$ is comoving the BAO radius,
$j_{0,2}(x)$ are spherical Bessel functions of the 0th and 2nd order, 
while $\Lambda_{\rm IR}$ denotes
the formal
scale separating the domain of IR modes. 
We have tested that the our final 
results do not depend on the particular choice of $\Lambda_{\rm IR}$. In practice, 
we use $\Lambda_{\rm IR}=0.2~\hMpc$ as in~\cite{Blas:2016sfa}.
As for the one-loop power
spectrum, we approximate
the last term in 
Eq.~\eqref{eq:irres_rsd}
as
\be 
\label{eq:apr1loop}
\begin{split}
&P^{FF}_{\rm 1-loop}[e^{-\mathcal{S}(k,\mu)}P_{ w} + P_{nw}]\approx P^{FF}_{\rm 1-loop}[P_{nw}]+P^{FF}_{\rm 1-loop}\Big|_{w}\,,\\
& P^{FF}_{\rm 1-loop}\Big|_{w}\equiv e^{-\mathcal{S}(k,\mu)}(P^{FF}_{\rm 1-loop}[P_{\rm lin}] -P^{FF}_{\rm 1-loop}[P_{nw}])\,,
\end{split}
\ee 
which allows us to significantly 
speed up the calculation
of loop integrals. 
The approximation in Eq.~\eqref{eq:apr1loop}
is within the regime 
of validity of the assumptions 
made to derive Eq.~\eqref{eq:damp_rsd}
so formally it does not 
introduce any additional errors
at the given 
order in the IR resummation 
power counting. Let us note that in principle,
one could derive analogs of
Eqs.~\eqref{eq:apr1loop}
without explicitly 
separating the linear power
spectrum into wiggly and 
non-wiggly components, see the discussion in~\cite{2024arXiv240910609I}.

Our expression for the 
IR resummed 
cross-spectrum is completely 
analogous to Eq.~\eqref{eq:irres_rsd},
with the obvious substitution
$P^{FF}\to P^{FQ}$.

To derive the BAO parameters
from the EFT power spectrum (in Sec.~\ref{sec:EFT_AP}), we adopt a fitting methodology
analogous to \cite{DESI:2024uvr}: we AP rescale only the 
arguments of the 
wiggly part of the total
power spectrum $P^{\rm tree}_w+P^{\rm 1-loop}\Big|_w$, which ensures
that the information
we extract 
comes specifically 
from the BAO wiggles, 
and not from the broadband power. 
Any shift
in $\alpha_\parallel$,
$\alpha_\perp$
obtained this way 
should be interpreted as a shift
of the BAO scale. 

\subsection{Fitting methodology} \label{sec:EFT_likelihood}
We calibrate the EFT bias parameters for the one-loop \Lya power spectrum and BAO scaling parameters by fitting our model, $P^{\rm model}$, introduced in Secs.~\ref{sec:EFT_theory}-\ref{sec:EFT_theory_cross} to the measured power spectra from \abacus simulations, $P_i^{\rm data}$, by sampling the $\chi^2$ function\footnote{Note that we do not include a noise floor as in e.g.,~\citet{Givans:2022qgb} to use the full sensitivity of the simulations to constrain the bias parameters.}
\begin{equation} \label{eq:chi2}
    \chi^2 = \sum_i \frac{\left[P_i^{\rm data}-P^{\rm model}(k_i,\mu_i)\right]^2}{2 \left(P_i^{\rm data}\right)^2/N_i}\,,
\end{equation}
where $N_i$ are the Fourier modes per bin, $k_i$ the Fourier wavenumbers with the cosine of the angle to the line-of-sight, $\mu=\kpar/k$. The data spectra are measured with $\Delta k = 0.005 \hMpcinv$ and in ten angular bins with $\Delta \mu =0.1$. The data vector consists of the flux-flux spectrum, $P_{FF}$, and, when including the cross-correlation, the flux-quasar spectrum, $P_{FQ}$. For the fits we assume a Gaussian diagonal covariance: $\mathbf{C} = \text{diag}(\mathbf{C}^{\rm FF}, \mathbf{C}^{\rm FQ})$. The auto covariance is given by $C_{ii}^{\rm FF} = 2(P_i^{FF})^2/N_{i}$ and for the cross-spectrum covariance we use the form $C_{ii}^{\rm FQ} = N_{i}^{-1}\left[(P_{i}^{FQ})^2+P_i^{FF}P_i^{QQ}\right]$ where the quasar-quasar power spectrum is denoted by $P^{\rm QQ}$.\footnote{We leave including the cross-covariance $C_{ii}^{\times} = 2(P_i^{\rm FF}P_i^{\rm FQ})^2/N_{i}$ to future work \citep[see, e.g.,~][]{Chudaykin:2025gsh}.} For our baseline results, we use $\kmax=0.6\hMpcinv$ for the auto-correlation and $k_{\rm max}^{\times}=0.3\hMpcinv$ for the cross-correlation.\footnote{We verified that varying the scale cuts by $\Delta k_{\rm max}= \pm 0.2 \hMpcinv$ yielded consistent constraints on our final result on $\Delta \alpha_{\{\perp, \parallel\}}$, which following the scaling Universe argument \citep{Ivanov2024} and the non-linear scale of the simulations \citep{deBelsunce:2024rvv}, defined as $\Delta^2(k) = k^3P_{\rm lin}(k,z)/(2\pi^2) \approx 1$, is located at $\knl\approx2.2 \hMpcinv$, is a conservative scale cut. Beyond which the EFT should be treated as a purely phenomenological model.}

For the fits to the auto-correlation, we sample over ten parameters, eight of which are EFT parameters and the two BAO scaling parameters, and analytically marginalize over the remaining 11 parameters. For jointly fitting the auto- and cross-correlation, we sample over 12 parameters (two of which are the BAO scaling parameters) and analytically marginalize over 18 parameters. The priors are given in Sec.~\ref{sec:EFT_theory} for the auto and Sec.~\ref{sec:EFT_theory_cross} for the analysis including the cross-correlation. For the BAO scaling parameters $\apar$ and $\aperp$ we use $\apar,\aperp \in \mathcal{N}(1.0,0.5^2)$ as  priors. 

In the present analysis, we fix the cosmological parameters of our EFT model to the input values of the \abacus simulation, described in Sec.~\ref{sec:abacus} for which the linear input power spectrum is generated using the Boltzmann solver \texttt{CLASS-PT}~\citep{Diego_Blas_2011,Chudaykin:2020aoj}. We explore the parameter space using the Markov Chain Monte Carlo (MCMC) sampler \texttt{Cobaya}~\citep{Torrado:2020dgo}. Each fit to an \abacus simulation takes $\sim 0.20$ CPU-hours (using one AMD Milan CPU on the Perlmutter computer at NERSC) and is considered converged once the Gelman-Rubin diagnostic reaches $R -1 < 0.01$ for all parameters~\citep{1992StaSc...7..457G}.

\subsection{BAO and EFT parameters from \abacus} \label{sec:EFT_alpha}
\begin{table*}
    \begin{center}
    \begin{tabular}{c c c c c}
     \hline\hline
    $b_{\mathcal{O}}$ & Model 1 & Model 2 & Model 3 & Model 4  \\[1ex]
     \hline
    $b_1$ & $-0.1463^{+0.0016}_{-0.0014}$ & $-0.1298^{+0.0015}_{-0.0013}$ & $-0.1311 \pm 0.0012$ & $-0.1273 \pm 0.0011$ \\[1ex]
    $b_{\eta}$ & $0.1448 \pm 0.0027$           & $0.1298 \pm 0.0026$           & $0.2734 \pm 0.0030$ & $0.3073 \pm 0.0031$ \\[1ex]
    $b_2$ & $-0.0292^{+0.1963}_{-0.1749}$ & $-0.1211^{+0.1861}_{-0.1823}$ & $0.0304^{+0.1145}_{-0.1050}$ & $-0.0061 \pm 0.1039$ \\[1ex]
    $b_{\mathcal{G}_2}$ & $-0.0364 \pm 0.1421$          & $0.0241^{+0.1346}_{-0.1782}$  & $0.0326 \pm 0.0784$ & $0.0175 \pm 0.0735$ \\[1ex]
    $b_{\eta^2}$ & $-0.1199 \pm 0.0556$          & $-0.3537^{+0.0778}_{-0.0546}$ & $-0.1253 \pm 0.0539$ & $-0.0800 \pm 0.0543$ \\[1ex]
    $b_{\delta \eta}$ & $-0.3543^{+0.1384}_{-0.1096}$ & $-0.6961^{+0.1612}_{-0.0513}$ & $-0.1398^{+0.0965}_{-0.0786}$ & $-0.1195^{+0.0978}_{-0.0786}$ \\[1ex]
    $b_{(KK)_\parallel}$ & $-0.0403^{+0.3311}_{-0.2811}$ & $-0.3424^{+0.4923}_{-0.2546}$ & $-0.0785 \pm 0.1947$ & $0.0007^{+0.1799}_{-0.1986}$ \\[1ex]
    $b_{\Pi^{[2]}_\parallel}$ & $-0.3146^{+0.0583}_{-0.0469}$ & $-0.2479 \pm 0.0531$          & $-0.2950^{+0.0590}_{-0.0527}$ & $-0.3425 \pm 0.0588$ \\[1ex]
    $\apar$ & $1.0040 \pm 0.0094$           & $1.0020 \pm 0.0102$           & $0.9994 \pm 0.0084$ & $0.9994 \pm 0.0081$ \\[1ex]
    $\aperp$ & $1.0011^{+0.0039}_{-0.0034}$  & $1.0004^{+0.0048}_{-0.0035}$  & $1.0014^{+0.0038}_{-0.0033}$ & $1.0005 \pm 0.0032$ \\[1ex]
    \hline
    $b_{\Pi^{[3]}_\parallel}$ & $0.0420 \pm 0.0486$           & $-0.0392 \pm 0.0443$          & $0.7338 \pm 0.0447$ & $0.8240 \pm 0.0450$ \\[1ex]
    $b_{\delta\Pi^{[2]}_\parallel}$ & $-1.6886 \pm 0.1442$          & $-1.4214 \pm 0.1335$          & $-0.9226 \pm 0.1722$ & $-1.1403 \pm 0.1795$ \\[1ex]
    $b_{(K\Pi^{[2]})_\parallel}$ & $-1.1586 \pm 0.1654$          & $-0.8222 \pm 0.1479$          & $-1.5634 \pm 0.1531$ & $-1.9723 \pm 0.1530$ \\[1ex]
    $b_{\eta\Pi^{[2]}_\parallel}$ & $-0.9338 \pm 0.4428$          & $-0.4698 \pm 0.4156$          & $-1.9866 \pm 0.5684$ & $-2.2649 \pm 0.5922$ \\[1ex]
    $b_{\GG}$ & $0.4228 \pm 0.0836$           & $0.4802 \pm 0.0763$           & $-0.3260 \pm 0.0947$ & $-0.3680 \pm 0.0971$ \\[1ex]
    $P_{\rm shot}$ & $-0.3066 \pm 0.0970$          & $0.1080 \pm 0.0757$           & $0.3826 \pm 0.0379$ & $0.2364 \pm 0.0328$ \\[1ex]
    $a_0$ & $-0.0109 \pm 0.8768$          & $-0.2250 \pm 0.7127$          & $-0.5768 \pm 0.7067$ & $-0.0894 \pm 0.6710$ \\[1ex]
    $a_2$ & $-0.6411 \pm 2.6056$          & $-2.6171 \pm 2.2136$          & $-4.3836 \pm 2.9013$ & $-5.6704 \pm 2.9223$ \\[1ex]
    $c_0$ & $0.1425 \pm 0.0351$           & $-0.0449 \pm 0.0301$          & $-0.1689 \pm 0.0171$ & $-0.1158 \pm 0.0163$ \\[1ex]
    $c_{2}$ & $-0.0882 \pm 0.0602$          & $0.1304 \pm 0.0558$           & $0.3216 \pm 0.0578$ & $0.3072 \pm 0.0573$ \\[1ex]
    $c_{4}$ & $-0.1559 \pm 0.0278$          & $-0.2293 \pm 0.0259$          & $-0.4255 \pm 0.0334$ & $-0.4418 \pm 0.0346$ \\[1ex]
    $\chi^2$ & $1202 $         & $1255$         & $1276 $ & $1281$ \\
     \hline \hline
    \end{tabular}
    \end{center}
    \caption{Mean best-fit values for the one-loop EFT parameters obtained from the \textsc{AbacusSummit} simulation ``one'' and line-of-sight parallel to the y-axis using models one to four. The default fit is performed with $\kmax=0.6\hMpcinv$ for the auto-correlation. The last two rows of the upper section of the table quotes the BAO scaling parameters which are consistent with unity, \textit{i.e.} the EFT model debiases the BAO measurement. We analytically marginalize over the parameters shown in the bottom part of the table and recover their posteriors from the chains \textit{a posteriori}.  Note that the counter terms ($c_{0,2,4}$) are divided by $(\hMpcinv)^2$.  The resulting $\chi^2$ and reduced $\chi^2_{\nu}$ for the best-fit linear parameters are quoted in the last two rows for 1175 data points and 18 degrees of freedom. In App.~\ref{app:triangle_plots} we show the triangle plot for the sampled parameters for the first column. In Table~\ref{tab:EFT_bias_meandatavector} we show the corresponding EFT bias parameters for the fits to the mean data vector.}
    \label{tab:EFT_bias}
\end{table*}
In this section, we present results on fits to the \Lya forest auto-correlation and \Lya--quasar cross-correlation measured on \textsc{AbacusSummit} simulations, introduced in Sec.~\ref{sec:lya_mocks}. The large size of the simulation boxes at redshift $z=2.5$ gives access to many quasi-linear modes which, in turn, yield tight constraints on the (linear) bias parameters. We present the results for all four models for simulation ``one'' and line-of-sight $y$ on the auto-correlation in Table~\ref{tab:EFT_bias} and for the cross-correlation in Table~\ref{tab:EFT_bias_cross}. Our measured linear bias parameters are consistent with the linear-theory results presented in \cite{2023MNRAS.524.1008H} at the $\sim 1 \sigma$ level. For both series of fits, we find BAO scaling parameters that are consistent with unity showing that the EFT model debiases the BAO fit. In agreement with the auto-correlation analysis on ACCEL$^2$ hydrodynamical simulations of box size $L=160\hinvMpc$ with a physical resolution down to $25\kpch$ in \citet{deBelsunce:2024rvv}, we detect the shot noise for all four models whilst we do not find evidence for the Wilson coefficients $a_0$ and $a_2$ which cancel  the UV sensitivity of the loop integrals. However, we detect the $c_{0,2,4}$  counter terms which account for the short range non-locality \citep{McDonald:2009dh}. We note that since the quasar field does not change between the four models, we would expect the counter terms $c_{0,2,4}^q$ to be in agreement with each other. Whilst the counter terms between models 1 and 2 as well as 3 and 4 are consistent with each other, we find a tension between those two sets of models (up to $\sim 3\sigma$). The presence of this tension suggests that the counter terms are affected by degeneracies which are not (fully) removed by including the \Lya - quasar cross-correlation. We leave including the quasar-autocorrelation to future work. 

For the auto-correlation the quality of the fits ($\chi^2=1202,\,1255,\,1276,\,1281$) is acceptable for 1175 data points with an increasing $\chi^2$ for the models one to four. The quality of the fits for the cross-correlation is somewhat worse ($\chi^2=1791,\,1839,\,1894,\,1901$) for 1755 data points.\footnote{We have verified that removing the counter terms from the cross-correlation fits did not substantially affect the quality of the fits.}\footnote{We caution the reader that since we are analytically marginalizing over the parameters that linearly enter the EFT model (see Eq.~\eqref{eq:ct_param}) the counting of degrees of freedom requires some care \citep{Bridle:2001zv,Taylor2010}. }

For completeness, we give the numerical values of the EFT fits to the auto-correlation mean data vector in Table~\ref{tab:EFT_bias_meandatavector} and for the joint fits of the auto-and cross-correlation in Table~\ref{tab:EFT_bias_meandatavector_cross}. We note that we find consistent, yet tighter, results than fits to individual simulations given in Table~\ref{tab:EFT_bias}. The fits to the mean data vector can not only differentiate between the linear biases but also between the non-linear bias parameters. In particular, we find clear differences in $b_2$ between the four models which, in turn, is a driver for the non-linear shift of the BAO peak.\footnote{Since the error bars are divided by an additional factor of $\sqrt{6}$, the resulting $\chi^2$ values should be taken with a grain of salt.}

\begin{figure*}
    \centering
    \includegraphics[width=0.49\linewidth]{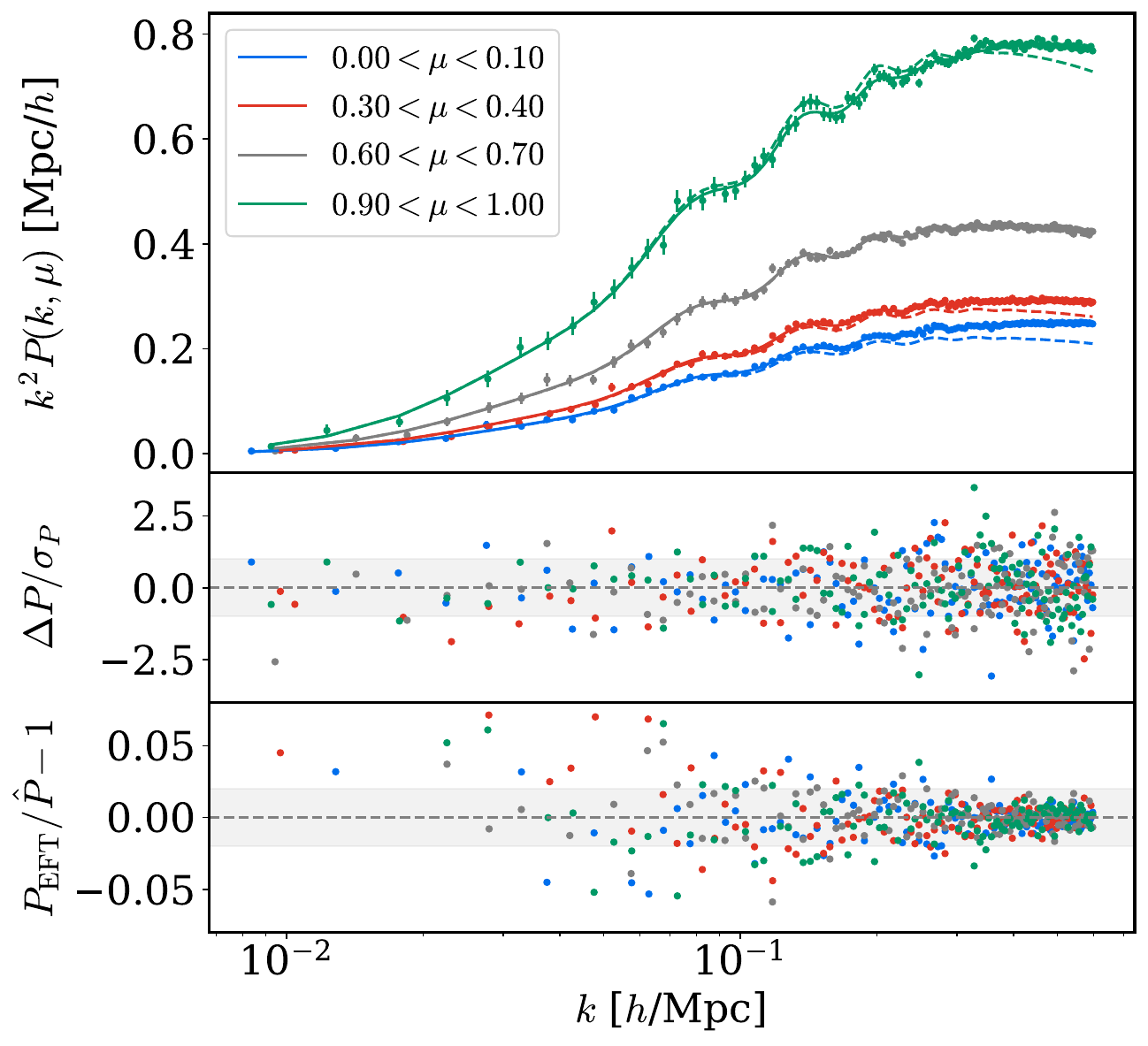} \hfill
    \includegraphics[width=0.49\linewidth]{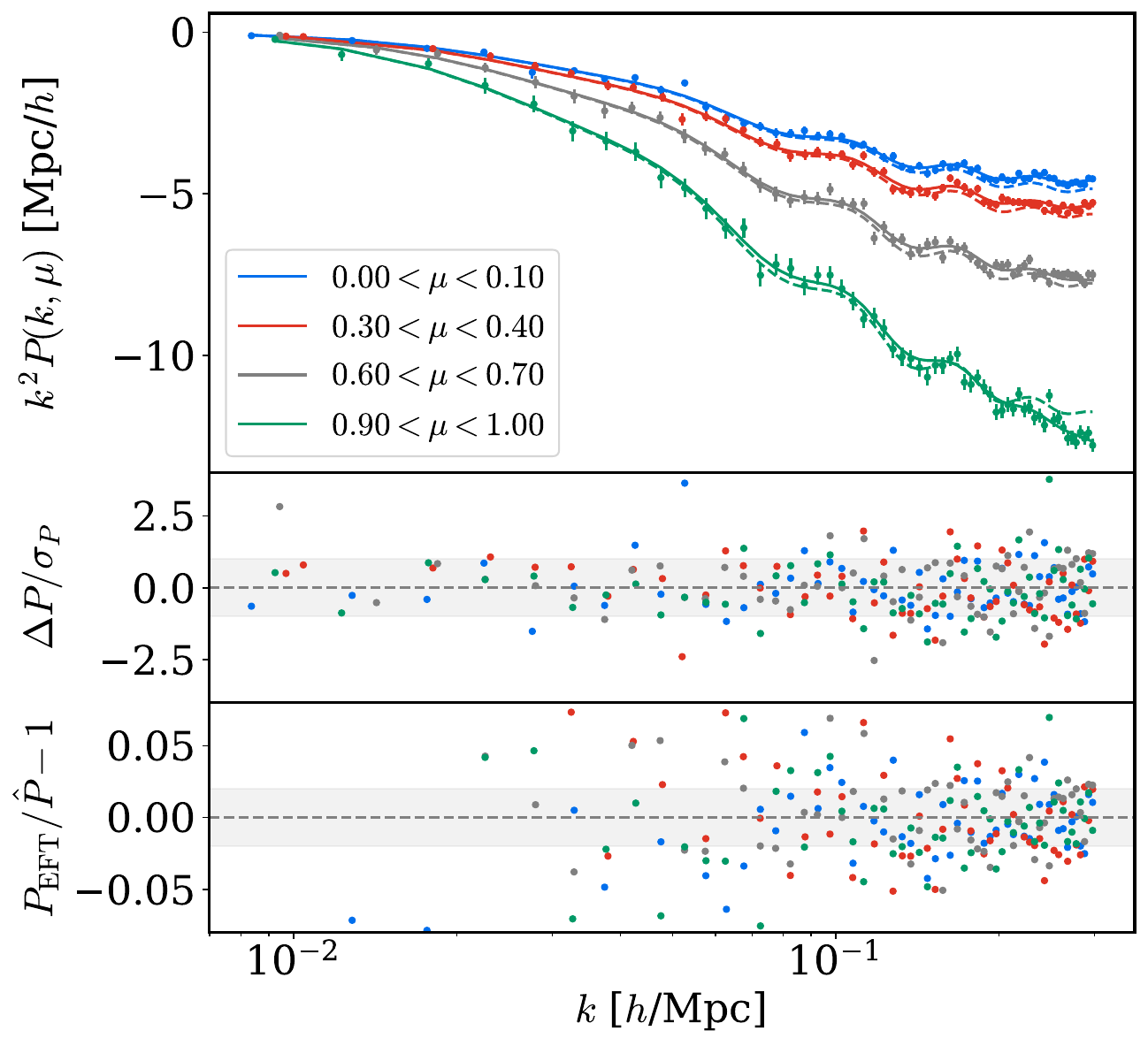} \hfill
    \vspace{-0.1in}
    \caption{EFT fits to the \Lya auto-correlation (\textit{left panel}) and \Lya-quasar cross-correlation (\textit{right panel}) function in four (out of ten) angular bins, $\mu$, with $\kmax =0.6\hMpcinv$ for the first simulation of model one along the y-axis. The best-fit parameters are shown in the first column of Table~\ref{tab:EFT_bias}. Best-fit EFT model (solid line) compared to $P^{\rm tree}$ (dashed line) and data points. The error bars are obtained assuming a diagonal Gaussian covariance based on the number of expected Fourier modes per bin $P(k,\mu)\sqrt{2/N(k,\mu)}$. The residuals between the model and the data are shown in the bottom two panels. The upper one shows the difference between model and data divided by the error bars with a gray band indicating the $1\sigma$ error region to guide the eye. The bottom panel shows the ratio of the two, with a 2\% error band to guide the eye.}
    \label{fig:EFT_bestfit_pk}
\end{figure*}

The best-fit auto spectra are shown in the left panel of Fig.~\ref{fig:EFT_bestfit_pk} for four angular bins. The corresponding best-fit cross-spectrum obtained from the joint fit of the auto-and cross spectrum is shown in the right panel. (For conciseness, we omit the auto spectrum from the joint fit as the quality is identical by eye.) The bottom two panels show the residuals between the model and the data: (i) the middle panel shown the residuals divided by the error bars, illustrating that, even with the small error bars from \abacus, we find agreement at the $\sim 2\sigma$-level at \textit{all} scales; (ii) the bottom panel shows that the deviations at small scales are for the auto alone sub-$2\%$ whilst for the joint auto- and cross-correlation fit around $3-5\%$. Note that the quality of the fit degrades along the line-of-sight due to the velocity field. 

\begin{figure*}
    \centering
    \includegraphics[width=0.49\linewidth]{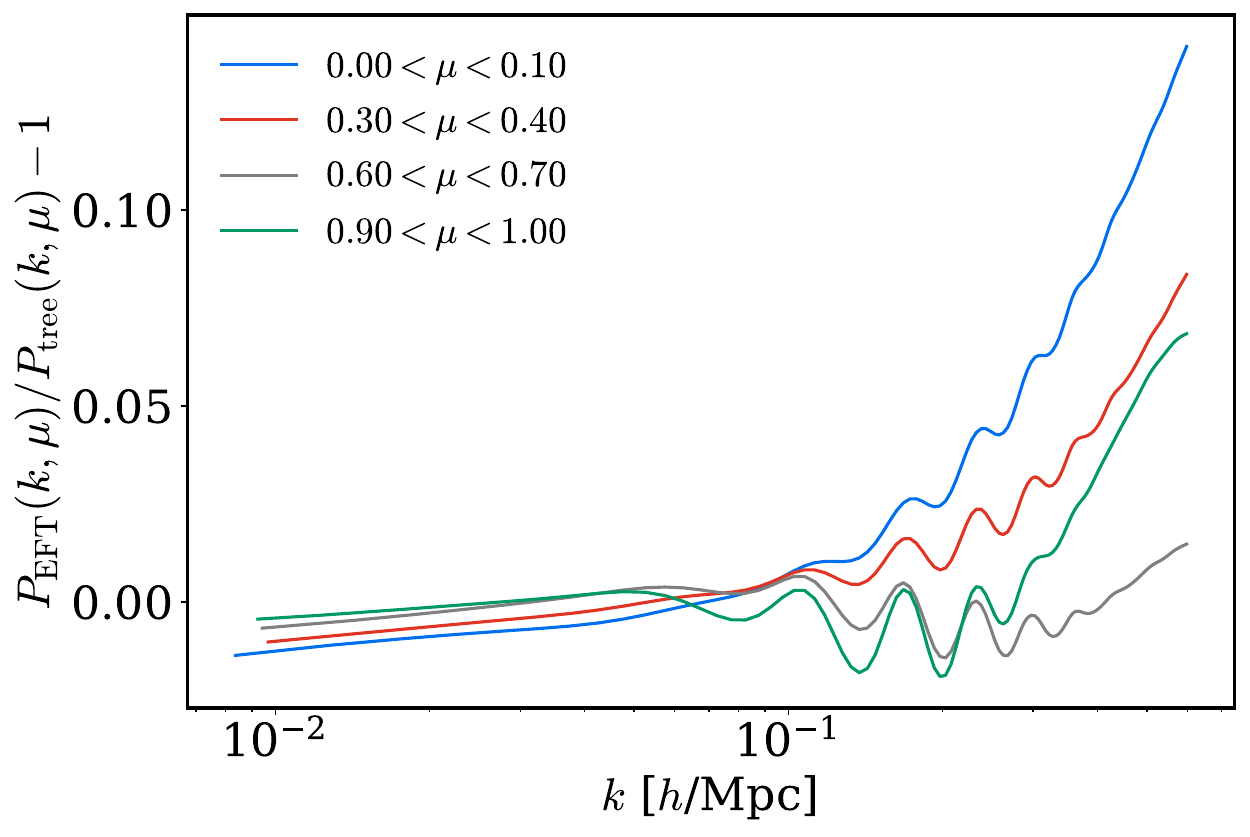}\hfill
    \includegraphics[width=0.49\linewidth]{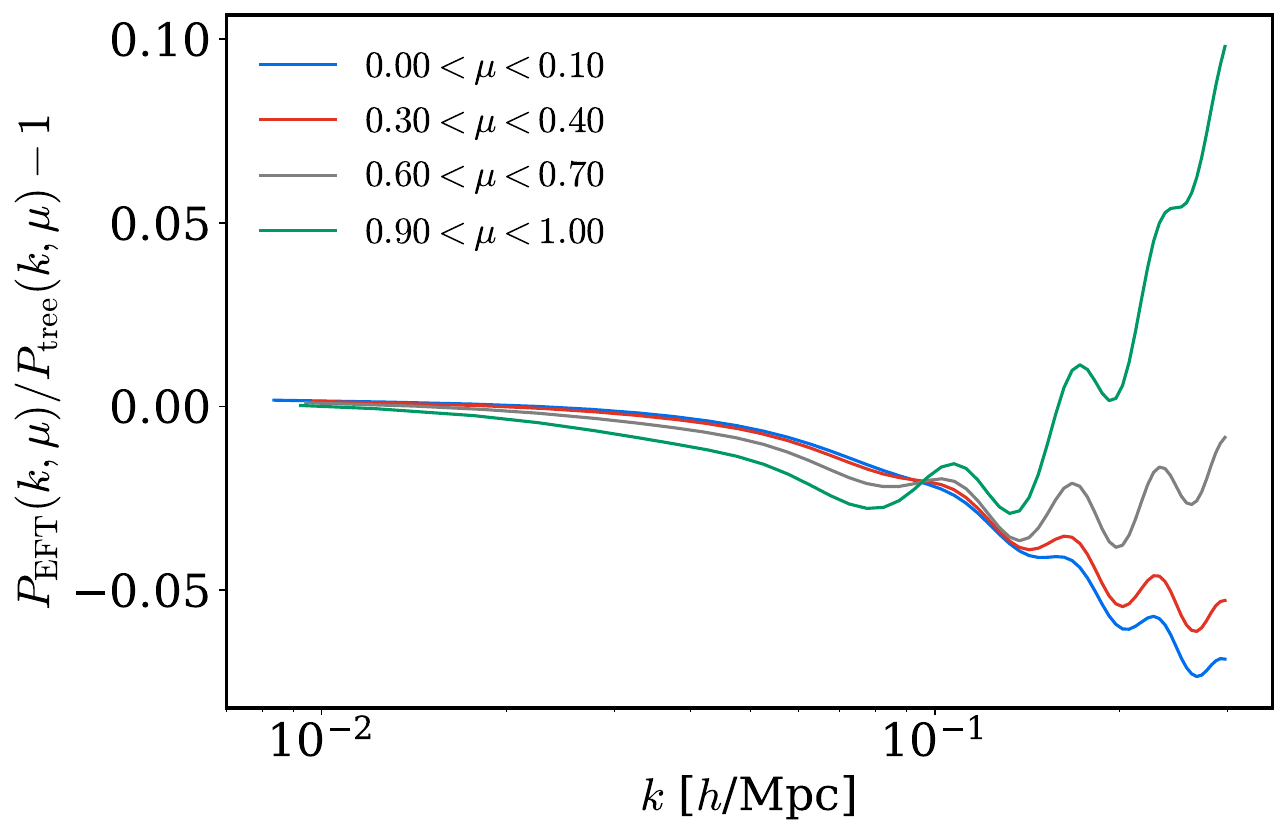}
    \vspace{-0.1in}
    \caption{Size of the one-loop corrections for the \Lya auto-correlation (\textit{left panel}) and \Lya-quasar cross-correlation (\textit{right panel}) model obtained from Fig.~\ref{fig:EFT_bestfit_pk} by comparing the IR-resummed linear theory power spectrum, $P_{\rm tree}$, to the best-fit one-loop power spectrum, $P_{\rm EFT}$.}
    \label{fig:EFT_bestfit_pk_loop_correction}
\end{figure*}

To assess at which scales the EFT corrections become important, we show the loop corrections, the ratio of the EFT model to the tree-level power spectrum, in the left for the auto and in the right panel the cross-spectrum obtained from the joint auto- and cross-correlation fit of the  of Fig.~\ref{fig:EFT_bestfit_pk_loop_correction}. Following baseline expectation, the large amount of available (quasi-) linear modes results in the EFT corrections to the tree-level power spectrum becoming relevant at $k\simgt 0.2 \hMpcinv$, especially for transverse modes (blue). As expected the perturbative reach of the EFT models is smaller for quasars (or halos) than for the \Lya forest (see, e.g.,~\citet{Maus:2024dzi} for a discussion of EFT full-shape analyses in the context of DESI for low-redshift galaxies and quasars). It is interesting to note that for the cross-correlation the loop corrections are more pronounced for the line-of-sight modes than for the transverse ones. Note that the unity-crossing of the loop corrections is an indicator as to when EFT formally breaks down which we do not reach here.\footnote{See, however, the discussion in \cite{deBelsunce:2024rvv}.} 

\begin{figure*}
    \centering
    \includegraphics[width=0.49\linewidth]{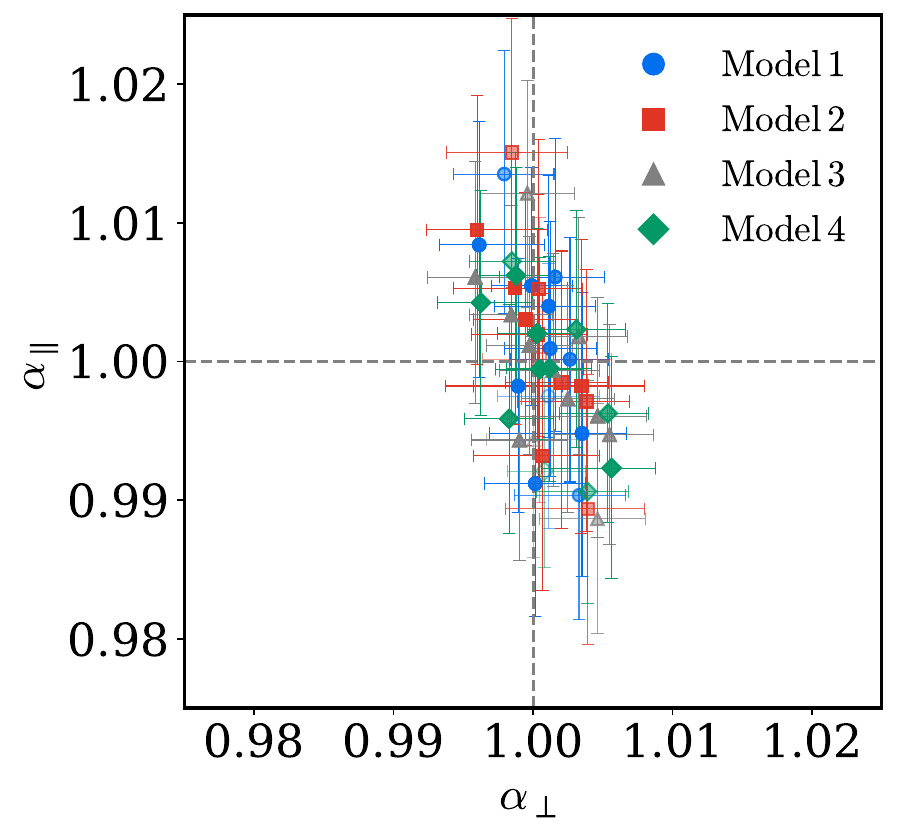}\hfill
    \includegraphics[width=0.49\linewidth]{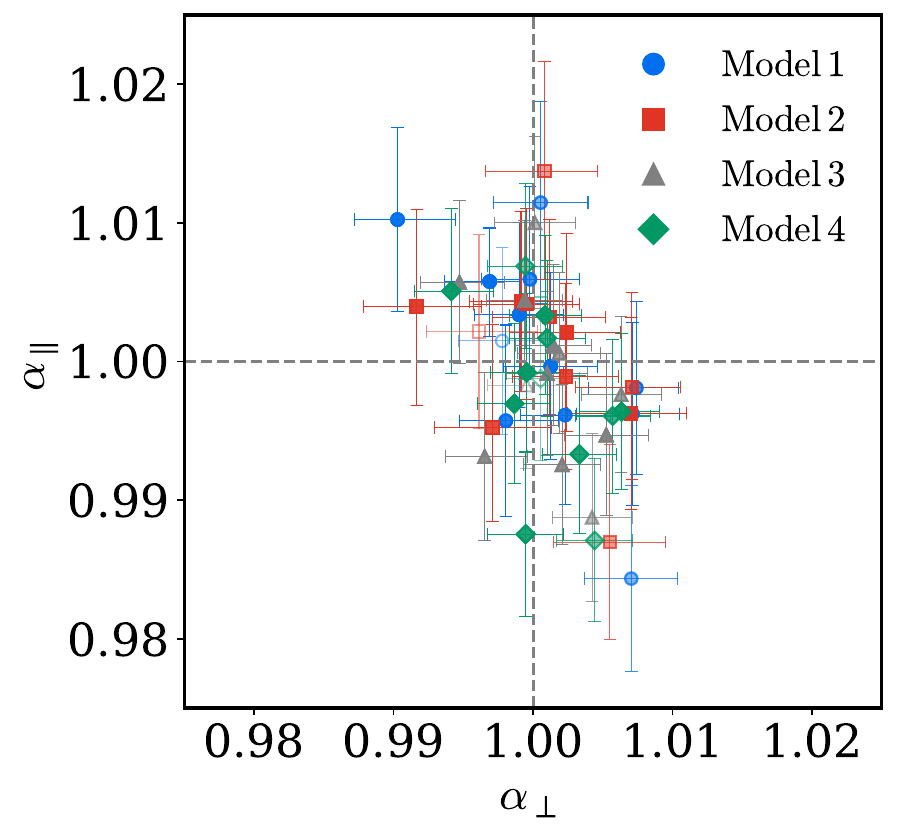}
    \vspace{-0.1in}
    \caption{BAO scaling parameters obtained from a joint fit of the $\alpha$'s and EFT bias parameters to each simulation, model and line-of-sight which are represented by a different color-shape combination: 1 (blue, circle), 2 (red, square), 3 (gray, triangle), and 4 (green, diamond). Filled (empty) markers represent the line-of-sight $y$-axis ($z$-axis). We show the \Lya auto-correlation in the left and the joint fit to the \Lya auto and \Lya--quasar cross-correlation in the right panel. The average cross-correlation coefficients are $\overline{\rho} = -0.39 \pm 0.04$ and $\overline{\rho}^{\times} = -0.42 \pm 0.03$, for auto alone and auto and cross jointly, respectively. The EFT model debiases the BAO fit and we find a value for BAO scaling parameters consistent with unity with the corresponding mean values for the $\alpha$'s in Table~\ref{tab:EFT_alpha_joint}.}
    \label{fig:EFT_alpha_joint}
\end{figure*}

The measured values for the BAO scaling parameters for the \Lya auto-correlation (joint auto- and cross-correlation) are shown in the left (right) panel of Fig.~\ref{fig:EFT_alpha_joint} with the corresponding mean values of the BAO parameters tabulated in Table~\ref{tab:EFT_alpha_joint}. We find that the EFT model debiases the $\alpha$ parameters and our fits are consistent with a value of $\alpha_{\parallel,\perp}=1$ across all four models. We find a mean cross-correlation coefficient over the 48 realizations for the $\alpha$'s of $\rho=-0.39\pm0.04$. Similar to the case for the \Lya auto-correlation, we find for the cross-correlation that the EFT debiases the BAO fit with a mean cross-correlation coefficient of $\rho^{\times}=-0.43\pm0.03$. 

\begin{table*}
    \begin{center}
    \begin{tabular}{c c c c c}
     \hline\hline 
    $b_{\mathcal{O}}$ & Model 1 & Model 2 & Model 3 & Model 4  \\[1ex]
     \hline
    $b_1^{\rm q}$ & $3.3105 \pm 0.0427$ & $3.3461 \pm 0.0441$ & $3.3906 \pm 0.0365$ & $3.3666 \pm 0.0356$ \\[1ex]
    $b_2^{\rm q}$ & $1.6487 \pm 0.9992$ & $1.3933 \pm 1.0026$ & $2.3774^{+1.0202}_{-1.0408}$ & $1.2312^{+0.9905}_{-1.0822}$ \\[1ex]
    $b_1$ & $-0.1433 \pm 0.0011$ & $-0.1264 \pm 0.0011$ & $-0.1277 \pm 0.0010$ & $-0.1234 \pm 0.0009$ \\[1ex]
    $b_{\eta}$ & $0.1492 \pm 0.0018$ & $0.1352 \pm 0.0017$ & $0.2818 \pm 0.0023$ & $0.3180 \pm 0.0024$ \\[1ex]
    $b_2$ & $-0.2162^{+0.1696}_{-0.1281}$ & $-0.5015^{+0.1268}_{-0.0849}$ & $-0.1706^{+0.1168}_{-0.1064}$ & $-0.1298 \pm 0.1058$ \\[1ex]
    $b_{\mathcal{G}_2}$ & $-0.1581^{+0.1141}_{-0.0908}$ & $-0.3255^{+0.1287}_{-0.0814}$ & $-0.0912 \pm 0.0710$ & $-0.0718 \pm 0.0668$ \\[1ex]
    $b_{\eta^2}$ & $-0.1786 \pm 0.0529$ & $-0.3289^{+0.0524}_{-0.0455}$ & $-0.2074 \pm 0.0489$ & $-0.1712 \pm 0.0503$ \\[1ex]
    $b_{\delta \eta}$ & $-0.4206^{+0.1422}_{-0.1148}$ & $-0.5092^{+0.1589}_{-0.0873}$ & $-0.1835^{+0.0811}_{-0.0715}$ & $-0.1084^{+0.0737}_{-0.0657}$ \\[1ex]
    $b_{(KK)_\parallel}$ & $0.3246^{+0.1906}_{-0.2469}$ & $0.5375^{+0.2356}_{-0.2848}$ & $0.4644^{+0.1303}_{-0.1508}$ & $0.4604^{+0.1242}_{-0.1478}$ \\[1ex]
    $b_{\Pi^{[2]}_\parallel}$ & $-0.2596^{+0.0573}_{-0.0479}$ & $-0.2056^{+0.0536}_{-0.0437}$ & $-0.1921 \pm 0.0536$ & $-0.2034 \pm 0.0551$ \\[1ex]
    $\apar$ & $0.9997 \pm 0.0068$ & $1.0021 \pm 0.0072$ & $0.9991 \pm 0.0059$ & $0.9992 \pm 0.0057$ \\[1ex]
    $\aperp$ & $1.0013 \pm 0.0034$ & $1.0024^{+0.0041}_{-0.0038}$ & $1.0010 \pm 0.0029$ & $0.9995 \pm 0.0026$ \\[1ex]
    \hline
    $b_{\Pi^{[3]}_\parallel}$ & $-0.1537 \pm 0.0316$ & $-0.1089 \pm 0.0299$ & $0.2844 \pm 0.0340$ & $0.2747 \pm 0.0350$ \\[1ex]
    $b_{\delta\Pi^{[2]}_\parallel}$ & $-1.2028 \pm 0.1269$ & $-0.7561 \pm 0.1163$ & $-1.1539 \pm 0.1269$ & $-1.4458 \pm 0.1297$ \\[1ex]
    $b_{(K\Pi^{[2]})_\parallel}$ & $-1.5418 \pm 0.1351$ & $-2.0036 \pm 0.1225$ & $-1.9400 \pm 0.1102$ & $-2.2706 \pm 0.1089$ \\[1ex]
    $b_{\eta\Pi^{[2]}_\parallel}$ & $0.2632 \pm 0.3354$ & $1.9228 \pm 0.3112$ & $-2.5832 \pm 0.3933$ & $-3.0661 \pm 0.4080$ \\[1ex]
    $b_{\GG}$ & $0.7170 \pm 0.0629$ & $1.4109 \pm 0.0581$ & $-0.1916 \pm 0.0687$ & $-0.2057 \pm 0.0703$ \\[1ex]
    $P_{\rm shot}$ & $-0.4064 \pm 0.0735$ & $-0.2247 \pm 0.0567$ & $0.4116 \pm 0.0212$ & $0.2750 \pm 0.0175$ \\[1ex]
    $a_0$ & $1.1312 \pm 3.1873$ & $0.4227 \pm 2.7668$ & $-4.1192 \pm 2.9160$ & $-2.6957 \pm 2.8277$ \\[1ex]
    $a_2$ & $0.5094 \pm 4.8389$ & $-1.3978 \pm 4.6306$ & $-1.5596 \pm 4.7806$ & $-1.3415 \pm 4.8413$ \\[1ex]
    $c_0$ & $0.1863 \pm 0.0316$ & $0.1179 \pm 0.0272$ & $-0.1787 \pm 0.0141$ & $-0.1207 \pm 0.0134$ \\[1ex]
    $c_{2}$ & $-0.2061 \pm 0.0365$ & $-0.1210 \pm 0.0341$ & $0.1979 \pm 0.0313$ & $0.1442 \pm 0.0316$ \\[1ex]
    $c_{4}$ & $-0.0610 \pm 0.0221$ & $-0.0862 \pm 0.0203$ & $-0.3384 \pm 0.0235$ & $-0.3535 \pm 0.0243$ \\[1ex]
    \hline
    $b_{\mathcal{G}_2}^{\rm q}$ & $0.1427 \pm 0.6497$ & $-0.3223^{+0.6085}_{-0.7342}$ & $0.1116 \pm 0.5026$ & $0.0458 \pm 0.4734$ \\[1ex]
    $P_{\rm shot}^{\rm q}$ & $-0.8379 \pm 2.8662$ & $2.1759 \pm 2.6402$ & $-5.1053 \pm 2.2466$ & $-5.9015 \pm 2.0804$ \\[1ex]
    $c_{0}^{\rm q}$ & $0.3062 \pm 0.4629$ & $-0.1038 \pm 0.4762$ & $1.7626 \pm 0.3687$ & $1.8137 \pm 0.3436$ \\[1ex]
    $c_{2}^{\rm q}$ & $1.2548 \pm 0.5366$ & $0.7582 \pm 0.5564$ & $1.8479 \pm 0.6097$ & $1.6705 \pm 0.6144$ \\[1ex]
    $c_{4}^{\rm q}$ & $0.3197 \pm 0.3279$ & $1.2225 \pm 0.3336$ & $0.0871 \pm 0.3936$ & $0.0509 \pm 0.4096$ \\[1ex]
    $b_{\GG}^{\rm q}$ & $0.0361 \pm 0.0102$ & $0.0514 \pm 0.0091$ & $-0.1879 \pm 0.0098$ & $-0.2424 \pm 0.0098$ \\[1ex]
    $c_x$ & $0.2753 \pm 0.9774$ & $0.2030 \pm 0.9631$ & $0.6029 \pm 0.9657$ & $0.5385 \pm 0.9779$ \\[1ex]
    $\chi^2$ & $1791$ & $1839$ & $1894$ & $1901$ \\
    \hline \hline
    \end{tabular}
    \end{center}
    \caption{Same as Table~\ref{tab:EFT_bias} but for the joint fits of the auto- and cross-correlation. The top section shows the sampled parameters and the middle (bottom) section the ones that we analytically marginalize over for the auto- (cross-) correlation. We use 1755 data points, sample over 12 parameters and analytically marginalize over 18.}
    \label{tab:EFT_bias_cross}
\end{table*}

\subsubsection{EFT fits to the mean data vector}\label{sec:EFT_mean_fits}
\begin{figure}
    \centering
    \includegraphics[width=\linewidth]{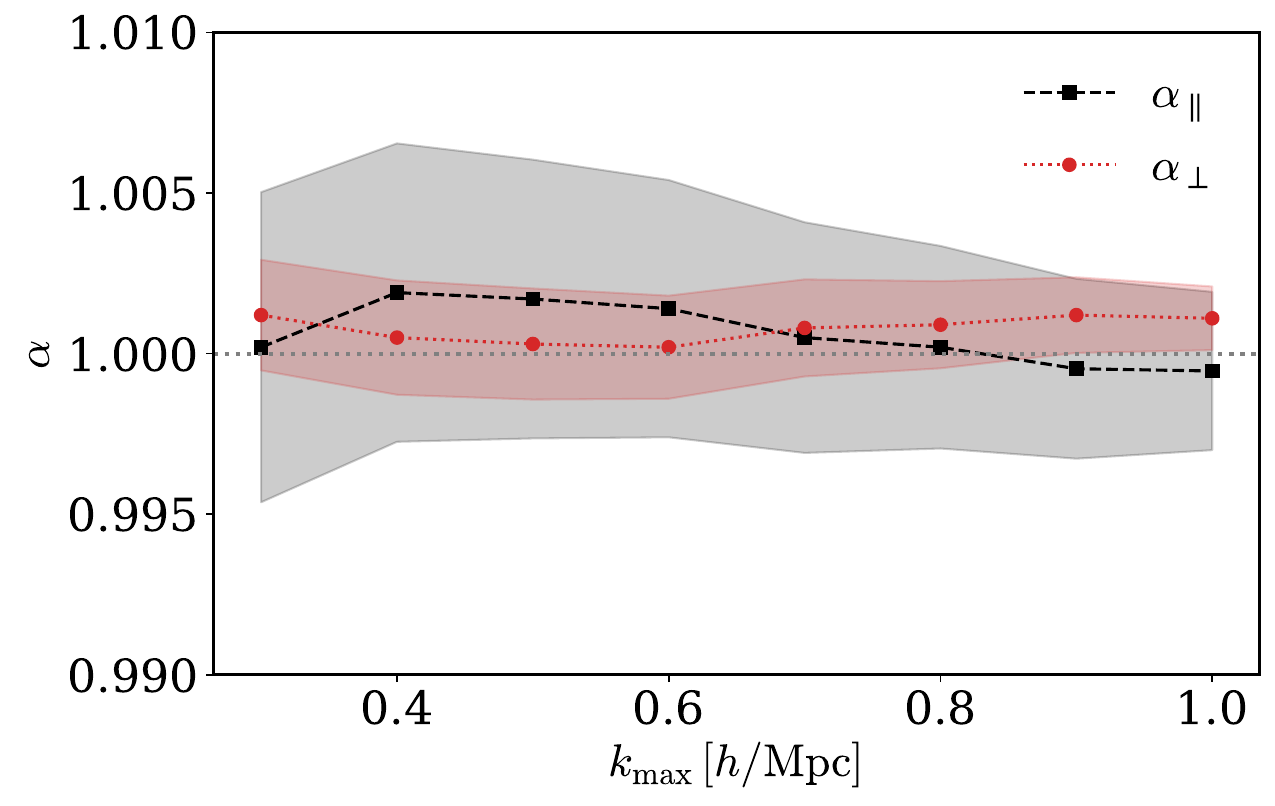}
    \vspace{-0.1in}
    \caption{Mean values of $\apar$ (black dashed line) and $\aperp$ (red dotted line) obtained from fits to the mean of six simulations along the y-axis for model 1. The fits are performed for increasing values of $\kmax=0.3-1.0 \hMpcinv$. We give the numerical values for our baseline value of $\kmax=0.6 \hMpcinv$ in the first row of the bottom panel in Table~\ref{tab:EFT_alpha_joint}. This plot shows that the EFT fits can isolate the BAO information and we use $\kmax=0.6\hMpcinv$ as a conservative scale cut for our analysis.}
    \label{fig:EFT_meandata_alpha_joint}
\end{figure}

To stress-test the fitting of the EFT model using a low(er) noise power spectrum, we average the measured spectra over simulation realizations and lines-of-sight, resulting in reduced error bars by a factor of $\sqrt{6}$. Again, we jointly fit the BAO scaling and EFT bias parameters and find no evidence for a shift in the BAO peak. For the auto-correlation, we show for model 1 the evolution of the BAO parameter as a function of scale-cut $\kmax$ in Fig.~\ref{fig:EFT_meandata_alpha_joint}. It is interesting (and reassuring) to note that the errorbars only slightly reduce above $k\sim 0.3 \hMpcinv$ as this indicates that we can isolate the information stemming from the BAO feature and not pick up full-shape information. This figure is an additional confirmation of our conservative choice of scale cut of $\kmax=0.6\hMpcinv$. The corresponding numerical values for all four models are given in Table~\ref{tab:EFT_alpha_joint} for our baseline scale cut. The results for the joint fit to the auto- and cross-correlation  are given in the bottom panel of the same table. Both series of fits are consistent with BAO scaling parameters of $\apar=\aperp=1$. 

\begin{table}
\centering
\setlength{\tabcolsep}{4pt}
\begin{tabular}{ccccc}
\hline \hline
& \multicolumn{2}{c}{Auto} & \multicolumn{2}{c}{Auto + Cross}\\[0.5ex]
\cmidrule(lr){2-3} \cmidrule(lr){4-5}
Model & $\apar$ & $\aperp$ & $\apar$ & $\aperp$\\[0.5ex]
\hline
\multicolumn{5}{l}{\textbf{Mean of fits}}\\[0.5ex]
1 & $1.0009_{-0.0028}^{+0.0027}$ & $1.0006_{-0.0016}^{+0.0014}$ & $1.0007_{-0.0019}^{+0.0019}$ & $1.0006_{-0.0010}^{+0.0010}$ \\[1ex]
2 & $1.0014_{-0.0029}^{+0.0029}$ & $1.0006_{-0.0020}^{+0.0016}$ & $1.0008_{-0.0020}^{+0.0020}$ & $1.0008_{-0.0012}^{+0.0011}$ \\[1ex]
3 & $0.9992_{-0.0024}^{+0.0024}$ & $1.0012_{-0.0016}^{+0.0014}$ & $0.9988_{-0.0017}^{+0.0017}$ & $1.0010_{-0.0008}^{+0.0008}$ \\[1ex]
4 & $0.9989_{-0.0023}^{+0.0023}$ & $1.0010_{-0.0013}^{+0.0013}$ & $0.9979_{-0.0017}^{+0.0017}$ & $1.0011_{-0.0008}^{+0.0008}$ \\[1ex]
\hline
\multicolumn{5}{l}{\textbf{Fits to stacked spectra}}\\[0.5ex]
1 & $1.0014_{-0.0041}^{+0.0039}$ & $1.0002_{-0.0017}^{+0.0015}$ &$1.0015_{-0.003}^{+0.003}$ & $1.0003_{-0.0013}^{+0.0013}$ \\[1ex]
2 & $1.0025_{-0.0039}^{+0.0038}$&  $1.0016_{-0.0025}^{+0.0024}$ & $1.0006_{-0.0031}^{+0.003}$&  $1.0015_{-0.0018}^{+0.0019}$ \\[1ex]
3 & $0.9994_{-0.0034}^{+0.0034}$ & $1.0014_{-0.0015}^{+0.0015}$ &$1.0000_{-0.003}^{+0.0029}$ & $1.0008_{-0.0013}^{+0.0014}$ \\[1ex]
4 & $0.9988_{-0.0034}^{+0.0033}$ & $1.0011_{-0.0014}^{+0.0015}$ &$0.9994_{-0.0028}^{+0.0028}$ & $1.0007_{-0.0014}^{+0.0013}$ \\[1ex]
\hline
\end{tabular}
\caption{Radial ($\apar$) and transverse ($\aperp$) BAO scaling parameters obtained from jointly fitting the BAO and EFT bias parameters. \textit{Top panel}: mean of fits to $N_{\rm sim} = 12$ simulations of each model, illustrated in Fig.~\ref{fig:EFT_alpha_joint}. The radial (transverse) mean error bars are divided by $\sqrt{N_{\rm sim}}$ ($\sqrt{N_{\rm sim}}/2$).  \textit{Bottom panel}: Fits to the mean data vector averaged over 12 realizations for each model. The left (right) columns are the values for the auto (auto+cross) correlation.  We apply scale cuts of $\kmax=0.6~\hMpcinv$ to $P^{\rm FF}$ and $\kmax=0.3~\hMpcinv$ to $P^{\rm FQ}$.}
\label{tab:EFT_alpha_joint}
\end{table}

\subsection{Consistency tests} \label{sec:EFT_consistency}
\subsubsection{Quasi-linear theory fits to the EFT best-fit data vector}\label{sec:EFT_consistency_1}
\begin{figure}
    \centering
    \includegraphics[width=\linewidth]{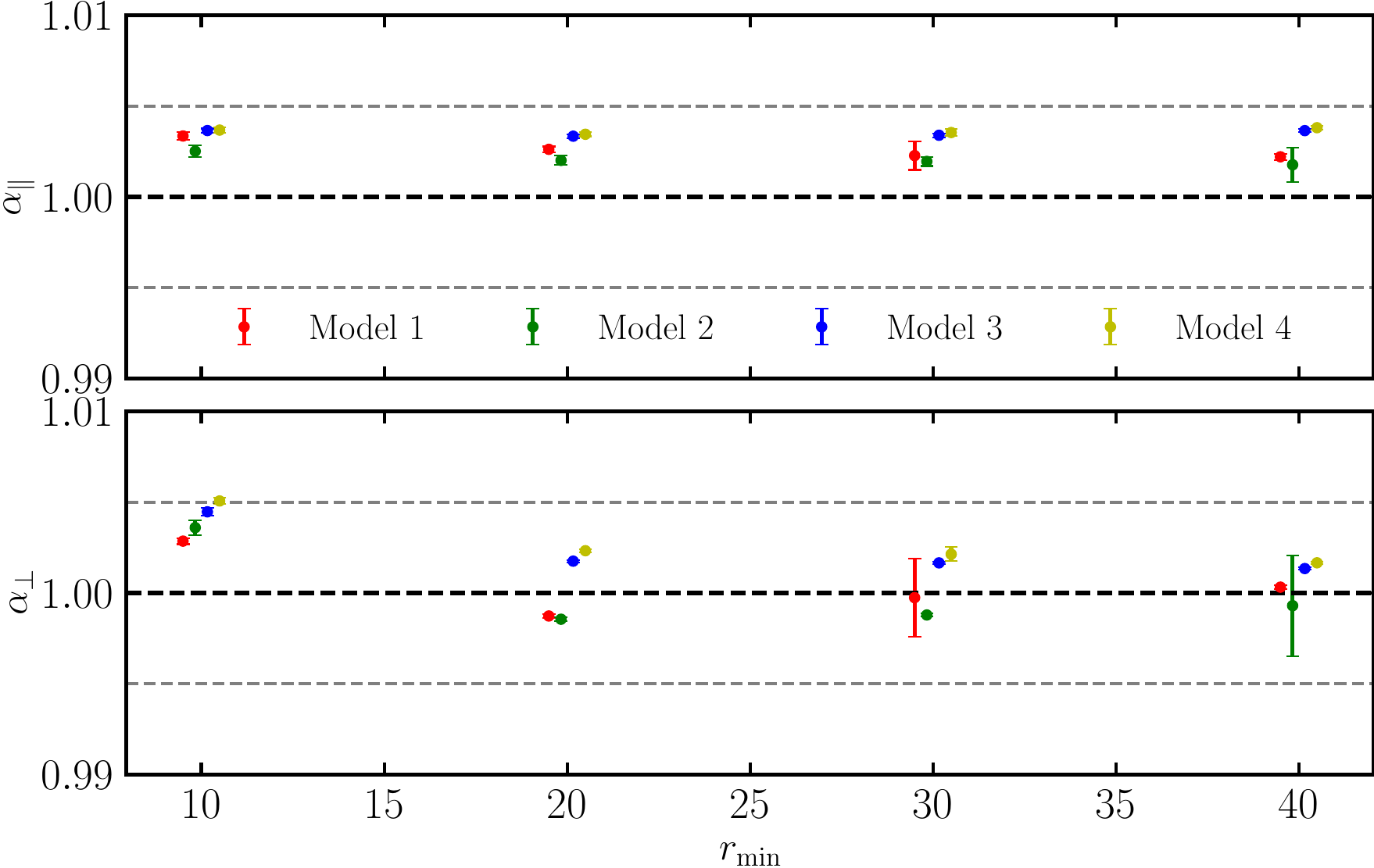}
    \vspace{-0.1in}      
    \caption{Fitting the EFT best-fit \Lya auto-correlation spectra (with BAO scaling parameters fixed to unity) instead of \abacus simulation data. We show the BAO parameters as a function of minimum separation, $r_{\rm min}$, whilst the EFT templates are obtained from fits with $\kmax=0.6\hMpcinv$ corresponding to $r \approx 10.5 \hinvMpc$. Wo obtain consistent results when replacing the \abacus data with the EFT templates, \textit{i.e.}~we detect a shift of 0.3-0.4\% for $\apar$, and a smaller shift of $\aperp$ which is consistent with zero within $\sim$1-2$\sigma$ for $r_{\rm min}\geq 20\hinvMpc$. We attribute the larger measured shift at $r_{\rm min}=10\hinvMpc$ to issues with the small-scale modeling when using the quasi-linear \textsc{Vega} model.}
    \label{fig:EFT_fits_vega_apat}
\end{figure}

As a consistency check and to reconcile the analyses in the first part of this work to the present section, we replace the \abacus data vector with the EFT best-fit spectra using the \textsc{Vega} fitting pipeline introduced in Sec.~\ref{sec:vega}.\footnote{Note that we fix $\apar\equiv \aperp\equiv 1$.} First, we transform the EFT best-fit theory predictions for $P(k,\mu)$ to anisotropic spectra using the standard multipole decomposition 
\begin{equation}
    P_{\ell}(k)\equiv \frac{2 \ell+1}{2}\int_{-1}^{1} \mathrm{d}\,\mu \mathcal{L}_\ell(\mu) P(k,\mu)\,,
\end{equation} 
where $\mathcal{L}_\ell$ are the Legendre moments in multipoles $\ell$. Second, we perform the same analysis as in Sec.~\ref{sec:fits_auto} and measure the BAO scaling parameters from our EFT templates
using the empirical model of~\cite{Arinyo-i-Prats:2015}. The results are shown in Fig.~\ref{fig:EFT_fits_vega_apat}. 
Following the baseline expectations, we obtain consistent results, \textit{i.e.} the \textsc{Vega} pipeline measures the same biases in $\alpha$ when replacing the \abacus data by the best-fit EFT spectra -- this confirms that the EFT accurately \textit{describes} the data. 

\subsubsection{Fisher forecasts for the non-linear BAO shift} \label{sec:EFT_Fisher}
Another avenue to quantify the magnitude of the BAO shift stemming from non-linearities in the three-dimensional clustering of the \Lya forest, is to theoretically predict it using perturbation theory. We follow the approach developed in the context of galaxies \citep{Chen:2024tfp} and in the context of the \Lya forest \citep{deBelsunce:2024rvv} and predict the shift in the BAO scaling parameters when using a linear theory model. Whilst the \textsc{AbacusSummit} simulations resolve the BAO feature, they have much lower physical resolution than hydrodynamical simulations \citep{2017MNRAS.464..897B, Chabanier2024}. Thus, the accuracy of the \abacus simulations should be confined to larger scales and the exact magnitude of the non-linear bias parameters should be treated as indicatively, e.g.,~\citet{Chabanier2024} showed that the physical resolution of the simulation affects the velocity bias. Therefore, this section serves to generate some intuition for the reader and we compare the Fisher prediction, briefly introduced in the following, to fits of the \abacus spectra using a linear theory model in Fourier space (i.e., this amounts to setting all the higher order bias parameters to zero bar $b_1$ and $b_{\eta}$ in Sec.~\ref{sec:EFT_theory}). 

Conceptually, in real space, the BAO feature contracts isotropically around overdensities and expands around underdensities. However, the \Lya  forest is suppressed around overdensities resulting in an expanded BAO radius. In general, the shift is sourced by long-wavelength modes (wavelengths longer than the distance a sound wave travels prior to the baryon drag epoch, post recombination, denoted by $r_d$), and is captured in  perturbation theory by isolating the $P^{(22)}$ term \citep[see, e.g.,~][and references therein]{Chen:2024tfp}. Schematically, the BAO shift in the auto-correlation is driven by (i) the relative amplitude of the linear ($b_1$) and quadratic contributions ($b_2$) to the \Lya clustering; and (ii) the amplitude of density fluctuations on BAO scales which is reduced by the growth factor $D(z)^2$.\footnote{The picture for the cross-correlation with quasars is analogous. The relative amplitude of the linear and quadratic contribution is added the one of the auto-correlation \citep{deBelsunce:2024rvv,Chudaykin:2025gsh}.} 

We use a Fisher formalism to predict the BAO shift using measurements of the bias parameters on \abacus
\begin{equation}
    \Delta \{\apar,\aperp\} = F^{-1}_{\{\apar,\aperp\}a b} t^b_i C^{-1}_{ij} \epsilon_j\, ,
    \label{eqn:fisher_shifts}
 \end{equation}
where $C$ is the covariance matrix, $t^b_i$ is the linear template $dP_i/d\theta_b$ corresponding to fitting parameter $\theta_b$, and $F_{ab}$ is the Fisher matrix. This setup assumes that, in addition to the BAO scales, we also marginalize over a set of other parameters including the linear density and velocity bias, and BAO damping parameters. For the covariance matrix $C$, we use the Gaussian covariance matrix from Sec.~\ref{sec:EFT_likelihood}, and $\epsilon$, are the ``wedges'' defined in total separation $k=(k_{\parallel}^2 +\mathbf{k}_{\perp}^2)^{1/2}$ and in $\mu$ bins for the analytic description of the BAO shift, multiplied by the mean-square linear density fluctuation in a sphere of radius $r_d$ and the response of the power spectrum, $\frac{\dd P}{\dd \ln k}$. The fit is performed with $k_{\rm min}=0.02 \hMpcinv$ and $\kmax = 0.3 \hMpcinv$ with a linear $\Delta k=0.005 \hMpcinv$ spacing.

\begin{table}
\centering
\setlength{\tabcolsep}{4pt}
\begin{tabular}{ccccc}
\hline \hline
& \multicolumn{2}{c}{Fisher prediction} & \multicolumn{2}{c}{Linear theory fit (auto)}\\[0.5ex]
\cmidrule(lr){2-3} \cmidrule(lr){4-5}
Model & $\Delta\apar\, [\%]$ & $\Delta\aperp\, [\%]$ & $\Delta\apar\, [\%]$ & $\Delta\aperp\, [\%]$\\[0.5ex]
\hline
1 & $-0.08\pm 0.07$ & $          -0.01 \pm 0.01$ & $-0.08\pm 0.19$ & $\phantom{-}0.16 \pm 0.14$ \\ [1ex]    
2 & $-0.12\pm 0.09$ & $\phantom{-}0.04 \pm 0.02$ & $-0.23\pm 0.19$ & $\phantom{-}0.22 \pm 0.15$ \\ [1ex]
3 & $-0.33\pm 0.11$ & $\phantom{-}0.03 \pm 0.01$ & $-0.63\pm 0.19$ & $          -0.07 \pm 0.14$ \\ [1ex]
4 & $-0.43\pm 0.13$ & $\phantom{-}0.07 \pm 0.04$ & $-0.59\pm 0.18$ & $          -0.22 \pm 0.14$ \\
\hline
\end{tabular}
\caption{Comparison of the Fisher prediction (left column) for the BAO parameter in radial ($\apar$) and transverse ($\aperp$) direction to measurements of the BAO parameters using linear theory (right column) for the \Lya auto-correlation. The corresponding bias parameters are given in Table~\ref{tab:EFT_bias_meandatavector} and obtained from fitting the mean data vector (averaged over 12 simulations per model).} \label{tab:EFT_Fisher_BAO_meandatavector}
\end{table}

In Table~\ref{tab:EFT_Fisher_BAO_meandatavector} we compare the results on the biases of the BAO scaling parameters obtained when performing linear theory fits in Fourier space (first column) to the Fisher forecasted BAO shifts (second column). As expected, both approaches agree at the 1-2 $\sigma$ level. The linear theory fits show a negatively biased scaling parameter in radial direction (-$0.2-0.5\%$) and a shift that is consistent with zero in the transverse direction at the $\sim 1.5 \sigma$ level, depending on the chosen model. The predictions and linear theory fits are obtained using $\kmax=0.3 \hMpcinv$. Varying $\kmax$ does not change our conclusions. We emphasize that these BAO shifts cannot directly be extrapolated to observational data, since we do not expect the (higher-order) biases to  agree with the data. Therefore, we refer the reader to \citet{deBelsunce:2024rvv} where they use hydrodynamical simulations which are expected to better represent the underlying physics. 

To understand the differences between the results obtained using the \texttt{Vega} fitting pipeline in Sec.~\ref{sec:lcv_res} and the linear theory fits in Fourier space we perform a series of tests: (i) To assess that both models agree, we perform a similar fit as in Fig.~\ref{fig:EFT_fits_vega_apat}. We fit with \vega the linear theory best-fit power spectrum  (with BAO scaling parameters set to unity) instead of the \abacus data. We find constraints on the radial and transverse scaling parameters that are consistent with zero at the $\sim 1\sigma$ level. This test confirms that the models are consistent with each other. (ii) To assess the validity of the covariance matrix, we compute the Jacobian matrix that converts the Gaussian covariance from Fourier to configuration space in $\{r_p,\textbf{r}_t\}$ space and fit the LCV \abacus data vector. We obtain consistent results with the eBOSS covariance matrix results, following baseline expectation. (iii) Further, to be closer to the implementation in \vega we remove the next-to-leading order correction of the one-loop IR resummation in time-sliced perturbation theory, discussed in Sec.~\ref{sec:EFT_IR}. We find that the radial BAO scaling parameter shifts by $+(0.1-0.2)\%$. This series of tests demonstrates how sensitive the results are to exact analysis choices and we leave exploring the optimal modeling for linear theory analyses of the BAO scaling parameters to future work. 

\section{Discussion and conclusions}
\label{sec:disc}

\subsection{Related literature}

Adopting the optimal combination for the isotropic parameter \citep{DESI:2024Lya}:
\begin{equation}
    \alpha_{\rm iso} = \alpha_\perp^{{9}/{20}} \alpha_\parallel^{{11}/{20}}
\end{equation}
and the Alcock-Paczy\'nski parameter:
\begin{equation}
    \alpha_{\rm AP} = \alpha_\perp/\alpha_\parallel ,
\end{equation}
we can convert our main result, the LCV-reduced $\alpha_\parallel$ and $\alpha_\perp$ constraints from the auto-correlation function into constraints on $\alpha_{\rm iso}$ and $\alpha_{\rm AP}$, to obtain shifts of $0.18 \pm 0.05\%$ and $-0.8 \pm 0.2\%$, respectively. These are consistent with the measurements we make from the cross-correlation with quasars, namely: $0.25 \pm 0.12\%$ and $-0.85 \pm 0.25\%$, respectively.

We remark that these values of $\alpha_{\rm iso}$ are markedly different from those reported in another recent work studying the shift of the BAO peak in the \lyaf\ from the isotropic part (monopole) of the signal \citep{2024ApJ...971L..22S}. The authors of that work find a 1\% negative shift to the $\alpha_{\rm iso}$ parameter (albeit with a slightly different parametrization) confirmed at the $\gtrsim 3$$\sigma$ level. While our work is broadly inconsistent with this result, we note that there are several important differences that could play a role in causing this discrepancy. \citet{2024ApJ...971L..22S} adopt a modification of the FGPA approach used to create the $\textsc{AbacusSummit}$ \lyaf\ mocks, which assigns a different connection between gas and dark matter depending on the cosmic web classification of the region. This novel method is motivated by the fact that the different phases of the neutral gas exhibit a different relation with the underlying matter \citep{2009ApJ...701...94F} and could therefore yield more accurate behavior on small scales where these astrophysical effects are highly important. At the same time, it could induce sharp edges in the neutral gas density field, thus 
causing a distortion at the BAO peak scale\footnote{Through private correspondence, it has been established that a standard FGPA implementation of the gas-painting method yields considerably smaller shifts to the $\alpha$ parameter.}. We remind the reader that the shifts we are looking for in this work are at the 0.1\% level, so any small systematic effects could potentially lead to large differences at that level. Related to this is the fact that both works use different fitting techniques to obtain the $\alpha$ distortions. We leave for future work the exploration of a more sophisticated small-scale model.


Our work is similar in spirit also to \citet{deBelsunce:2024rvv}, in which the authors utilize a state-of-the-art hydrodynamical simulation to measure the EFT bias parameters and translate them into $\alpha$ parameters via the EFT of \lyaf\ \citep{Ivanov2024} (a direct fit of the BAO parameters is not possible due to the limited volume of the hydro simulation). They find a -0.2\% ($-0.3\%$) deviation in $\alpha_\parallel$ ($\alpha_\perp$). Whilst the magnitude of the BAO shift is similar to the present work, we note that their sign is opposite to ours. As the BAO parameter shifts are (at leading order) dictated by the values of the EFT bias parameters $b_1$ and $b_2$, it is not too surprising that two very different realizations of the \lyaf\ field would yield different predictions for the $\alpha$ parameters. A more direct connection with this line of work and how to self-consistently model the BAO shift is presented in Section~\ref{sec:eft}. In the future, we plan to address the limitation of employing a single hydro simulation and the FGPA approach when making this comparison.

\subsection{Summary of findings}

In this work, we aim to constrain the size of the shift of the BAO peak due to non-linear structure growth as seen in the \lyaf. In the era of DESI, where the error bars on the BAO parameters from the \lyaf\ measurements will be at the subpercent level, it is of utmost importance to place constraints on the error budget due to various systematic effects. One of these is the intrinsic shift to the $\alpha$ parameters, which if unaccounted for, could lead to a biased inference on the cosmology. In this work, we utilize the large \lyaf\ mocks produced on the $N$-body simulation suite \textsc{AbacusSummit} to measure the BAO shift, which features four different models with different prescriptions. Specifically, we make measurements of the auto- and cross- (with quasars) correlation function, $\xi(r_p, \pi)$, for the six available simulations, each with two lines of sight, along the $z$ and the $y$ directions, for a total of 12 measurements assumed to be independent. We employ the software developed specifically for fitting the DESI \lyaf\ correlation functions as well as the publicly available covariance matrix from the eBOSS DR16 analysis \citep{dMdB2020}. In that way, our work validates both the analysis pipeline of DESI \lyaf\ data and also tests for possible shifts to the BAO parameters. Our main findings can be summarized as follows:
\begin{itemize}
\item To mitigate the noise on these measurements, we also adopt a linear control variates (LCV) technique meant to reduce the noise on large scales for which linear theory yields an accurate prediction of the two-point correlation function. Applying the LCV method to our problem yields tighter measurements of the $\alpha$ parameters both in the auto- and in the cross-correlation case. From the auto-correlation, we detect a small 0.35\% shift at the 3$\sigma$ level to $\alpha_\parallel$ and virtually no shift to $\alpha_\perp$. 
From the cross-correlation, we see a similar shift to $\alpha_\parallel$, albeit with larger error bars, and a small negative shift, $\sim$0.25\%, at the 2$\sigma$ level. Similarly to the raw measurements, we see the same structure of $r_{\rm min}$ dependence that is attributable to the accuracy of the model on small scales. For the explicit values of the shifts, we direct the reader to Table~\ref{tab:alpha_auto_lcv} and Table~\ref{tab:alpha_cross_lcv}.
\item We find that the one-loop \Lya forest EFT model accurately fits the \abacus data. When jointly fitting the BAO and EFT parameters, we find unbiased fits for $\apar$ and $\aperp$ that are consistent with one, i.e., no shift in the BAO parameter. Crucially, a full one-loop EFT treatment mitigates and \emph{automatically includes} this shift into the final parameter constraints, rendering unnecessary to account for this shift in the systematic error budget. 
\end{itemize}

Overall, in this work we confirm the assumption of a number of \lyaf\ analyses \citep[e.g.,][]{DESI:2024Lya} that the BAO peak is shifted by less than 0.5\%. Nonetheless, we recommend that a 0.35\% shift be accounted for in the systematic errors budgeted by the \lyaf\ analysis. Doing more detailed tests with a wide range of \lyaf\ mocks (both applied to $N$-body as well as hydro simulations \citep{deBelsunce:2024rvv}) and improved models that provide an accurate description of the clustering down to small scales (e.g., EFT) will bring great insight and prove invaluable for the analysis of large-scale structure surveys in the future. 

The validation of the EFT fitting procedure on large simulations paired with a quantification that the EFT debiases the BAO fitting paves the way to a number of future analyses such as full-shape analyses of the \Lya forest in Fourier space using the three-dimensional power
spectrum \citep{Font-Ribera:2018, deBelsunce:2024knf,Horowitz:2024nny} with field-level simulation-based priors \citep{2024arXiv240910609I, 2024PhRvD.110f3538I}, beyond BAO analyses using the configuration space correlation function \citep{Cuceu:2021hlk}, or (near optimal compression techniques of) the 3D bispectrum \citep{Belsunce_skewspectrum}.

\section*{Acknowledgements}
The authors thank Stephen Chen, Pat McDonald, Martin White, Mark Maus, and Nathalie Palanque-Delabrouille for fruitful discussions. 

This research used resources of the National Energy Research Scientific Computing Center (NERSC), a U.S. Department of Energy Office of Science User Facility operated under Contract No.~DE–AC02–05CH11231. 

\section*{Data Availability}

We make all our synthetic maps and catalogs publicly available on Globus through NERSC SHARE at this link: \url{https://app.globus.org/file-manager?origin_id=9ce29982-eed1-11ed-9bb4-c9bb788c490e&path=\%2F} under the name ``AbacusSummit Lyman Alpha Forest''. 



\bibliographystyle{mnras}
\bibliography{example} 



\appendix

\section{Control variates validation}
\label{app:lcv}

\begin{figure*}
    \centering
    \includegraphics[width=.32\textwidth]{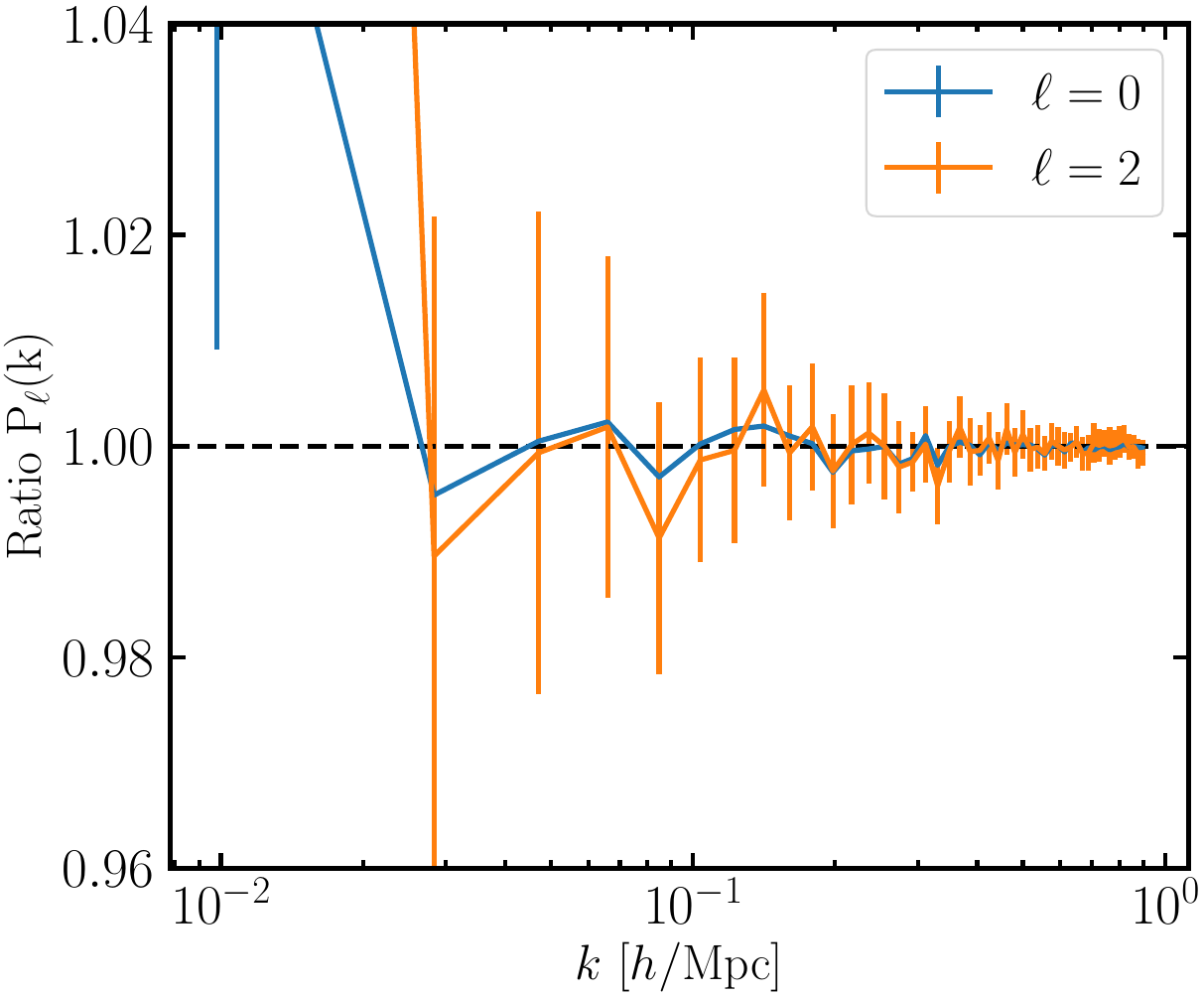}
    \includegraphics[width=.32\textwidth]{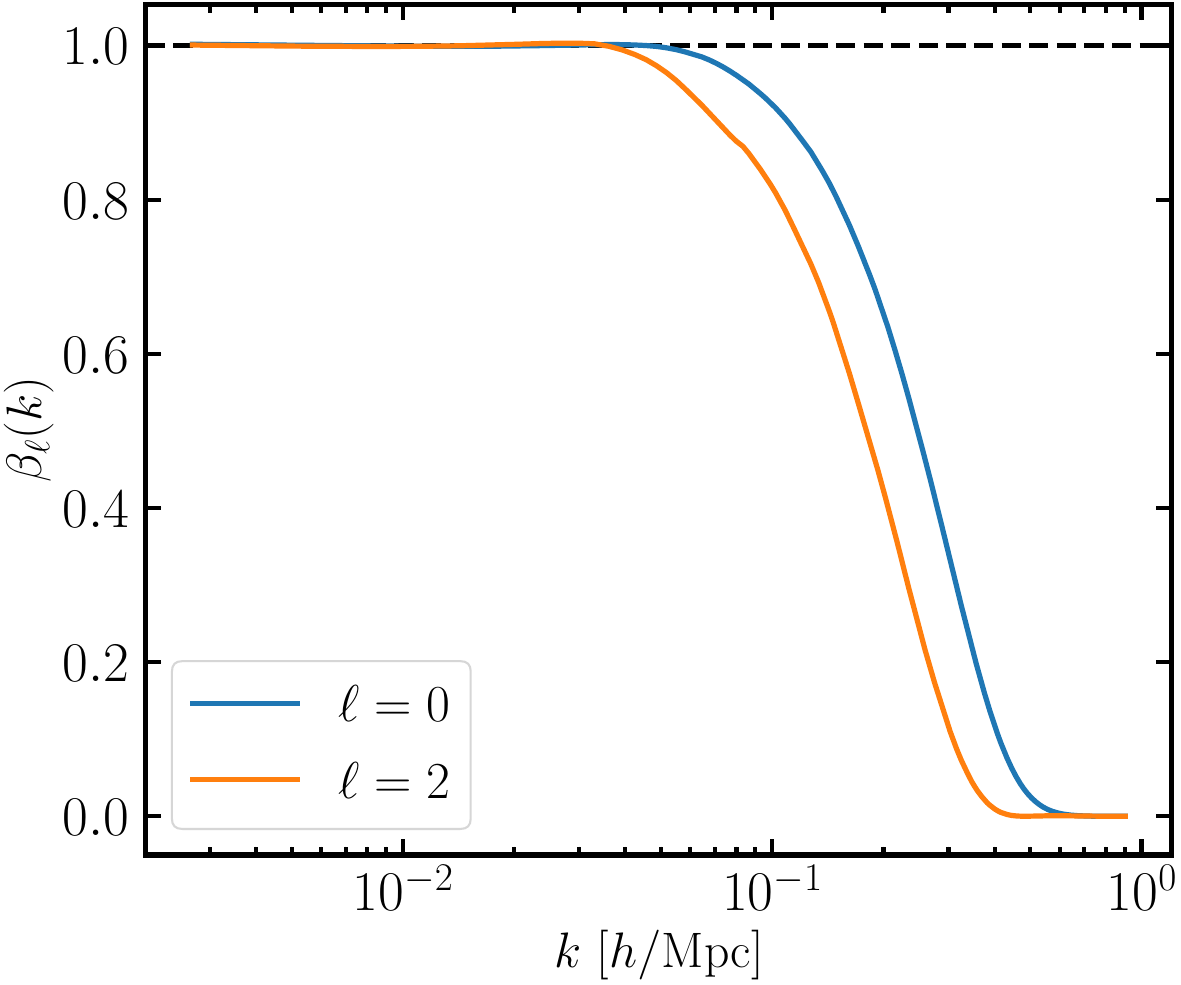}
    \includegraphics[width=.32\textwidth]{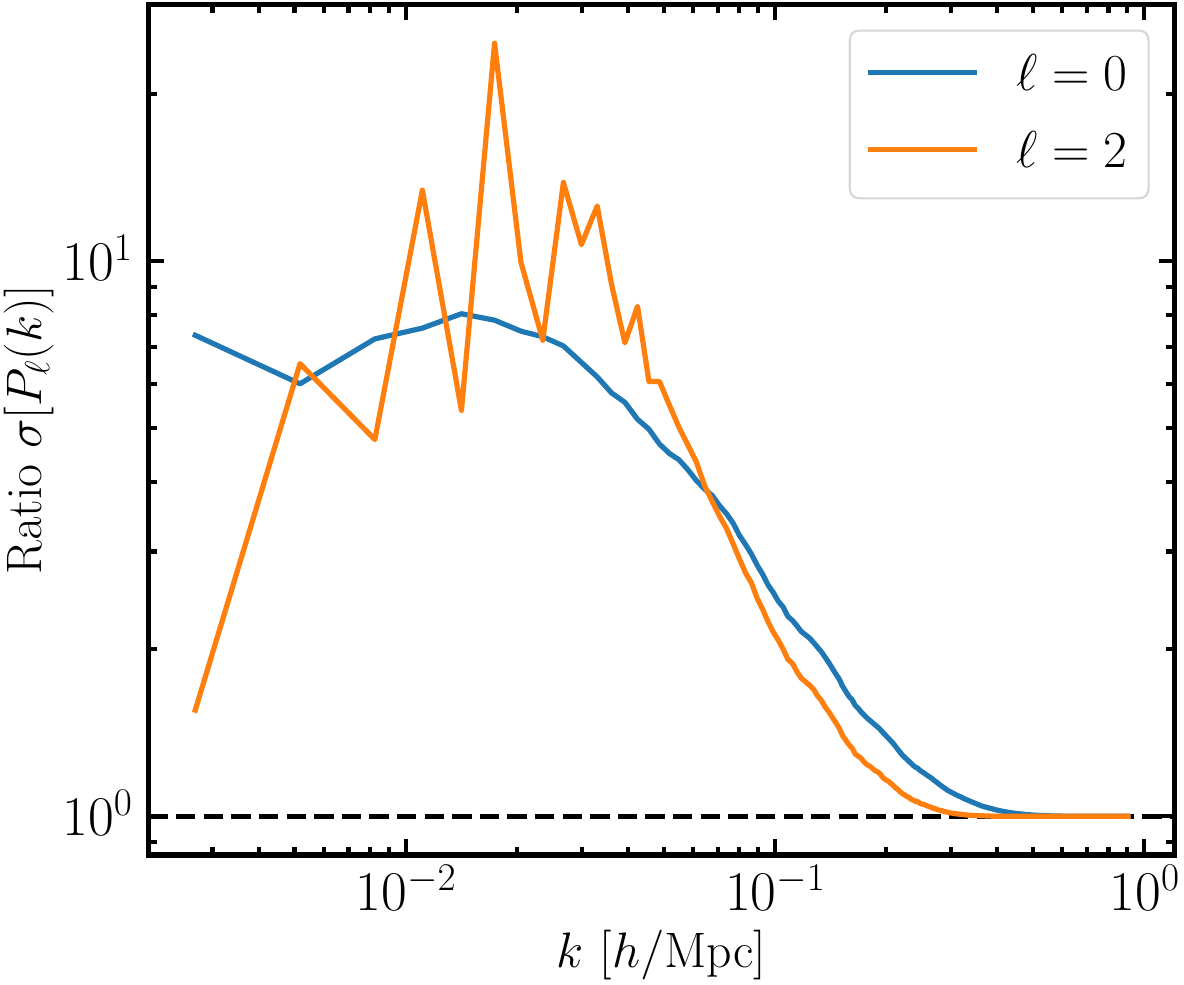}
    \caption{Validation of the linear control variate technique for the \lyaf\ auto-power spectrum. \textit{Leftmost panel:} Ratio of the analytical model fo the control variable (i.e., the linear theory approximation to the auto-power spectrum of the \lyaf). Sub-percent level agreement is seen down to small scales. \textit{Middle panel:} The $\beta$ parameter (see Eq.~\ref{eq:beta}) quantifying the amount of correlation between the control variable and the variable of interest (the auto-power spectrum of the \lyaf\ at $z = 2.5$). The correlation is strongest on large scales up to $k \sim 0.1 \ h/{\rm Mpc}$. \textit{Rightmost panel:} Factor by which the error bars on the monopole and quadrupole of the \lyaf\ auto-power spectrum decrease as a function of scale. The largest reduction of error is on large scales, where the improvement factor is about 7 or 8. The error on the power spectrum is virtually unchanged beyond $k \gtrsim 2 \ h/{\rm Mpc}$.}
    \label{fig:lcv}
\end{figure*}

In this section, we validate the linear control variate (LCV) technique by studying three important metrics in the auto-correlation function: the ratio between theory and simulation, which needs to be unbiased for the technique to work; the coefficient $\beta_\ell(k)$, which gives us the scales over which LCV error mitigation is applied (for more details, see Section~\ref{sec:lcv}); the error reduction as a result of adopting LCV.  

The left panel of Fig.~\ref{fig:lcv} shows the ratio of the analytical model fo the control variable (i.e., the linear theory approximation to the auto-power spectrum of the \lyaf). The fact that we see perfect sub-percent level agreement down to very small scales ($k \sim 1 \ h/{\rm Mpc}$) assures us that no bias to the two-point correlation function can be introduced through the LCV method. 

The middle panel displays the $\beta$ parameter (see Eq.~\ref{eq:beta}) quantifying the amount of correlation between the control variable and the variable of interest (the auto-power spectrum of the \lyaf\ at $z = 2.5$). We see that the correlation is strongest on large scales up to $k \sim 0.1 \ h/{\rm Mpc}$. Beyond that the linear theory approximation breaks down and we can no longer rely on this technique to reduce the variance fo the measured signal. Additionally, we see that the theoretical model does a slightly better job of reproducing the monopole, which is expected as redshift-space effects due to the non-linear clustering can propagate to larger scales (as seen in the $\beta_{\ell=2}(k)$ curve). 

Finally, in the rightmost panel, we show the factor by which the error bars on the monopole and quadrupole of the \lyaf\ auto-power spectrum decrease as a function of scale. Looking at the monopole, we see that the largest reduction of error is on large scales, where the improvement factor is about 7 or 8. The error on the power spectrum is virtually unchanged beyond $k \gtrsim 2 \ h/{\rm Mpc}$. The quadrupole curve, on the other hand, is very noisy, but it follows the same trend. This is expected as linear theory describes the clustering most accurately on large scales.

\section{Effect of adopting broadband parameters}
\label{app:bb}

As specified in Section~\ref{sec:vega}, for our default analysis we perform a fit to the clustering data with a model featuring broadband polynomial parameters, aimed to make our model immune to limitations on small scales. In particular, the model employed in this study is the so-called `Arinyo model,' which is motivated by measurements of the clustering performed in a hydro simulation. In the main body of the paper, we fix the Arinyo parameters to their fiducial values and include a comprehensive set of broadband parameters, which we vary. In this Appendix, we are interested in exploring the case where we do not include broadband polynomial templates and instead simply perform our fits with the default Arinyo parameters.

\begin{figure*}
    \centering
    \includegraphics[width=\textwidth]{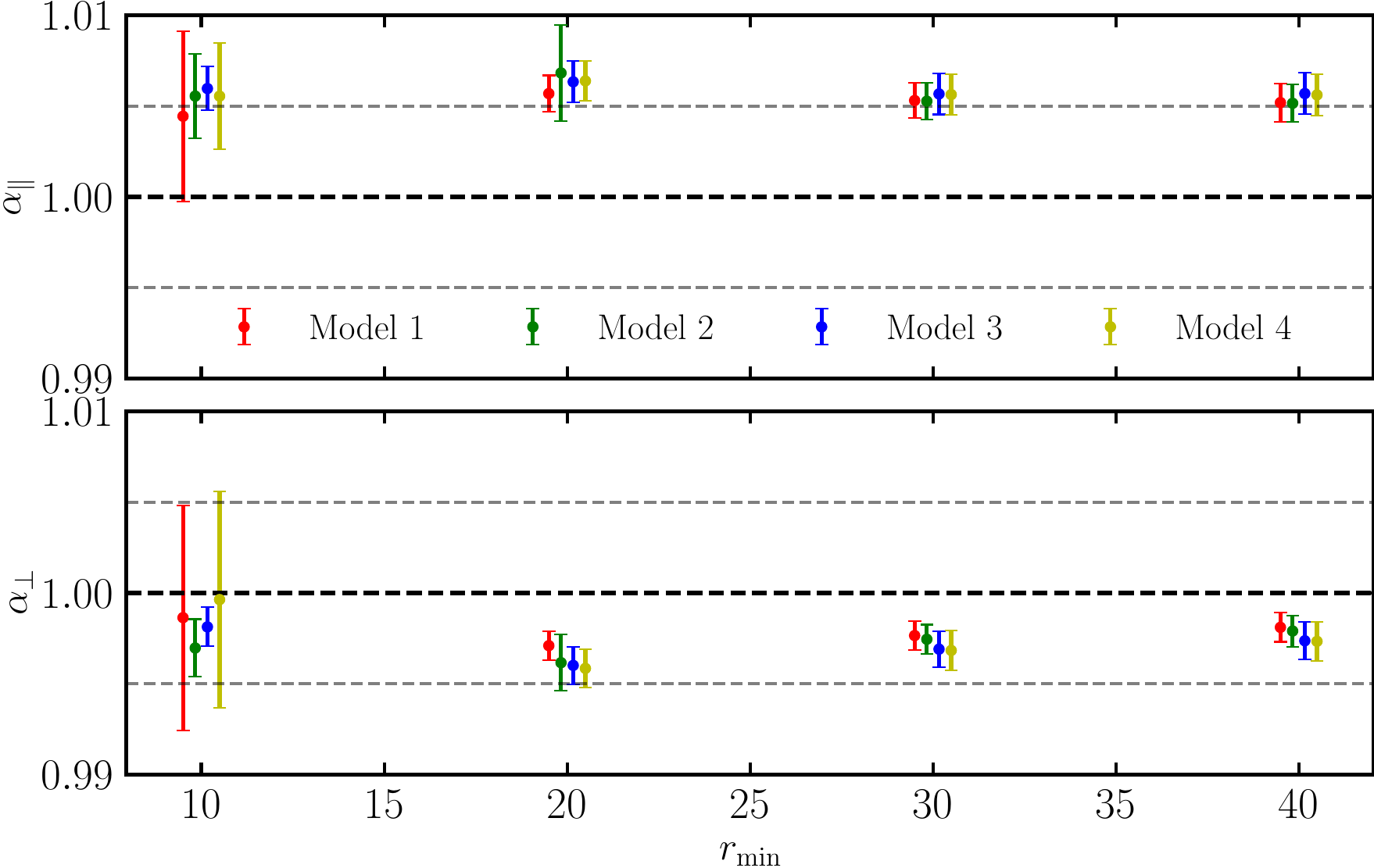}
    \caption{Similar to Fig.~\ref{fig:rmin_auto_lcv}. The difference is that here, we have switched off the broadband parameters, and are instead adopting the `Arinyo' model with parameters of the model fixed to their fiducial values. We see that the values of the $\alpha$ parameters are largely consistent with the fiducial analysis, which is encouraging. At the same time, the fits shown here exhibit a stronger dependence on $r_{\rm min}$ and are more noisy.}
    \label{fig:rmin_auto_lcv_nobb}
\end{figure*}

In Fig.~\ref{fig:rmin_auto_lcv_nobb}, we see that the values of $\alpha_\parallel$ and $\alpha_\perp$ are broadly consistent, both in terms of the mean values as well as  their error bars, with the values in Fig.~\ref{fig:rmin_auto_lcv}, which is reassuring. Nonetheless, the fits do appear to be more noisy and also show a stronger dependence on $r_{\rm min}$ than their counterparts using the broadband parametrization. This shows that the broadband model manages to absorb well many of the issues on small scales. 
Upon visual inspection of the fitted correlation functions, we furthermore find that when using only the Arinyo parameters without the broadband templates, the fits are noticeably poorer. This is another reason we opt to keep the Arinyo parameters fixed and vary the broadband parameters. We also tried varying the Arinyo model parameters, dropping the broadband parameters, and found very similar results to our default case. However, we stress that this model is not available for modeling the cross-correlation function, and for this reason, we opt not to use it. In the future, we plan to study extensions and alternatives to the Arinyo model in more detail, exploring comparisons with more physically motivated frameworks such as EFT \citep{Ivanov2024}.

\section{Triangle plots for EFT fits to auto- and cross-correlation}\label{app:triangle_plots}
\begin{figure*}
    \centering
    \includegraphics[width=\linewidth]{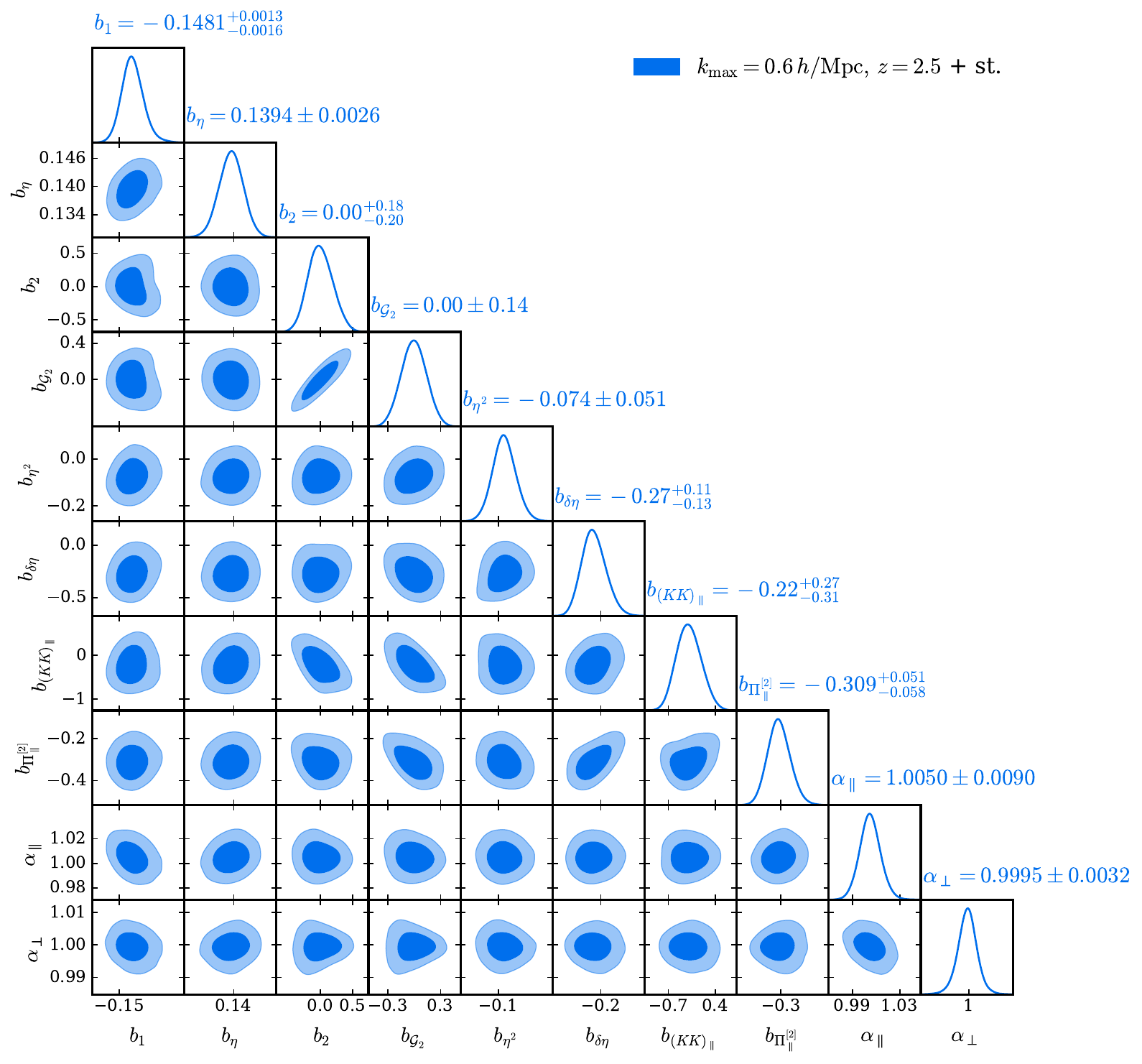}
    \caption{Triangle plot of the EFT bias and BAO scaling parameters obtained from fitting simulation ``one'' and model 1. The corresponding numerical values are given in Table~\ref{tab:EFT_bias}. We show the best-fit 1D and 2D marginalized posteriors for the \textit{sampled} parameters. } \label{fig:EFT_triangle}
\end{figure*}

\begin{figure*}
    \centering
    \includegraphics[width=\linewidth]{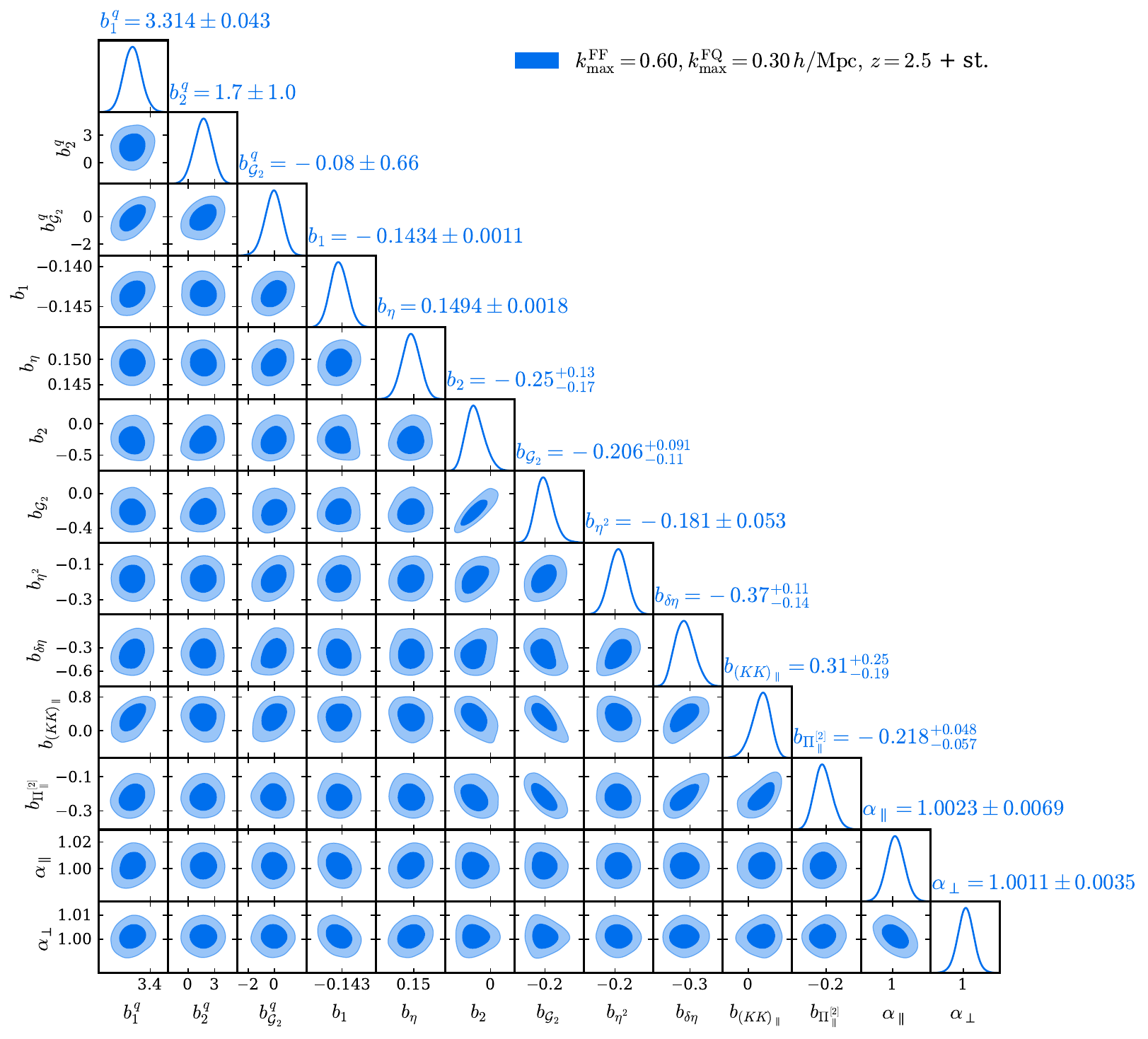}
    \caption{Triangle plot of the EFT bias and BAO scaling parameters obtained from jointly fitting the auto- and cross-correlation of simulation ``one'' and model 1. The corresponding numerical values are given in Table~\ref{tab:EFT_bias_cross}. We show the best-fit 1D and 2D marginalized posteriors for the \textit{sampled} parameters. }
    \label{fig:EFT_triangle_cross}
\end{figure*}

In this appendix, we show the full marginalized posteriors for the sampled BAO and EFT bias parameters for model 1, simulation one and line-of-sight $y$. In Fig.~\ref{fig:EFT_triangle} we show the results for the auto-correlation. The 1D marginalized posteriors are highly Gaussian and we only see clear degeneracies between $b_{\mathcal{G}_2}-b_2$ and  $b_{\Pi_{\parallel}^{[2]}}-b_{\delta \eta}$. The picture is consistent when jointly fitting the auto- and cross-correlation, shown in Fig.~\ref{fig:EFT_triangle_cross}.

\begin{table*}
\centering
\begin{tabular}{lcccc}
\hline \hline
$b_{\mathcal{O}}$ & Model 1 & Model 2 & Model 3 & Model 4 \\ [0.5ex]
\hline
$b_1$ & $-0.1485 \pm 0.0006$ & $-0.1314^{+0.0007}_{-0.0006}$ & $-0.1337 \pm 0.0005$ & $-0.1295 \pm 0.0005$ \\[1ex]
$b_{\eta}$ & $0.1423 \pm 0.0012$ & $0.1295 \pm 0.0011$ & $0.2705 \pm 0.0013$ & $0.3041 \pm 0.0014$ \\[1ex]
$b_2$ & $-0.1231 \pm 0.0942$ & $-0.3316^{+0.0897}_{-0.0770}$ & $-0.0353 \pm 0.0505$ & $-0.0400 \pm 0.0483$ \\[1ex]
$b_{\mathcal{G}_2}$ & $-0.0847 \pm 0.0701$ & $-0.2765^{+0.0651}_{-0.0482}$ & $-0.0383 \pm 0.0361$ & $-0.0320 \pm 0.0351$ \\[1ex]
$b_{\eta^2}$ & $-0.1088 \pm 0.0300$ & $-0.2631^{+0.0244}_{-0.0223}$ & $-0.0811 \pm 0.0271$ & $-0.0385 \pm 0.0275$ \\[1ex]
$b_{\delta\eta}$ & $-0.3771^{+0.0557}_{-0.0568}$ & $-0.3511^{+0.0610}_{-0.0743}$ & $-0.0633 \pm 0.0429$ & $-0.0167 \pm 0.0422$ \\[1ex]
$b_{KK_\parallel}$ & $-0.0751^{+0.1499}_{-0.1329}$ & $0.4841^{+0.1285}_{-0.1773}$ & $-0.0164 \pm 0.0868$ & $0.0327^{+0.0883}_{-0.0864}$ \\[1ex]
$b_{\Pi^{[2]}_\parallel}$ & $-0.3046 \pm 0.0339$ & $-0.2943 \pm 0.0473$ & $-0.2429 \pm 0.0353$ & $-0.2578 \pm 0.0379$ \\[1ex]
\hline
$b_{\Pi^{[3]}_\parallel}$ & $0.1155 \pm 0.0216$ & $0.4038 \pm 0.0192$ & $0.7470 \pm 0.0200$ & $0.7984 \pm 0.0200$ \\[1ex]
$b_{\delta\Pi^{[2]}_\parallel}$ & $-1.1613 \pm 0.0614$ & $1.5483 \pm 0.0553$ & $-1.5111 \pm 0.0800$ & $-2.0542 \pm 0.0846$ \\[1ex]
$b_{(K\Pi^{[2]})_\parallel}$ & $-0.7105 \pm 0.0694$ & $-1.2533 \pm 0.0622$ & $-1.1988 \pm 0.0675$ & $-1.7604 \pm 0.0676$ \\[1ex]
$b_{\eta\Pi^{[2]}_\parallel}$ & $-0.8287 \pm 0.1952$ & $6.0565 \pm 0.1771$ & $-5.5202 \pm 0.2661$ & $-6.8086 \pm 0.2816$ \\[1ex]
$b_{\GG}$ & $0.1699 \pm 0.0365$ & $1.3522 \pm 0.0329$ & $-1.0584 \pm 0.0429$ & $-1.1835 \pm 0.0445$ \\[1ex]
$P_{\rm shot}$ & $-0.6533 \pm 0.0439$ & $-0.1932 \pm 0.0337$ & $0.2082 \pm 0.0181$ & $0.0885 \pm 0.0160$ \\[1ex]
$a_0$ & $3.3854 \pm 0.3902$ & $2.2565 \pm 0.3119$ & $2.2597 \pm 0.3040$ & $2.5358 \pm 0.2903$ \\[1ex]
$a_2$ & $-3.4240 \pm 1.2475$ & $-4.9955 \pm 1.0051$ & $-11.3607 \pm 1.4209$ & $-12.4223 \pm 1.4650$ \\[1ex]
$c_0$ & $0.1763 \pm 0.0150$ & $0.0155 \pm 0.0129$ & $-0.1859 \pm 0.0073$ & $-0.1412 \pm 0.0069$ \\[1ex]
$c_{2}$ & $-0.0342 \pm 0.0277$ & $0.1095 \pm 0.0248$ & $0.5097 \pm 0.0275$ & $0.5065 \pm 0.0277$ \\[1ex]
$c_{4}$ & $-0.1394 \pm 0.0123$ & $-0.1347 \pm 0.0110$ & $-0.4875 \pm 0.0156$ & $-0.5160 \pm 0.0164$ \\[1ex]
$\chi^2$ & $726.4$ & $933.1$ & $1032.7$ & $1043.3$ \\[1ex]
\hline
\end{tabular}
\caption{EFT bias parameters for each model obtained from fits to the stacked data vector of 12 simulations for $1175$ data points. Table~\ref{tab:EFT_bias} is the corresponding table for fits to the individual simulations. Note that the average error bars are divided by $\sqrt{6}$. Given the small error bars, we can differentiate between the linear bias parameters in addition to the $b_2$ parameter  }\label{tab:EFT_bias_meandatavector}
\end{table*}

\begin{table*}
\centering
\begin{tabular}{lcccc}
\hline
\hline
$b_{\mathcal{O}}$ & Model 1 & Model 2 & Model 3 & Model 4 \\ [0.5ex]
\hline
$b_1^{\rm q}$           & $3.3160 \pm 0.0435$ & $3.3146 \pm 0.0441$ & $3.3730 \pm 0.0368$ & $3.3744 \pm 0.0359$ \\[1ex]
$b_2^{\rm q} $          & $1.5437 \pm 0.9904$ & $1.1257 \pm 0.9936$ & $1.8083 \pm 1.0180$ & $1.9024^{+0.9954}_{-1.0921}$ \\[1ex]
$b_{\mathcal{G}_2}^{\rm q}$          & $0.0605 \pm 0.6385$ & $-0.6220^{+0.6124}_{-0.7404}$ & $0.1710 \pm 0.4942$ & $-0.0383 \pm 0.4716$ \\[1ex]
$b_1$              & $-0.1465^{+0.0012}_{-0.0011}$ & $-0.1294 \pm 0.0011$ & $-0.1316 \pm 0.0009$ & $-0.1284 \pm 0.0009$ \\[1ex]
$b_{\eta}$          & $0.1448 \pm 0.0018$ & $0.1315 \pm 0.0017$ & $0.2745 \pm 0.0022$ & $0.3102 \pm 0.0024$ \\[1ex]
$b_2$          & $-0.3613^{+0.1664}_{-0.1298}$ & $-0.4283^{+0.1288}_{-0.0855}$ & $-0.1961^{+0.1188}_{-0.1048}$ & $-0.0885 \pm 0.1060$ \\[1ex]
$b_{\mathcal{G}_2}$           & $-0.2598^{+0.1140}_{-0.0897}$ & $-0.2186^{+0.1304}_{-0.0833}$ & $-0.0862 \pm 0.0705$ & $-0.0455 \pm 0.0671$ \\[1ex]
$b_{\eta^2}$          & $-0.2032 \pm 0.0525$ & $-0.3448^{+0.0531}_{-0.0452}$ & $-0.1961 \pm 0.0483$ & $-0.1837 \pm 0.0498$ \\[1ex]
$b_{\delta\eta}$    & $-0.4041^{+0.1402}_{-0.1163}$ & $-0.6267^{+0.1596}_{-0.0856}$ & $-0.2108^{+0.0811}_{-0.0710}$ & $-0.1393^{+0.0763}_{-0.0681}$ \\[1ex]
$b_{KK_\parallel}$        & $0.4594^{+0.1919}_{-0.2422}$ & $0.3074^{+0.2361}_{-0.2866}$ & $0.4538^{+0.1286}_{-0.1548}$ & $0.3918^{+0.1279}_{-0.1427}$ \\[1ex]
$b_{\Pi^{[2]}_\parallel}$      & $-0.2123^{+0.0577}_{-0.0486}$ & $-0.2376^{+0.0532}_{-0.0429}$ & $-0.2407 \pm 0.0531$ & $-0.2078 \pm 0.0571$ \\[1ex]
\hline
$b_{\Pi^{[3]}_\parallel}$ 
        & $-0.2223 \pm 0.0313$ & $-0.2287 \pm 0.0295$ & $0.2143 \pm 0.0341$ & $0.2681 \pm 0.0353$ \\[1ex]
$b_{\delta\Pi^{[2]}_\parallel}$     & $-1.7482 \pm 0.1283$ & $-1.3608 \pm 0.1184$ & $-1.4153 \pm 0.1267$ & $-1.6799 \pm 0.1298$ \\[1ex]
$b_{(K\Pi^{[2]})_\parallel}$       & $-1.7220 \pm 0.1328$ & $-1.5296 \pm 0.1244$ & $-1.8891 \pm 0.1107$ & $-2.2383 \pm 0.1098$ \\[1ex]
$b_{\eta\Pi^{[2]}_\parallel}$     & $-1.2416 \pm 0.3383$ & $-0.2563 \pm 0.3164$ & $-3.2567 \pm 0.3889$ & $-3.5330 \pm 0.4075$ \\[1ex]
$b_{\GG}$          & $0.7481 \pm 0.0628$ & $0.9753 \pm 0.0589$ & $-0.2344 \pm 0.0680$ & $-0.2553 \pm 0.0706$ \\[1ex]
$P_{\rm shot}$     & $-0.3132 \pm 0.0732$ & $-0.0226 \pm 0.0569$ & $0.4352 \pm 0.0214$ & $0.3174 \pm 0.0177$ \\[1ex]
$a_0$     & $-0.8737 \pm 3.1579$ & $-1.4900 \pm 2.7583$ & $-4.6427 \pm 2.9631$ & $-3.6087 \pm 2.8547$ \\[1ex]
$a_2$           & $0.4249 \pm 4.7041$ & $-0.8907 \pm 4.6365$ & $-0.9101 \pm 4.7460$ & $-1.2032 \pm 4.8208$ \\[1ex]
$c_0$            & $0.1571 \pm 0.0314$ & $0.0252 \pm 0.0275$ & $-0.1853 \pm 0.0142$ & $-0.1364 \pm 0.0135$ \\[1ex]
$c_2$            & $-0.1610 \pm 0.0362$ & $-0.0280 \pm 0.0341$ & $0.2055 \pm 0.0310$ & $0.1768 \pm 0.0316$ \\[1ex]
$c_4$            & $-0.1091 \pm 0.0222$ & $-0.1539 \pm 0.0205$ & $-0.3657 \pm 0.0232$ & $-0.3803 \pm 0.0243$ \\[1ex]
$c_x$            & $-0.9703 \pm 2.8711$ & $1.6564 \pm 2.6310$ & $-6.0331 \pm 2.2181$ & $-6.3074 \pm 2.0911$ \\[1ex]
$c_0^{\rm q}$           & $0.0726 \pm 0.0102$ & $0.0961 \pm 0.0092$ & $-0.1810 \pm 0.0099$ & $-0.2006 \pm 0.0097$ \\[1ex]
$c_2^{\rm q}$         & $0.2795 \pm 0.9659$ & $0.1858 \pm 0.9687$ & $0.5637 \pm 0.9728$ & $0.5738 \pm 0.9829$ \\[1ex]
$\chi^2$     & $1790.6621$ & $1842.2431$ & $1892.3472$ & $1898.7141$ \\
\hline
\end{tabular}
\caption{Same as Table~\ref{tab:EFT_bias_meandatavector} for the joint fits of the auto- and cross-correlation with 1,755 data points.}
\label{tab:EFT_bias_meandatavector_cross}
\end{table*}

\bsp	
\label{lastpage}
\end{document}